\documentclass[imslayout,preprint]{imsart} 
\def\journal@name{} 
\usepackage[margin=1.5in]{geometry}

\usepackage{mathtools}
\usepackage[colorlinks,citecolor=blue,urlcolor=black,linkcolor=blue,pdfborder={0 0 0}]{hyperref}
\usepackage{url}
\usepackage{comment}
\usepackage{graphicx}
\usepackage{enumitem}
\usepackage{framed}
\usepackage{dcolumn}
\usepackage{subcaption}
\usepackage[sort]{natbib}
\usepackage[capitalize]{cleveref}  
\usepackage{datetime}
\usepackage{algorithmic}
\usepackage[linesnumbered,ruled,vlined]{algorithm2e}
\DontPrintSemicolon
\usepackage[normalem]{ulem}
\usepackage{booktabs}
\usepackage{rotating}

\crefname{nlem}{Lemma}{Lemmas}
\crefname{nprop}{Proposition}{Propositions}
\crefname{ncor}{Corollary}{Corollaries}
\crefname{nthm}{Theorem}{Theorems}
\crefname{nnon}{Conjecture}{Conjectures}
\crefname{assumption}{Assumption}{Assumptions}
\crefname{algorithm}{Algorithm}{Algorithms}
\crefname{exa}{Example}{Examples}

\RequirePackage[usenames,dvipsnames]{color}

\usepackage{jhh-misc2}
\usepackage{mathrsfs}
\usepackage{autonum}

\usepackage{crossreftools}
\newcommand{\logistic}{\phi}

\newcommand{\littleo}{o}
\newcommand{\littleoP}{\littleo_{P}}

\newcommand{\bigo}{O}
\newcommand{\bigoP}{\bigo_{P}}
\newcommand{\bigoPouter}{\bigo_{P^{+}}}

\newcommand{\optsym}{\circ}

\newcommand{\obsspace}{\mathbb X}
\newcommand{\numobs}{N}
\newcommand{\data}{x}
\newcommand{\datarv}{X}

\newcommand{\datarvarg}[1]{X_{1:#1}}

\newcommand{\alldatarv}{\datarvarg{\infty}}

\newcommand{\obs}[1]{x_{#1}}
\newcommand{\obsrv}[1]{X_{#1}}

\newcommand{\postdist}[1]{\Pi_{#1}}
\newcommand{\postdistfull}[2]{\Pi(#1 \given #2)}

\newcommand{\priordist}{\postdist{0}}

\newcommand{\likdist}[1]{P_{#1}}
\newcommand{\likfun}[1]{p_{#1}}
\newcommand{\lik}[2]{\likfun{#2}(#1)}
\newcommand{\loglik}[2]{\ell_{#2}(#1)}
\newcommand{\loglikfun}[1]{\ell_{#1}}

\newcommand{\marginallikfull}[1]{p(#1)}

\newcommand{\param}{\theta}
\newcommand{\paramspace}{\Theta}

\newcommand{\optparam}{\param_{\optsym}}
\newcommand{\paramsample}{\vartheta}

\newcommand{\bbparamsample}{\paramsample^{\bbsym}}

\newcommand{\modelspace}{\mathfrak{M}}
\newcommand{\model}{\mathfrak{m}}

\newcommand{\modelmarginallik}[2]{\marginallikfull{#1 \given #2}}
\newcommand{\modelpriordist}{Q_{0}}
\newcommand{\modelpostdistfull}[2]{Q(#1 \given #2)}

\newcommand{\lmldiff}[1]{\Lambda_{#1}}
\newcommand{\optmodelparam}[1]{\param_{#1 \optsym}}

\newcommand{\Ehessloglik}[1]{J_{#1}}

\newcommand{\obsdist}{P_{\optsym}}

\newcommand{\bbsym}{*} 
\newcommand{\bsnumobs}{M}

\newcommand{\bscount}[1]{K_{#1}}

\newcommand{\bsobs}[1]{\obs{#1}^{\bbsym}}
\newcommand{\bsobsrv}[1]{\obsrv{#1}^{\bbsym}}
\newcommand{\bsdata}{\data^{\bbsym}}
\newcommand{\bsdatarv}{\datarv^{\bbsym}}
\newcommand{\bsdatasample}[1]{\data^{\bbsym}_{(#1)}}

\newcommand{\bsdatarvarg}[1]{\bsdatarv_{1:#1}}

\newcommand{\bbpostdistfull}[2]{\Pi^{\bbsym}(#1 \given #2)}

\newcommand{\bbmodelpostdistfull}[2]{Q^{\bbsym}(#1 \given #2)}

\newcommand{\bsscale}{c}

\newcommand{\modelmismatch}{\mathcal{I}}
\newcommand{\nan}{\textsf{NA}}
\newcommand{\concconst}{C_{\numobs}}
\newcommand{\effectsize}{\eta_{\infty}}

\newcommand{\dsupconvex}{d_{\mathcal{C}}}

\def\norm#1{\left\|{#1}\right\|} 
\newcommand{\twonorm}[1]{\norm{#1}_2} 

\graphicspath{{figures/}}

\allowdisplaybreaks[2]

\begin{document}

\begin{frontmatter}

\title{Reproducible Model Selection \\ Using Bagged Posteriors}
\runtitle{~Reproducible Model Selection Using Bagged Posteriors}
\runauthor{J.\ H.\ Huggins and J.\ W.\ Miller~}

\begin{aug}
\author[A]{\fnms{Jonathan H.} \snm{Huggins}\ead[label=e1]{huggins@bu.edu}} 
\and
\author[B]{\fnms{Jeffrey W.} \snm{Miller}\ead[label=e2]{jwmiller@hsph.harvard.edu}}

\address[A]{Department of Mathematics \& Statistics, Boston University, \printead{e1}}
\address[B]{Department of Biostatistics, Harvard University, \printead{e2}}

\end{aug}

\begin{abstract}
Bayesian model selection is premised on the assumption that the data are generated from one of the postulated models.
However, in many applications, all of these models are incorrect (that is, there is misspecification). 
When the models are misspecified, two or more models can provide a nearly equally good fit to the data,
in which case Bayesian model selection can be highly unstable, potentially leading to self-contradictory findings.
To remedy this instability, we propose to use bagging on the posterior distribution (``BayesBag'') --
that is, to average the posterior model probabilities over many bootstrapped datasets.
We provide theoretical results characterizing the asymptotic behavior of the posterior and the bagged posterior 
in the (misspecified) model selection setting.
We empirically assess the BayesBag approach on synthetic and real-world data in
(i) feature selection for linear regression and (ii) phylogenetic tree reconstruction.
Our theory and experiments show that, when all models are misspecified, BayesBag (a) provides greater reproducibility
and (b) places posterior mass on optimal models more reliably, compared to the usual Bayesian posterior;
on the other hand, under correct specification, BayesBag is slightly more conservative than the usual posterior, 
in the sense that BayesBag posterior probabilities tend to be slightly farther from the extremes of zero and one. 
Overall, our results demonstrate that BayesBag provides an easy-to-use and widely applicable approach that improves upon
 Bayesian model selection by making it more stable and reproducible. 
\end{abstract}


\end{frontmatter}

\section{Introduction} \label{sec:introduction}

In Bayesian statistics, the usual method of quantifying uncertainty in the choice of model is simply to use the posterior distribution over models.
An implicit assumption of this approach is that one of the assumed models is exactly correct.
But it is widely recognized that in practice, this assumption is typically unrealistic.
When all of the models are incorrect (that is, they are \emph{misspecified}), the posterior concentrates on the model that 
provides the best fit in terms of Kullback--Leibler divergence \citep{Berk:1966}. %
However, when two or more models can explain the data almost equally well, 
the posterior becomes unstable and can yield contradictory results when seemingly inconsequential 
changes are made to the models or to the data \citep{Yang:2018,Meng:2019,Oelrich:2020}. 
For instance, as the size of the data set grows, the posterior probability of a given model may oscillate between values very close to 1 and very close to 0, \textit{ad infinitum}.
In short, Bayesian model selection can be unreliable and non-reproducible. 

This article develops the theory and practice of \emph{BayesBag}, a simple and widely applicable approach to stabilizing Bayesian model selection. 
Originally suggested by \citet{Waddell:2002} and \citet{Douady:2003} in the context of phylogenetic inference and then independently proposed by~\citet{Buhlmann:2014} (who coined the name), the idea of BayesBag is to apply bagging~\citep{Breiman:1996} to the Bayesian posterior. 
Let $\modelpostdistfull{\model}{\data} \propto \modelmarginallik{\data}{\model}\modelpriordist(\model)$ denote the posterior probability of model $\model \in \modelspace$
given data $\data$, where $\modelspace$ is a finite or countably infinite set of models,  $\modelmarginallik{\data}{\model}$ is the marginal likelihood, 
and $\modelpriordist(\model)$ is the prior probability. 
We define the \emph{bagged posterior} $\bbmodelpostdistfull{\model}{\data}$ by taking bootstrapped copies $\bsdata \defined (\bsobs{1},\ldots,\bsobs{\bsnumobs})$ 
of the original dataset $\data \defined (\obs{1},\ldots,\obs{\numobs})$ and averaging over the posteriors obtained 
by treating each bootstrap dataset as the observed data -- that is, 
\[ \label{eq:bayesbag-definition-model-selection}
\bbmodelpostdistfull{\model}{\data} 
\defined \frac{1}{\numobs^{\bsnumobs}}\sum_{\bsdata} \modelpostdistfull{\model}{\bsdata},
\]
where the sum is over all possible $N^M$ bootstrap datasets of $\bsnumobs$ samples drawn 
with replacement from the original dataset.
The BayesBag approach is to use $\bbmodelpostdistfull{\model}{\data}$ to quantify uncertainty in the model $\model$.
In practice, we can approximate $\bbmodelpostdistfull{\model}{\data}$ by generating $B$ bootstrap datasets $\bsdatasample{1}$, $\dots$, $\bsdatasample{B}$,
where each $\bsdatasample{b}$ consists of $\bsnumobs$ samples drawn with replacement from $\data$, yielding the approximation
\[
\bbmodelpostdistfull{\model}{\data}  \approx \frac{1}{B} \sum_{b=1}^{B}  \modelpostdistfull{\model}{\bsdatasample{b}}.  \label{eq:bayesbag-approximation}
\]
Hence, BayesBag is easy to use since the bagged posterior model probability is simply an average over Bayesian model probabilities.
No additional algorithmic tools are needed beyond what a data analyst would normally use for posterior inference. 
Implementing BayesBag via \cref{eq:bayesbag-approximation} does require more computation since one 
must approximate $B$ posteriors (one for each bootstrap dataset), where typically $B \approx$ 100.
However, this drawback is minimized by the fact that each posterior can be approximated in parallel, which is ideal for modern cluster-based high-performance computing environments. 

Despite its attractive features, there has been limited methodological or theoretical work on BayesBag prior to the present paper.
\citet{Buhlmann:2014} and \citet{Huggins:2019:BayesBagI} consider BayesBag in the parameter inference and prediction setting.
In this paper, we focus on the use of BayesBag for model selection, which has been explored empirically in an application to phylogenetic tree reconstruction \citep{Waddell:2002,Douady:2003}.
Building off this previous work, our primary contributions are:
\benum
\item	We develop a rigorous asymptotic theory showing that, when all models are misspecified and two or more models have similar predictive accuracy,
	Bayesian model selection is unstable, while BayesBag model selection remains stable. 
	Our analysis quantifies the effects of the relevant factors such as the mean and variance of the log-likelihood ratios and the correlation structure of the log-likelihoods. 
\item We provide concrete guidance on selecting the bootstrap dataset size $\bsnumobs$ and, via our theory, we clarify the effect of $\bsnumobs$ on the stability of BayesBag model selection. 
\item We verify through numerical experiments on synthetic and real data that, when all of the models are misspecified, BayesBag model selection leads to more
	stable inferences across datasets and small model changes, while Bayesian model selection is unstable. 
	When one of the models is correctly specified, BayesBag is slightly more conservative than Bayesian model selection, 
	in the sense that the bagged posterior probabilities tend to be slightly farther from zero and one. 
\eenum
In short, we find that in the presence of misspecification, model selection with the bagged posterior has appealing statistical properties 
while also being easy to use and computationally tractable on practical problems. 

The paper is organized as follows.
\cref{sec:overview} provides an overview of our theory, methodology, and experiments, and how they relate to previous work. 
In \cref{sec:model-selection}, we present our theoretical results, illustrate the theory graphically, discuss the use of BayesBag for model criticism,
and outline our recommended workflow.
\cref{sec:linreg-model-selection-sim} contains a simulation study using BayesBag for feature selection in linear regression.
In \cref{sec:experiments}, we evaluate BayesBag on real-world data in applications involving
(i) feature selection for linear regression and (ii) phylogenetic tree reconstruction. 
We conclude in \cref{sec:discussion} with a discussion of current limitations and future directions.

\section{Summary of results} \label{sec:overview}

\subsection{Theory} 
It has long been known that when the best fit to the data distribution is attained by more than one model,
the posterior typically does not converge on a single model \citep{Berk:1966}.
In \cref{thm:model-selection-asymptotic-posteriors,thm:many-model-selection-asymptotic-posteriors},
we characterize the asymptotic distribution of the posterior on models
in this setting, for both the usual posterior (``Bayes'') and the bagged posterior (``BayesBag'').
More generally, our theory covers the case of multiple misspecified models with approximately equally good fit.

Suppose the observed data $\obs{1},\ldots,\obs{\numobs}$ are realizations of 
independent and identically distributed (i.i.d.) random variables $\obsrv{1},\ldots,\obsrv{\numobs}\in\obsspace$,
and denote $\datarvarg{\numobs} = (\obsrv{1},\ldots,\obsrv{\numobs})$.
First, consider the special case of two distinct models, $\modelspace = \{\model_{1}, \model_{2}\}$.
Assume these models are asymptotically equally misspecified in the sense that 
\[ 
\lim_{\numobs \to \infty}\numobs^{-1/2}\EE\{\log\modelmarginallik{\datarvarg{\numobs}}{\model_{1}} - \log\modelmarginallik{\datarvarg{\numobs}}{\model_{2}}\} = 0.
\]
Then under mild conditions, \cref{thm:model-selection-asymptotic-posteriors} (part 1) shows that the Bayes posterior mass on model $\model_{1}$ 
converges in distribution to a $\distBern(1/2)$ random variable:
\[
\modelpostdistfull{\model_{1}}{\datarvarg{\numobs}} \xrightarrow[\numobs\to\infty]{\mathcal{D}} \distBern(1/2). \label{eq:bayes-model-asymptotics}
\]
In other words, when $\numobs$ is large, with probability 1/2 model $\model_{1}$ has posterior probability $\approx 1$ and otherwise it has posterior probability $\approx 0$.
Since, asymptotically, both models provide equally good fit to the true data-generating distribution,  
one might hope that $\modelpostdistfull{\model_{1}}{\datarvarg{\numobs}} \to 1/2$.
However, \cref{eq:bayes-model-asymptotics} describes the opposite behavior: a single arbitrary model has posterior probability 1.

We show that BayesBag model selection does not suffer from this pathological behavior (\cref{thm:model-selection-asymptotic-posteriors}, part 2).
In the special case above (two models with asymptotically equally good fit), when $\bsnumobs = \numobs$, 
the bagged posterior probability of model $\model_{1}$ converges in distribution to a uniform random variable on the interval from 0 to 1:
\[
\bbmodelpostdistfull{\model_{1}}{\datarvarg{\numobs}} \xrightarrow[N\to\infty]{\mathcal{D}} \distUnif(0,1). \label{eq:bayesbag-model-asymptotics}
\]
Alternatively, if we choose $\bsnumobs$ such that $\bsnumobs/\numobs\to 0$ and $\bsnumobs/\numobs^{1/2}\to \infty$, 
then the bagged posterior mass on model $\model_{1}$ has the appealing behavior of converging to 1/2:
\[
\bbmodelpostdistfull{\model_{1}}{\datarvarg{\numobs}} \xrightarrow[N\to\infty]{P} 1/2. \label{eq:bayesbag-model-asymptotics-c-zero}
\]
This is not simply due to the bagged posterior reverting to the prior; this result holds for any prior giving positive mass to both models.  
\cref{thm:many-model-selection-asymptotic-posteriors} extends \cref{thm:model-selection-asymptotic-posteriors} to the case of more than two models, in which case the 
asymptotic distribution depends on the covariance structure of the log marginal likelihoods of the models. 
\cref{cor:exchangeable-model-selection-asymptotic-posteriors} extends \cref{thm:model-selection-asymptotic-posteriors} to the case of models with non-trivial parameter spaces.

In practice, it is unlikely that two models would fit the true data-generating distribution \emph{exactly} equally well. 
However, even if, say, model $\model_{1}$ has posterior probability tending to $1$ asymptotically, 
for a finite sample size it may be that $\numobs^{-1/2}\EE\{\log\modelmarginallik{\datarvarg{\numobs}}{\model_{1}} - \log\modelmarginallik{\datarvarg{\numobs}}{\model_{2}}\} \approx 0$,
such that with probability $\approx 1/2$, model $\model_{2}$ has posterior probability near $1$. 
Indeed, the analysis of  \citet{Yang:2018} was motivated by observations of this phenomenon in Bayesian phylogenetic tree 
reconstruction \citep{Douady:2003,Wilcox:2002,Alfaro:2003},
though it occurs more generally~\citep{Meng:2019}, such as in neuroscience and economic modeling \citep{Oelrich:2020}.

To understand this kind of finite-sample behavior via an asymptotic analysis,
\cref{thm:model-selection-asymptotic-posteriors,thm:many-model-selection-asymptotic-posteriors}
are formulated for sequences of models for $\numobs=1,2,\ldots$
that are not exactly equally good, but are asymptotically comparable
in the sense that the expected log-likelihood ratios between models are $O(\numobs^{1/2})$.
In this way, our results provide insight into cases where the models are not dependent on $\numobs$
but the sample size is not yet large enough for the posterior to concentrate at the best fitting model(s).

\subsection{Methodology}

BayesBag requires the choice of a bootstrap dataset size $\bsnumobs$ and the number of bootstrap datasets $B$.
First, the choice of $B$ controls the accuracy of the Monte Carlo approximation to the bagged posterior; see \cref{eq:bayesbag-definition-model-selection,eq:bayesbag-approximation}.
It is straightforward to empirically estimate the error using the standard formula for the variance of a Monte Carlo approximation
\citep{Huggins:2019:BayesBagI}.
If $\bsnumobs = \numobs$, we have found $B = 100$ to be sufficient in all of the applications we have considered.
For $\bsnumobs < \numobs$, the following result provides a natural lower bound on $B$ to ensure 
all available data are used with high probability.

\bnprop \label{prop:min-B}
For $\numobs > 1$, if $B \ge (\numobs - 1/2)\log(\numobs/\delta)/\bsnumobs$ then the probability that all observations are
included in at least one bootstrap sample is greater than $1 - \delta$. 
\enprop

While \cref{prop:min-B} offers a minimum value for $B$, we still recommend checking that the Monte Carlo standard error is sufficiently small
for the application at hand. 
For the choice of $\bsnumobs$, 
our theoretical and empirical results indicate that $\bsnumobs = \numobs^{0.95}$ is a good default choice that will behave fairly well for model selection,
both in cases where one model is correctly specified and, at the opposite extreme, when multiple misspecified models explain the data-generating distribution equally well. 
If significant misspecification is likely and there is a sufficient amount of data,
a more aggressive choice such as  $\bsnumobs = \numobs^{0.75}$ could be appropriate. 
A recommended workflow is in \cref{sec:workflow}. 

\subsection{Experiments}
We validate our theory and proposed methods through simulations on feature selection for linear regression,
and we evaluate the performance of BayesBag on real-data applications involving feature selection and phylogenetic tree reconstruction.
Overall, our empirical results demonstrate that in the presence of significant misspecification, the bagged posterior produces more stable inferences and puts significant mass on optimal models more reliably than the usual Bayes posterior; on the other hand, when one of the models is correctly specified, 
the bagged posterior with $\numobs^{0.95} \le \bsnumobs \le \numobs$ is slightly more conservative than the posterior.
Thus, BayesBag leads to more stable model selection results that are robust to minor changes in the model or representation of the data. 

\subsection{Related work}

First, we discuss previous work in the parameter inference and prediction setting, with a model smoothly parameterized by $\param \in \paramspace \subseteq \reals^{D}$. 
In a short discussion paper, \citet{Buhlmann:2014} introduced the name ``BayesBag'' to refer to bagging the posterior, and he presented a few simulation results in a simple Gaussian location model. 
However, \citet{Buhlmann:2014} employed a parametric bootstrap, which does not provide much benefit in a misspecified setting.
In contrast, in recent work \citep{Huggins:2019:BayesBagI},
we found that using the nonparametric bootstrap to implement BayesBag yielded significant benefits for parameter inference and prediction under misspecification.
In that work, we developed asymptotic theory for uncertainty quantification of the Kullback--Leibler optimal parameter, providing insight into how to choose the bootstrap
dataset size ($\bsnumobs = 2\numobs$ if the model is correctly specified, and $\bsnumobs=\numobs$ if the model is misspecified). 
Neither paper considered model selection, which raises fundamentally different issues because 
it involves a discrete space where smoothness does not play a role.
Notably, our recommendation in this paper to take $\bsnumobs = o(\numobs)$ for model selection is very different from our recommendations for parameter inference and prediction.

The previous work most closely related to the present work is a mix of empirical investigation \citep{Waddell:2002,Douady:2003,Oelrich:2020} and theoretical work \citep{Buhlmann:2002,Yang:2018,Oelrich:2020}. 
The purely empirical papers undertake limited investigations in the setting of phylogenetic tree inference: \citet{Waddell:2002} focus primarily on 
speeding up model selection and \citet{Douady:2003} mainly aim to compare Bayesian inference to the bootstrap. 
Our \cref{thm:model-selection-asymptotic-posteriors} is similar in spirit to the bagging result of \citet[Proposition 2.1]{Buhlmann:2002}.
However, the \citet{Buhlmann:2002} result is not applicable in the model selection setting since it would require assigning probability 1 to whichever model 
has the larger marginal likelihood---which does not correspond to Bayesian model selection---and then applying bagging to this selection procedure. 
Our other results (\cref{thm:many-model-selection-asymptotic-posteriors,cor:exchangeable-model-selection-asymptotic-posteriors}) go well beyond 
the scope of the  \citet{Buhlmann:2002} result, covering three or more models as well as non-trivial parameter spaces. 

Regarding the behavior of Bayesian model selection under the usual posterior, \citet{Yang:2018} prove a result similar to \cref{eq:bayes-model-asymptotics} 
but more limited than our general versions in part 1 of \cref{thm:model-selection-asymptotic-posteriors,thm:many-model-selection-asymptotic-posteriors}.
Finally, \citet{Oelrich:2020} provide complementary results to our own:
they study additional real-world examples of overconfident model selection and, in the feature selection setting, analyze the mean and variance of the log marginal likelihood
ratio for a particular type of linear regression model with known variance, offering a more precise characterization of the posterior in that particular setting.
However, they do not analyze or consider using the bagged posterior.

\section{Theory and methodology} \label{sec:model-selection}

In this section, we present our theoretical results,
illustrate the theory with plots comparing the asymptotics of BayesBag versus Bayes (\cref{sec:asymptotic-analysis}),
discuss the use of BayesBag for model criticism (\cref{sec:model-criticism}),
and provide a recommended workflow (\cref{sec:workflow}).

\subsection{Asymptotic analysis} \label{sec:asymptotic-analysis}

In Bayesian model selection, we have a countable set of models $\modelspace$.
Assume that model $\model \in \modelspace$ has prior probability $\modelpriordist(\model) > 0$ and marginal likelihood 
\[
\modelmarginallik{\datarvarg{\numobs}}{\model} = \int \left\{\prod_{n=1}^{\numobs}\lik{\obsrv{n} \given \model}{\param_{\model}}\right\}\priordist(\dee\param_{\model} \given \model),
\]
where $\param_{\model} \in \paramspace_{\model}$ is an element of a model-specific parameter space with prior distribution $\priordist(\dee\param_{\model} \given \model)$. 
Assume $\obsrv{1},\obsrv{2},\ldots$ are i.i.d.\ from some unknown distribution $\obsdist$.
Further, for each $\model \in \modelspace$, assume there is a unique parameter 
\[
\optmodelparam{\model} \defined \argmin_{\param_{\model} \in \paramspace_{\model}}-\EE\{\log \lik{\obsrv{1} \given \model}{\param_{\model}}\}
\]
that 
minimizes the Kullback--Leibler divergence from $\obsdist$ to the model.
We say that \emph{model $\model$ is misspecified} if $\likdist{\optmodelparam{\model}} \ne \obsdist$.
The posterior probability of $\model \in \modelspace$ is 
$\modelpostdistfull{\model}{\datarvarg{\numobs}} 
\propto \modelmarginallik{\datarvarg{\numobs}}{\model}\modelpriordist(\model)$.
Let $\bsdatarvarg{\bsnumobs}$ denote a bootstrapped copy of $\datarvarg{\numobs}$ with $\bsnumobs$ observations;
that is, each observation $\obsrv{n}$ is replicated $\bscount{n}$ times in $\bsdatarvarg{\bsnumobs}$,
where $(\bscount{1},\ldots,\bscount{\numobs}) \dist \distMulti(\bsnumobs, 1/\numobs)$ is a multinomial-distributed count vector.
The bagged posterior probability of model $\model \in \modelspace$ is then
\[
\bbmodelpostdistfull{\model}{\datarvarg{\numobs}} 
\defined \EE\{\modelpostdistfull{\model}{\bsdatarvarg{\bsnumobs}} \given \datarvarg{\numobs} \}.
\]
Note that this is equivalent to the informal definition in \cref{eq:bayesbag-definition-model-selection}.

\begin{figure}[tbp]
\begin{center}
\begin{subfigure}[b]{0.49\textwidth}
\includegraphics[height=1.7in]{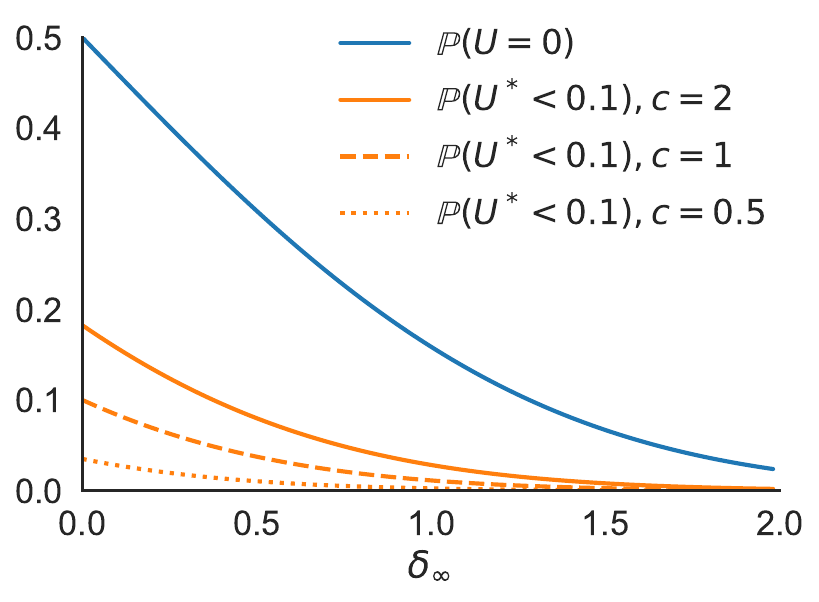}
\caption{}
\end{subfigure}
\begin{subfigure}[b]{0.49\textwidth}
\includegraphics[height=1.7in]{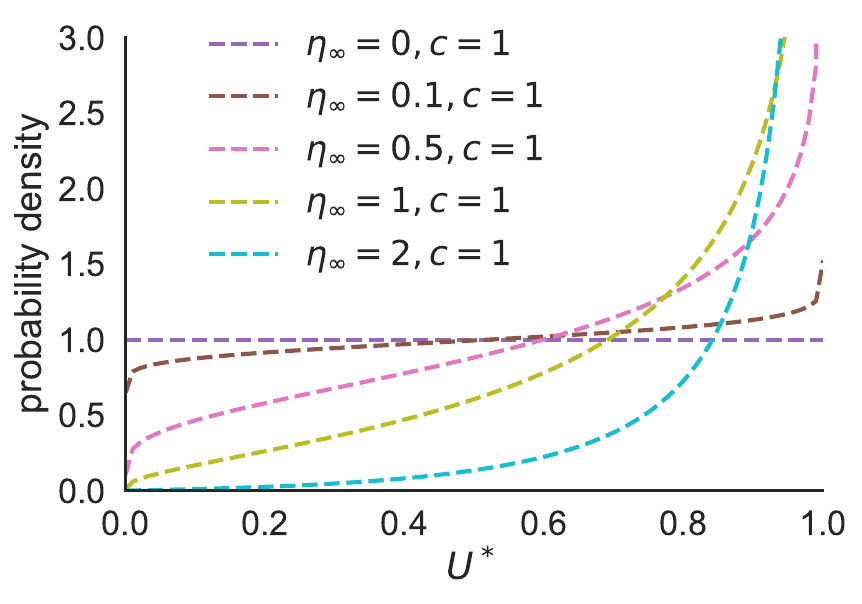}
\caption{}
\end{subfigure}
\begin{subfigure}[b]{0.7\textwidth}
\centering
\includegraphics[height=1.7in]{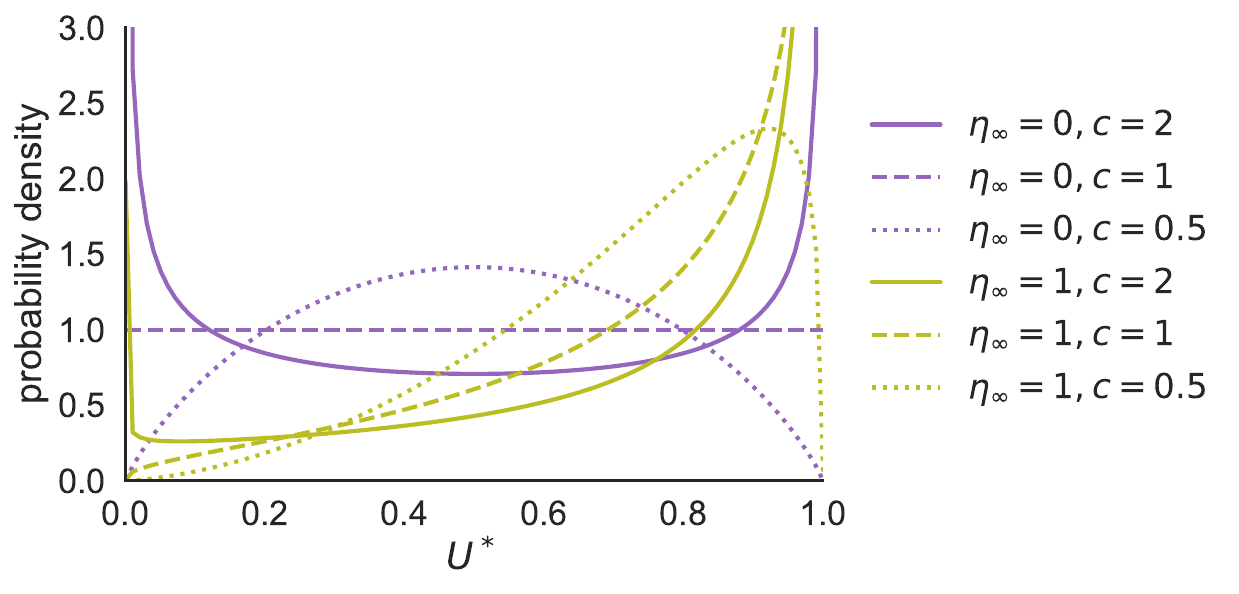}
\caption{}
\end{subfigure}
\caption{
The bagged posterior (BayesBag) is far less likely than the usual posterior (Bayes) to strongly favor the wrong model (or an arbitrary equally good model). 
When there are two models, the asymptotic posterior probability of model $\model_{1}$ is a random variable $U$ (for Bayes)
or $U^{\bbsym}$ (for BayesBag), where
$U \dist \distBern(\Phi(\effectsize))$, $U^{\bbsym} = \Phi(c^{1/2}W^{\bbsym})$, $W^{\bbsym} \dist \distNorm(\effectsize, 1)$,
and $\effectsize$ is the asymptotic effect size in favor of $\model_{1}$; see \cref{thm:model-selection-asymptotic-posteriors}.
\textbf{(a)} $U = 0$ represents the event that the Bayes posterior overwhelmingly favors the wrong model (or equally good model, if $\effectsize = 0$), that is, 
the model with lower (or equal) expected log-likelihood under the true distribution. 
Likewise, $U^{\bbsym} < 0.1$ is the event that the BayesBag posterior strongly favors the wrong (or equivalent) model. 
\textbf{(b)} $U^{\bbsym}$ is a continuous random variable on $[0,1]$.  The density of $U^{\bbsym}$ is shown
for a range of $\effectsize$ values, with $\bsscale = \lim \bsnumobs/\numobs$ fixed at $\bsscale=1$; see \cref{thm:model-selection-asymptotic-posteriors}.
Here, $\numobs$ is the dataset size and $\bsnumobs$ is the bootstrap dataset size.
\textbf{(c)} Densities of $U^{\bbsym}$ as both $\effectsize$ and $\bsscale$ vary.
}
\label{fig:model-selection-asymptotics}
\end{center}
\end{figure}

\paragraph{Two models with degenerate parameter spaces.}
We first state our asymptotic theory in the case of two misspecified models, $\modelspace = \{\model_{1},\model_{2}\}$,
since the results are more intuitive in this special case.
For the moment, we also assume that each model contains a single parameter value (that is, $|\paramspace_{\model}| = 1$). 
On the other hand, we allow the observation model $\lik{\obsrv{n} \given \model}{\numobs}$ to depend on the number of observations $\numobs$, so that
$\modelmarginallik{\datarvarg{\numobs}}{\model} = \prod_{n=1}^{\numobs} \lik{\obsrv{n} \given \model}{\numobs}$.
Let $Z_{\numobs} \defined \log\modelmarginallik{\datarvarg{\numobs}}{\model_{1}} - \log\modelmarginallik{\datarvarg{\numobs}}{\model_{2}}$ denote the model log-likelihood ratio and,
for $n=1,\dots,\numobs$, let $Z_{\numobs n} \defined \log\lik{\obsrv{n} \given \model_{1}}{\numobs} - \log\lik{\obsrv{n} \given \model_{2}}{\numobs}$ denote the log-likelihood ratio for each observation.

To perform an asymptotic analysis that captures the behavior of the nonasymptotic regime in which the mean of $Z_{\numobs}$ is comparable to its standard deviation, 
we assume that $\mu_{\infty} \defined \lim_{\numobs \to \infty} \numobs^{1/2}\EE(Z_{\numobs 1})$ and 
$\sigma_{\infty}^{2} \defined\lim_{\numobs \to \infty} \var(Z_{\numobs 1})$ exist.
Thus,  when $\numobs$ is large, $\EE(Z_{\numobs}) \approx \numobs^{1/2}\mu_{\infty}$ and $\std(Z_{\numobs}) \approx \numobs^{1/2}\sigma_{\infty}$. Consequently, $\EE(Z_{\numobs})$ does not overwhelm $\std(Z_{\numobs})$, even in the asymptotic regime.
The asymptotic effect size $\effectsize \defined  \mu_{\infty}/\sigma_{\infty}$ quantifies the amount of evidence in favor of $\model_{1}$ under the true distribution $\obsdist$. 
If $\effectsize > 0$, then $\model_{1}$ is favored, whereas $\model_{2}$ is favored if $\effectsize < 0$.

Our first result shows that (1) the posterior probability of model $\model_{1}$ converges to a Bernoulli random variable with parameter depending
on $\effectsize$, and (2) the bagged posterior probability of model $\model_{1}$ converges to a continuous random variable on $[0,1]$
with a distribution that depends on $\effectsize$ and $\lim_{\numobs\to\infty} \bsnumobs/\numobs$.
For $\mu \in \reals$ and $\sigma^2 > 0$, let $\distNorm(\mu, \sigma^2)$ denote the normal distribution with mean $\mu$ and variance $\sigma^2$,
and let $\Phi(t)$ denote the cumulative distribution function of the standard normal distribution $\distNorm(0,1)$.

\bnthm \label{thm:model-selection-asymptotic-posteriors}
Let $\obsrv{1},\obsrv{2},\ldots\;\iid\dist \obsdist$ for some distribution $\obsdist$ and
assume
\begin{enumerate}[label=(\roman*)]
\item $\mu_{\infty} \defined \lim_{\numobs \to \infty} \numobs^{1/2}\EE(Z_{\numobs 1})\in \reals$ exists,
\item $\sigma_{\infty}^{2} \defined\lim_{\numobs \to \infty} \var(Z_{\numobs 1}) \in (0,\infty)$ exists,
\item $\limsup_{\numobs \to \infty}\EE(|Z_{\numobs 1}|^{6}) < \infty$,
\item  $\bsnumobs=\bsnumobs(\numobs)$ satisfies $\lim_{\numobs \to \infty} \bsnumobs/\numobs^{1/2} = \infty$, and
\item $\bsscale \defined \lim_{\numobs \to \infty} \bsnumobs/\numobs \in [0,\infty)$.
\end{enumerate}
Then
\begin{enumerate}
\item for the usual posterior, $\modelpostdistfull{\model_{1}}{\datarvarg{\numobs}}  \convD U \dist \distBern(\Phi(\effectsize))$, where
$\effectsize = \mu_{\infty}/\sigma_{\infty}$;
\item for the bagged posterior, $\bbmodelpostdistfull{\model_{1}}{\datarvarg{\numobs}} \convD \Phi(c^{1/2}W^{\bbsym})$, where $W^{\bbsym} \dist \distNorm(\effectsize, 1)$.
\end{enumerate}
In particular, for the usual posterior, if $\effectsize = 0$ then $\modelpostdistfull{\model_{1}}{\datarvarg{\numobs}} \convD \distBern(1/2)$.
Meanwhile, for the bagged posterior, if $\effectsize = 0$ then $\bbmodelpostdistfull{\model_{1}}{\datarvarg{\numobs}} \convD \distUnif(0,1)$ when $\bsscale = 1$
and $\bbmodelpostdistfull{\model_{1}}{\datarvarg{\numobs}} \convP 1/2$ when $\bsscale = 0$. 
\enthm

\Cref{thm:model-selection-asymptotic-posteriors} will follow as an immediate corollary of \cref{thm:many-model-selection-asymptotic-posteriors} below.
Note that in part~2, when $\bsscale > 0$, the cumulative distribution function of the random variable $U^{\bbsym} \defined \Phi(\bsscale^{1/2}W_{\bbsym})$ is given by
$u \mapsto \Phi(\bsscale^{-1/2}\Phi^{-1}(u) - \effectsize)$ for $u\in (0,1)$.
Thus, by differentiating this function, we find that the density of $U^{\bbsym}$ is, for $u\in (0,1)$,
\[
f(u) = \Phi'\big(\bsscale^{-1/2}\Phi^{-1}(u) - \effectsize\big) \bsscale^{-1/2} / \Phi'(\Phi^{-1}(u)).
\]

\Cref{fig:model-selection-asymptotics} illustrates how \cref{thm:model-selection-asymptotic-posteriors} establishes
the greater stability of BayesBag versus Bayes for model selection. 
Even for effect sizes $\effectsize > 1$, which should strongly favor model $\model_{1}$, the Bayes posterior overwhelmingly favors model $\model_{2}$ with non-negligible probability -- that is, $\Pr\{\modelpostdistfull{\model_{1}}{\datarvarg{\numobs}} \approx 0\} \not\approx 0$. 
On the other hand, the probability that the BayesBag posterior strongly favors model $\model_{2}$ goes to zero more rapidly as $\effectsize$ increases
-- that is, $\Pr\{\bbmodelpostdistfull{\model_{1}}{\datarvarg{\numobs}} \approx 0\} \to 0$ more rapidly as $\effectsize$ grows. 
For example, when $\effectsize = 2$ and $\bsscale = 1$, $\Pr(U=0) > 0.02$ whereas $\Pr(U^{\bbsym} < 0.1) < 7 \times 10^{-5}$.
Thus, in this example, Bayes will overwhelmingly favor the ``wrong'' model in approximately $1$ out of $50$ experiments, 
whereas BayesBag will somewhat strongly favor the wrong model in only approximately 7 out of 100,000 experiments.

\begin{figure}[tbp]

\begin{subfigure}[b]{0.4\textwidth}
\includegraphics[height=1.7in,trim=0 0 10.35cm 0,clip]{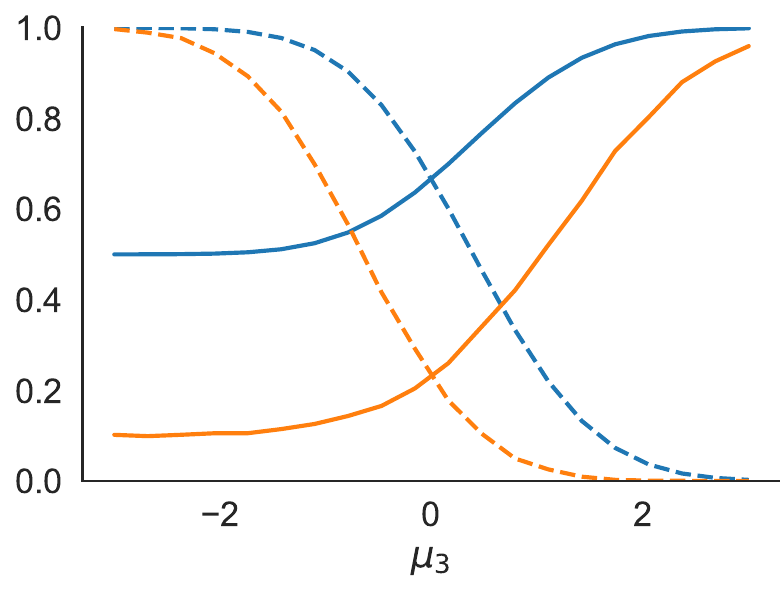}
\caption{}
\end{subfigure}
\begin{subfigure}[b]{0.4\textwidth}
\includegraphics[height=1.7in,trim=13.1cm 0 0 0,clip]{BayesBag-3-model-selection-asymptotic-posteriors-varying-mean}
\end{subfigure}

\begin{subfigure}[b]{0.4\textwidth}
\includegraphics[height=1.7in,trim=0 0 10.35cm 0,clip]{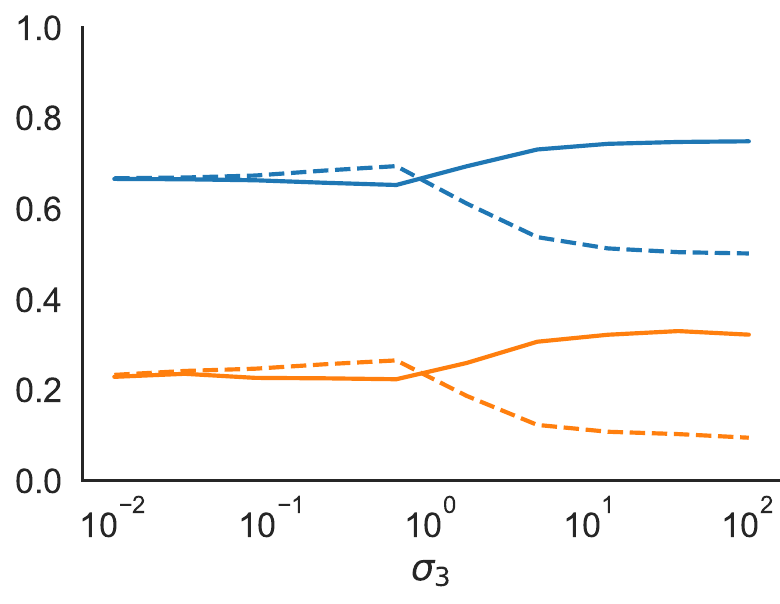}
\caption{}
\end{subfigure} 
\begin{subfigure}[b]{0.4\textwidth}
\includegraphics[height=1.7in,trim=0 0 10.35cm 0,clip]{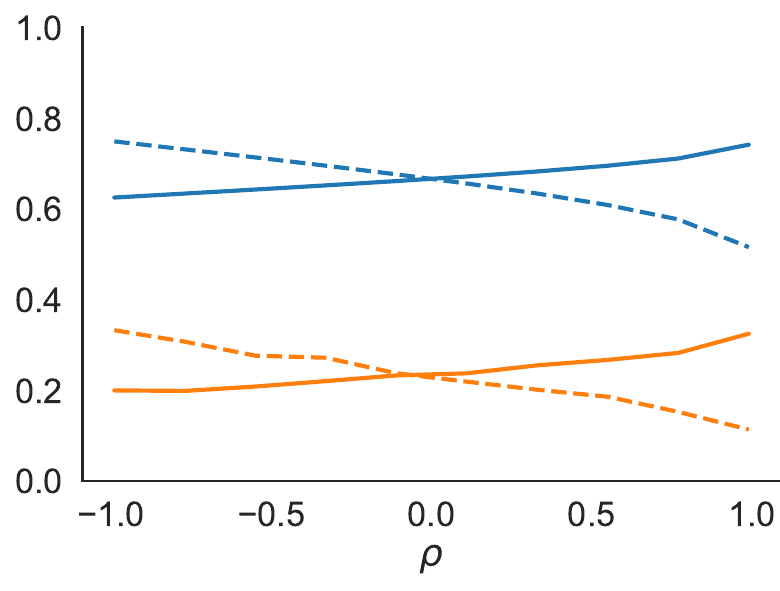}
\caption{}
\end{subfigure} \hspace{.1\textwidth}
\begin{center}
\caption{
When there are more than two models,
the bagged posterior (BayesBag) is far less likely than the usual posterior (Bayes) to strongly favor the wrong model (or an arbitrary equally good model) for a wide variety of mean and covariance structures of the asymptotic log-likelihoods. 
The asymptotic posterior probability of model $\model_{k}$ is a random variable $U_{k}$ (for Bayes)
or $U_{k}^{\bbsym}$ (for BayesBag), where
$U_{k} \dist \distBern(\Phi_{-\mu_{\infty}, \Sigma_{\infty}}(0))$, $U_{k}^{\bbsym} = \Phi_{0,\Sigma_{\infty}}(\bsscale^{1/2}W^{\bbsym})$, 
$W^{\bbsym} \dist \distNorm(\mu_{\infty}, \Sigma_{\infty})$, and $\mu_{\infty} \in \reals^{K-1}$ and $\Sigma_{\infty} = \reals^{K-1 \times K-1}$ are, respectively, the asymptotic mean and covariance
of the log-likelihood ratio of $\model_{k}$ versus each other model; see \cref{thm:many-model-selection-asymptotic-posteriors}.
$U_{k} = 0$ represents the event that the Bayes posterior overwhelmingly rejects model $\model_{k}$,
and $U_{k}^{\bbsym} < 0.1$ is the event that the BayesBag posterior strongly rejects model $\model_{k}$.
Three scenarios are shown for the case of $K=3$ models, for a range of values of $\mu_{\infty}' \in \reals^{3}$ and $\Sigma_{\infty}' \in \reals^{3 \times 3}$, the asymptotic mean and covariance of the log-likelihoods. 
\textbf{(a)} First, we vary $\mu_{3}$, where $\mu_{\infty}' = (0, 0, \mu_{3})^{\top}$ and the entries of $\Sigma_{\infty}'$
are given by $\Sigma_{\infty,i,j}' = 0.5^{\ind(i\ne j)}$. 
\textbf{(b)} Second, we vary $\sigma_{3}$, where $\mu_{\infty}' = (0, 0, 0)^{\top}$ and 
$\Sigma_{\infty,i,j}' = 0.5^{\ind(i\ne j)} \sigma_{3}^{\ind(i=3)} \sigma_{3}^{\ind(j=3)}$.
\textbf{(c)} Third, we vary $\rho$, where $\mu_{\infty}' = (0, 0, 0)^{\top}$ and $\Sigma_{\infty,i,j}' = \ind(i = j) + \rho \ind(i = 1, j=2) + \rho \ind(i=2, j=1)$.
}
\label{fig:many-model-selection-asymptotics}
\end{center}
\end{figure}

\paragraph{Extension to three or more models.}
In the case of three or more models,
the behavior of the posteriors is more complicated because there is  
dependence on both the correlation structure and the relative variances of the log-likelihood ratios between each pair of models.
Consider the case of $K < \infty$ models and enumerate them from $1$ to $K$, so that $\modelspace = \theset{\model_{1},\dots,\model_{K}}$. 
For $k = 1,\ldots,K$, define the individual model log-likelihood terms $Y_{\numobs n, k} \defined  \log\lik{\obsrv{n} \given \model_k}{\numobs}$,
and let $Y_{\numobs n} \defined (Y_{\numobs n,1},\dots,Y_{\numobs n,K})^{\top} \in \reals^{K}$. 
For $t, \mu \in \reals^{K-1}$ and $\Sigma \in \reals^{K - 1 \times K -1}$ positive definite,
let $\Phi_{\mu,\Sigma}(t)$ denote the cumulative distribution function of the 
$(K-1)$-dimensional normal distribution $\distNorm(\mu, \Sigma)$.
\bnthm \label{thm:many-model-selection-asymptotic-posteriors}
Let $\obsrv{1},\obsrv{2},\ldots\;\iid\dist \obsdist$ for some distribution $\obsdist$.
Defining $\mu_{\numobs}' \defined  \numobs^{1/2}\EE(Y_{\numobs 1})$ and $\Sigma_{\numobs}' \defined  \cov(Y_{\numobs 1})$, assume
\begin{enumerate}[label=(\roman*)]
\item $\mu_{\infty}' \defined \lim_{\numobs \to \infty}\mu_{\numobs}' \in\reals^{K}$,
\item $\Sigma_{\infty}' \defined \lim_{\numobs \to \infty} \Sigma_{\numobs}'$ positive definite,
\item $\limsup_{\numobs \to \infty}\EE(\|Y_{\numobs 1}\|_{2}^{6}) < \infty$,
\item $\bsnumobs=\bsnumobs(\numobs)$ satisfies $\lim_{\numobs \to \infty} \bsnumobs/\numobs^{1/2} = \infty$, and
\item $\bsscale \defined \lim_{\numobs \to \infty} \bsnumobs/\numobs \in [0,\infty)$.
\end{enumerate}
Without loss of generality, consider the probability of $\model_{1}$. Define
$\mu_{\infty,k} \defined \mu_{\infty,1}' - \mu_{\infty,k+1}'$,
$\Sigma_{\infty,k,\ell} \defined \Sigma_{\infty,1,1}' + \Sigma_{\infty,k+1,\ell+1}' - \Sigma_{\infty,1,k+1}'  - \Sigma_{\infty,1,\ell+1}'$
for $k,\ell\in\{1,\ldots,K-1\}$. Then
\begin{enumerate}
\item for the usual posterior, $\modelpostdistfull{\model_{1}}{\datarvarg{\numobs}}  \convD U \dist \distBern(\Phi_{-\mu_{\infty}, \Sigma_{\infty}}(0))$;
\item for the bagged posterior, $\bbmodelpostdistfull{\model_{1}}{\datarvarg{\numobs}} \convD \Phi_{0,\Sigma_{\infty}}(\bsscale^{1/2}W^{\bbsym})$,
where \\ $W^{\bbsym} \dist \distNorm(\mu_{\infty}, \Sigma_{\infty})$.
\end{enumerate}
\enthm
The proof is in \cref{sec:proof-of-many-model-selection-asymptotic-posteriors}.
\Cref{fig:many-model-selection-asymptotics} shows how \cref{thm:many-model-selection-asymptotic-posteriors} establishes
that across a range of mean and covariance structures of the log-likelihoods, BayesBag is more stable than Bayes. 
Indeed, both methods behave fairly consistently as the covariance varies. 

\paragraph{Extension to non-degenerate parameter spaces.}
We now extend
\cref{thm:model-selection-asymptotic-posteriors} to non-degenerate parameter spaces $\paramspace_{1} \subset \reals^{D_1}$ and  $\paramspace_{2} \subset \reals^{D_2}$ and we integrate over $\param_{\model}\in\paramspace_{\model}$ for each model $\model$.
To avoid tedious arguments, we only consider the case where $\mu_{\infty} = 0$.
For $\model \in \{\model_{1}, \model_{2}\}$, let $\loglik{\obsrv{n}}{\model,\param_{\model}}\defined \log\lik{\obsrv{n} \given \model}{\param_{\model}}$
and recall that the optimal parameter is 
$\optmodelparam{\model} = \argmin_{\param_{\model} \in \paramspace_{\model}}-\EE\{\loglikfun{\model,\param_{\model}}(\obsrv{1})\}$.
Let $\lmldiff{\datarvarg{\numobs}} \defined \log \modelmarginallik{\datarvarg{\numobs}}{\model_{1}}\modelpriordist(\model_{1}) -  \log \modelmarginallik{\datarvarg{\numobs}}{\model_{2}}\modelpriordist(\model_{2})$,
where $\modelmarginallik{\datarvarg{\numobs}}{\model} = \int \big\{\prod_{n=1}^{\numobs}\lik{\obsrv{n} \given \model}{\param_{\model}}\big\} \priordist(\dee\param_{\model} \given \model)$ denotes the marginal likelihood. 
Let $\alldatarv$ denote the infinite sequence $(\obsrv{1},\obsrv{2},\dots)$.
We will assume that conditionally on $\alldatarv$, for almost every $\alldatarv$,
\[
\lmldiff{\bsdatarvarg{\bsnumobs}} = \frac{1}{2}(D_{2} - D_{1}) \log \numobs + \sum_{m=1}^{\bsnumobs} \log \frac{\lik{\bsobsrv{m} \given \model_{1}}{\optmodelparam{1}}}{\lik{\bsobsrv{m} \given \model_{2}}{\optmodelparam{2}}} + \bigoPouter(1), \label{eq:lml-decomp}
\]
where $\bsdatarvarg{\bsnumobs}$ is bootstrapped from $\datarvarg{\numobs}$ and $\bigoPouter(1)$ denotes a random quantity which is bounded in (outer) probability. 
It is well known that \cref{eq:lml-decomp} holds with $\datarvarg{\numobs}$ in place of $\bsdatarvarg{\bsnumobs}$, under standard regularity assumptions~\citep{Clarke:1990}.
Thus, we expect \cref{eq:lml-decomp} to hold under similar but slightly stronger conditions, since we must consider a triangular array rather than a sequence of random variables. 

The posterior distribution given $\datarvarg{\numobs}$ and $\model$ is
\[ 
\postdistfull{\dee\param_{\model}}{\datarvarg{\numobs}, \model} 
\defined \frac{\prod_{n=1}^{\numobs}\lik{\obsrv{n} \given \model}{\param_{\model}}}{\modelmarginallik{\datarvarg{\numobs}}{\model}}\priordist(\dee\param_{\model} \given \model). \label{eq:iid-data-posterior}
\]
The \emph{bagged posterior} $\bbpostdistfull{\cdot}{\datarvarg{\numobs}, \model}$ given $\datarvarg{\numobs}$ and $\model$  is defined such that 
\[
\bbpostdistfull{A}{\datarvarg{\numobs}, \model} \defined \EE\{\postdistfull{A}{\bsdatarvarg{\bsnumobs}, \model } \given \datarvarg{\numobs}\}
\]
for all measurable $A \subseteq \paramspace$.
Let $\Ehessloglik{\param_{\model}} \defined -\EE\{\grad_{\param_{\model}}^{2}\loglikfun{\model,\param_{\model}}(\obsrv{1})\}$ denote the Fisher information matrix.
For a measure $\nu$ and function $f$, we use the shorthand $\nu(f) \defined \int f \dee \nu$.  

\bncor \label{cor:exchangeable-model-selection-asymptotic-posteriors}
Let $\obsrv{1},\obsrv{2},\ldots\;\iid\dist \obsdist$ and for $\model \in \{\model_{1},\model_{2}\}$,
assume that
\begin{enumerate}[label=(\roman*)]
\item $\param_{\model} \mapsto \loglik{\obsrv{1}}{\param_{\model}}$ is differentiable at $\optmodelparam{\model}$ in probability;
\item there is an open neighborhood $U$ of $\optmodelparam{\model}$
and a function $m_{\optmodelparam{\model}} : \obsspace \to \reals$ such that $\obsdist(m_{\optmodelparam{\model}}^{3}) < \infty$ and for all $\param_{\model},\param_{\model}' \in U$,
$|\loglikfun{\param_{\model}} - \loglikfun{\param_{\model}'}| \le m_{\optmodelparam{\model}}\twonorm{\param_{\model} - \param_{\model}'}$ a.s.$[\obsdist]$;
\item $-\obsdist(\loglikfun{\param_{\model}} - \loglikfun{\optmodelparam{\model}}) = \frac{1}{2}(\param_{\model} - \optmodelparam{\model})^{\top}\Ehessloglik{\optmodelparam{\model}}(\param_{\model} - \optmodelparam{\model}) + \littleo(\twonorm{\param_{\model} - \optmodelparam{\model}}^{2})$ as $\param_{\model} \to \optmodelparam{\model}$;
\item $\Ehessloglik{\optmodelparam{\model}}$ is an invertible matrix; and
\item letting $\bbparamsample_{\model}\dist \bbpostdistfull{\cdot}{\datarvarg{\numobs}, \model}$, 
it holds that conditionally on $\alldatarv$, for almost every $\alldatarv$, for every sequence of constants $\concconst\to\infty$,
\[
\EE\Big\{\postdistfull{\twonorm{\bbparamsample_{\model} - \optmodelparam{\model}} > \concconst/\bsnumobs^{1/2}}{\bsdatarvarg{\bsnumobs}, \model} \;\Big\vert\; \datarvarg{\numobs}\Big\} \to 0.
\]
\end{enumerate} 
Further, assume that \cref{eq:lml-decomp} holds, $\lim_{\numobs \to \infty} \bsnumobs/\numobs^{1/2} = \infty$, 
$\bsscale \defined \lim_{\numobs \to \infty} \bsnumobs/\numobs \in [0,\infty)$,
$\EE\{\loglik{\obsrv{1}}{\model_{1},\optmodelparam{1}} - \loglik{\obsrv{1}}{\model_{2},\optmodelparam{2}}\} = 0$,
and $\EE[\{\loglik{\obsrv{1}}{\model_{1},\optmodelparam{1}} - \loglik{\obsrv{1}}{\model_{2},\optmodelparam{2}}\}^{3}] \in (0,\infty)$.
Then the conclusions of \cref{thm:model-selection-asymptotic-posteriors} apply in the case of $\effectsize = 0$. 
\encor 

The proof is in \cref{sec:proof-of-exchangeable-model-selection-asymptotic-posteriors}.

\paragraph{Extension to dependent data.}

A further extension, which we will not pursue in detail, is to non-independent data such as those 
encountered in time-series and spatial data analysis. 
In principle the generalization to, for example, time-series using the block bootstrap (or another nonparametric estimator such as a Gaussian process) is straightforward.
However, the accompanying theory is much less straightforward since
we must (A) determine the asymptotic distribution of rescaled log marginal likelihoods $\numobs^{-\kappa}\log p(\model \given \datarvarg{\numobs})$ and
(B) show that a nonparametric estimator has the same asymptotic distribution.
More concretely, consider the two-model scenario and define 
$W(\datarvarg{\numobs}) \defined \numobs^{-\kappa}\{\log p(\model_{1} \given \datarvarg{\numobs}) - \log p(\model_{2} \given \datarvarg{\numobs})\}$.
Then we must determine an appropriate $\kappa$ such that $W(\datarvarg{\numobs}) \convD W_{\infty}$, where $W_{\infty}$ is a non-degenerate distribution. 
Moreover, for (A) we must show that $\lim_{\numobs \to \infty}\dsupconvex(\mathcal{L}\{W(\datarvarg{\numobs})\}, \mathcal{L}(W_{\infty})) = 0$,
where $\mcL(\xi)$ denotes the law of a random variable $\xi$ and the metric $\dsupconvex$ is defined in \cref{sec:proof-of-many-model-selection-asymptotic-posteriors}.
Then, for (B) we must show that for data $\bsdatarvarg{\bsnumobs}$ distributed according to the nonparametric estimator, 
\[
\dsupconvex(\mcL\{W(\bsdatarvarg{\bsnumobs}) - (\numobs/\bsnumobs)^{\kappa}W(\datarvarg{\numobs}) \mid \datarvarg{\numobs}\},\, W_{\infty} - \EE\{W_{\infty}\}) \convP 0.
\]
We leave a thorough investigation of models for dependent data to future work.

\subsection{Model criticism with BayesBag} \label{sec:model-criticism}

In the setting of parameter inference and prediction, \citet{Huggins:2019:BayesBagI} developed a measure 
quantifying the amount of misspecification exhibited by the usual Bayes posterior,
referred to as the \emph{model--data mismatch index}, based on comparing the BayesBag posterior to the Bayes posterior.
To define a mismatch index in the setting of model selection, 
we perform parameter inference in a special designated model that we refer to as the \emph{litmus model}.
Suppose $f(\optparam)$ is a selected quantity of inferential interest, where $f : \paramspace \to \reals$ and $\paramspace$ is the parameter space of the litmus model.
Let $v_{\numobs}$ and $v_{\bsnumobs}^{\bbsym}$ denote, respectively, the Bayes and BayesBag posterior variances of $f(\param)$
under the litmus model.
If the litmus model is well-specified, then asymptotically, $\bsnumobs v_{\bsnumobs}^{\bbsym} = 2\numobs v_{\numobs}$ \citep{Huggins:2019:BayesBagI}.
We define the asymptotic version of the mismatch index as
\[
\modelmismatch(f) \defined \begin{cases}
1 - {2 \numobs v_{\numobs}}/({\bsnumobs v_{\bsnumobs}^{\bbsym}}) & \text{if $\bsnumobs v_{\bsnumobs}^{\bbsym} > \numobs v_{\numobs}$} \\
\nan & \text{otherwise.}
\end{cases} 
\] 
The interpretation is as follows: $\modelmismatch(f) \approx 0$ indicates no evidence of mismatch;
$\modelmismatch(f) > 0$ (respectively, $\modelmismatch(f) < 0$) indicates the Bayes posterior is overconfident (respectively, under-confident);
$\modelmismatch(f)= \nan$ indicates that either there is severe model--data mismatch or the required asymptotic assumptions do not hold
(for example, due to multimodality in the posterior or small sample size).
We refer the interested reader to  \citet{Huggins:2019:BayesBagI} for more justification and description of a non-asymptotic version of $\modelmismatch$. 

The litmus model should be chosen such that if any model $\model\in\modelspace$ is well specified, then the litmus model is well specified.
One common case is a finite set of models with partial order $\prec$ based on inclusion such that there exists a unique maximal model;
in this case, the maximal model can be used as the litmus model.
More precisely, let $\mcP_{\model} \defined  \{ \lik{\cdot \given \model}{\param_{\model}} : \param_{\model} \in \paramspace_{\model} \}$. 
Then for models $\model, \model' \in \modelspace$, $\model \prec \model'$ if and only if $\mcP_{\model} \subseteq \mcP_{\model'}$,
and $\model$ is the unique maximal model if $\model' \prec \model$ for all $\model' \in \modelspace$. 
Feature selection (\cref{sec:linreg-model-selection-sim}) is an example of this type, where the maximal model includes all features. 
Another common situation is when all models have a set of shared, interpretable parameters,
in which case we can define the litmus model to be the disjoint union of all models $\model\in\modelspace$.
Phylogenetic tree reconstruction (\cref{sec:experiments}) is an example of this type.

When there is more than one univariate quantity of inferential interest, we consider a collection of functions $f\in\mcF$
and suggest taking the most pessimistic mismatch value: $\modelmismatch(\mcF) \defined \sup_{f \in \mcF}\modelmismatch(f)$. 
In general, $\mcF$ can be chosen to reflect the quantities of interest in the application at hand. 
When $\param \in \reals^{D}$, two natural choices for the collection $\mcF$ are $\mcF_{1} \defined \{ \param \mapsto w^{\top}\param : \twonorm{w} = 1 \}$ 
and $\mcF_{\text{proj}} = \{ \param \mapsto \param_{d} : d = 1,\dots,D\}$.
In our experiments, we use the latter and therefore adopt the shorthand notation $\modelmismatch \defined \modelmismatch(\mcF_{\text{proj}})$.

\subsection{Recommended workflow}\label{sec:workflow}

Algorithm~\ref{alg:using-BayesBag} outlines our recommended workflow for using BayesBag;
here, $\dim(\paramspace_{\model})$ is the dimensionality of the parameter space of model $\model$.
In steps 1--3, we suggest computing the mismatch index with $\bsnumobs = \numobs$ since the definition of the mismatch index
is based on asymptotics, and thus, it is desirable to make $\bsnumobs$ large in order to improve the accuracy of this asymptotic approximation.
The mismatch index assesses the fit of the usual posterior, so there is no reason to use the same value of $\bsnumobs$
that is used for robust inference with BayesBag.
For very large datasets, it may be preferable to compute the mismatch index using $\bsnumobs < \numobs$ in order to reduce 
the computation required. 

In steps 4--7, our recommendations for when to use $\bsnumobs =  \numobs^{\alpha}$ with $\alpha = 0.95$ versus $\alpha = 0.75$, and these particular values of $\alpha$, should be taken as rough guidelines.
The condition  $\sum_{\model \in \modelspace} \dim(\paramspace_{\model}) > \varrho \numobs^{0.75}$ is meant to capture being in the ``small-data'' regime, 
where using very small bootstrap dataset sizes may result in unsatisfactory estimation accuracy. 
However, this precise condition may not always be appropriate;
for example, it could be more appropriate to instead use $\max_{\model \in \modelspace} \dim(\paramspace_{\model}) > \varrho \numobs^{0.75}$ when the models are nested. 

\begin{algorithm}[tb] %
\SetKwInput{Input}{Input}
\Input{mismatch cutoff $\overline{\modelmismatch}$ (default: $0.3$), \newline
model size cutoff factor $\varrho$ (default: $1.0$).}
\IncMargin{1em}
Compute Bayes posterior on $\paramspace$ under the litmus model. \;
Compute BayesBag posterior on $\paramspace$ under the litmus model, using $\bsnumobs =  \numobs$. \;
Compute mismatch index $\modelmismatch$ using the results from steps 1 and 2. \; 
\eIf{$\modelmismatch < \overline{\modelmismatch}$ or $\sum_{\model \in \modelspace} \dim(\paramspace_{\model}) > \varrho \numobs^{0.75}$}{
	Compute BayesBag posterior on $\modelspace$ using $\bsnumobs =  \numobs^{0.95}$.
}{
	Compute BayesBag posterior on $\modelspace$ using $\bsnumobs =  \numobs^{0.75}$.
}
\caption{Recommended workflow for BayesBag model selection}
\label{alg:using-BayesBag}
\end{algorithm}
\DecMargin{1em}

\section{Simulation study} \label{sec:linreg-model-selection-sim}

To validate our theory and assess the performance of BayesBag for model selection, we carry out a simulation study in the setting of feature selection for linear regression.

\paragraph*{Model} 
The data consist of regressors $Z_{n} \in \reals^{D}$ and observations $Y_{n} \in \reals$ for $n=1,\dots,\numobs$, 
and the goal is to predict $Y_{n}$ given $Z_{n}$. 
For each $\gamma\in\{0,1\}^{D}$, define a model such that the $d$th regressor is included in the linear regression if and only if $\gamma_{d}= 1$.
Letting $D_{\gamma} \defined \sum_{d=1}^{D}\gamma_{d}$ denote the number of regressors in model $\gamma$ and 
$k^{\star} \in \{1,\dots,D\}$ denote the maximum number of regressors to include, we consider a collection of models 
$\modelspace_{k^{\star}} \defined \{ \gamma \in \{0,1\}^{D} \given D_{\gamma} \le k^{\star} \}$.
Let $Z \in \reals^{N \times D}$ denote the matrix with the $n$th row equal to $Z_{n}$ and let $Z_{\gamma}$ denote 
the submatrix of $Z$ that includes the $d$th column if and only if $\gamma_{d} = 1$. 
Conditional on $\gamma \in \modelspace_{k^{\star}}$, the assumed model is 
\[
\sigma^{2} &\dist \distInvGam(a_{0}, b_{0}) \\
\beta_{d} &\given \sigma^{2} \distiid \distNorm(0, \sigma^{2}/\lambda),  & d=1,\dots,D_{\gamma} \\
Y_{n} &\given Z_{\gamma}, \beta, \sigma^{2} \distind \distNorm(Z_{\gamma,n}^{\top}\beta, \,\sigma^{2}), & n=1,\dots,\numobs. 
\]
We parameterize the model as 
$\param = (\param_{0}, \dots, \param_{D_{\gamma}}) = (\log \sigma^{2}, \beta_{1},\dots,\beta_{D_{\gamma}}) \in \paramspace_{\gamma} = \reals^{D_{\gamma}+1}$.
To perform posterior inference for $\gamma$, we analytically compute the marginal likelihood for each $\gamma \in \modelspace_{k^{\star}}$,
integrating out $\sigma^{2}$ and $\beta$; specifically, for $Y \defined (Y_{1},\dots,Y_{\numobs})^{\top}$, we use
\[
\modelmarginallik{Y}{Z, \gamma}  
&= \frac{b_{0}^{a_{0}}\Gamma(a_{0} + \numobs/2)}{(2\pi)^{\numobs/2}\Gamma(a_{0})} \frac{\lambda^{D_{\gamma}/2}}{b_{\gamma}^{a_{0} + \numobs/2}|\Lambda_{\gamma}|^{1/2}},
\]
where $\Lambda_{\gamma} \defined Z_{\gamma}^{\top}Z_{\gamma} + \lambda I$
and $b_{\gamma} \defined b_{0} +  Y^{\top}(I - Z_{\gamma}\Lambda_{\gamma}^{-1}Z_{\gamma}^{\top})Y/2$.
For the prior on $\gamma \in \modelspace_{k^{\star}}$, we let $\modelpriordist(\gamma) \propto q_{0}^{D_{\gamma}}(1 - q_{0})^{D - D_{\gamma}}$,
where $q_{0} \in (0,1)$ is the prior inclusion probability of each component. 
Thus, the posterior probability of model $\gamma$ is  
\[
\modelpostdistfull{\gamma}{Y, Z} 
= \frac{\modelmarginallik{Y}{Z, \gamma}\modelpriordist(\gamma)}{\sum_{\gamma' \in \modelspace_{k^{\star}}}\modelmarginallik{Y}{Z, \gamma'}\modelpriordist(\gamma')}
\]
and the \emph{posterior inclusion probability} of the $d$th regressor is
\[
\modelpostdistfull{\gamma_{d}=1}{Y, Z}  \defined  \frac{\sum_{\gamma \in \modelspace_{k^{\star}}}\gamma_{d}\;\modelmarginallik{Y}{Z, \gamma}\modelpriordist(\gamma)}{\sum_{\gamma' \in \modelspace_{k^{\star}}}\modelmarginallik{Y}{Z, \gamma'}\modelpriordist(\gamma')}. \label{eq:pip}
\]

\paragraph*{Data}
We simulate data by generating 
$Z_{n} \distiid G$, $\eps_{n} \distiid \distNorm(0,1)$, and 
\[
Y_{n} = f(Z_{n})^{\top}\beta_{\dagger} + \eps_{n}  \label{eq:simulated-linreg-data}
\]
for $n =1,\dots,N$, with the regressor distribution $G$, the regression function $f$, and the coefficient vector $\beta_{\dagger} \in \reals^{D}$ as described next. 
Using the \textsf{linear} regression function $f(z) = z$ results in well-specified data. 
To generate misspecified data, we use the \textsf{nonlinear} component-wise cubic function $f(z) = (z_{1}^{3},\ldots,z_{D}^3)^{\top}$.
We choose $G$ and $\beta_{\dagger}$ in the spirit of genome-wide association study fine-mapping~\citep{Schaid:2018}
to simulate a scenario with many highly correlated regressors, of which only a few regressors are actually employed in the data-generating process. 
For $k \in \{1,2\}$, we use a \textsf{$k$-sparse} vector (that is, a vector with $k$ non-zero components)
defined by setting $\beta_{\dagger d} = 1$ if $d \in \{ \lfloor j (D+\tfrac{1}{2})/(k+1)\rfloor \given j = 1,\dots,k\}$ and $\beta_{\dagger d} = 0$ otherwise. 
For $h > 2$ and $\psi > 0$, $Z \dist G$ is defined by generating
$\xi\dist\chi^{2}(h)$ and then 
$Z \given \xi \dist\distNorm(0,\Sigma)$, where the $(d, d')$ entry of $\Sigma \in \reals^{D \times D}$ is given by $\Sigma_{dd'} = \exp\{-(d-d')^{2}/\psi^{2}\} / (\xi_{d}\xi_{d'})$, and
$\xi_{d} = \sqrt{\xi / (h-2)}$ if $d$ is odd and $\xi_{d} = 1$ otherwise. 
The motivation for this data simulation procedure is to generate correlated regressors that have different
tail behaviors while still having the same first two moments, since regressors are typically standardized to have mean 0 and variance 1. 
Note that, marginally, $Z_{1}, Z_{3}, \dots$ are each rescaled $t$-distributed random variables with $h$ degrees of freedom such that $\var(Z_{1}) = 1$,
and $Z_{2}, Z_{4},\dots$ are $\distNorm(0,1)$.

\paragraph*{Experimental conditions}
We generate datasets under the \textsf{$k$-sparse-linear} and \textsf{$k$-sparse-nonlinear} settings according to \cref{eq:simulated-linreg-data}
with $h = 10$, $\psi = 8$, and either $(D,k) = (10, 1)$ or $(D, k) = (20, 2)$. 
We set $q_{0} = k/D$ and the model hyperparameters to $a_{0} = 2$, $b_{0} = 1$, and $\lambda = 16$,
with the latter setting helping to penalize the addition of extraneous features. 
We consider $\bsnumobs = \numobs^{\alpha}$ for $\alpha \in \{1, 0.95, 0.75, 0.55 \}$.
We consider $k^{\star} \in \{1, 2\}$ for \textsf{$1$-sparse} data and $k^{\star} = 2$ for \textsf{$2$-sparse} data.
We then compute the posterior inclusion probabilities as defined in \cref{eq:pip}. 
Each experimental condition is replicated 50 times, resulting in 50 posterior inclusion probabilities for each regressor in each experimental setting. 

\begin{figure}[tbp]
\begin{center}
\begin{subfigure}[b]{\textwidth}
\centering
\includegraphics[height=1.05in]{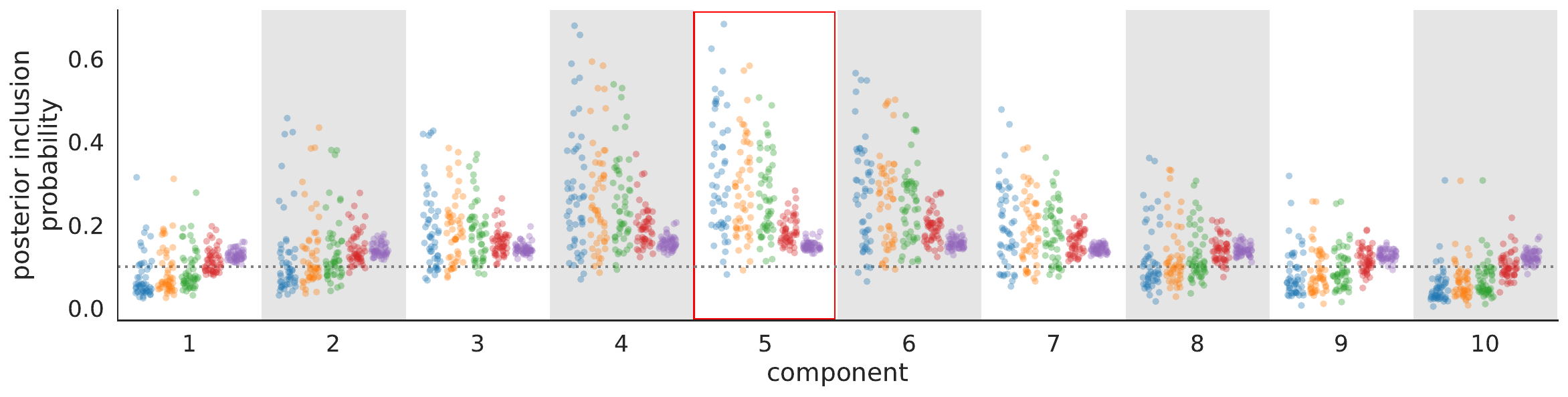}
\caption{\textsf{1-sparse-linear}, $N = 5 \times 10^{1}$}
\end{subfigure}
\begin{subfigure}[b]{\textwidth}
\centering
\includegraphics[height=1.05in]{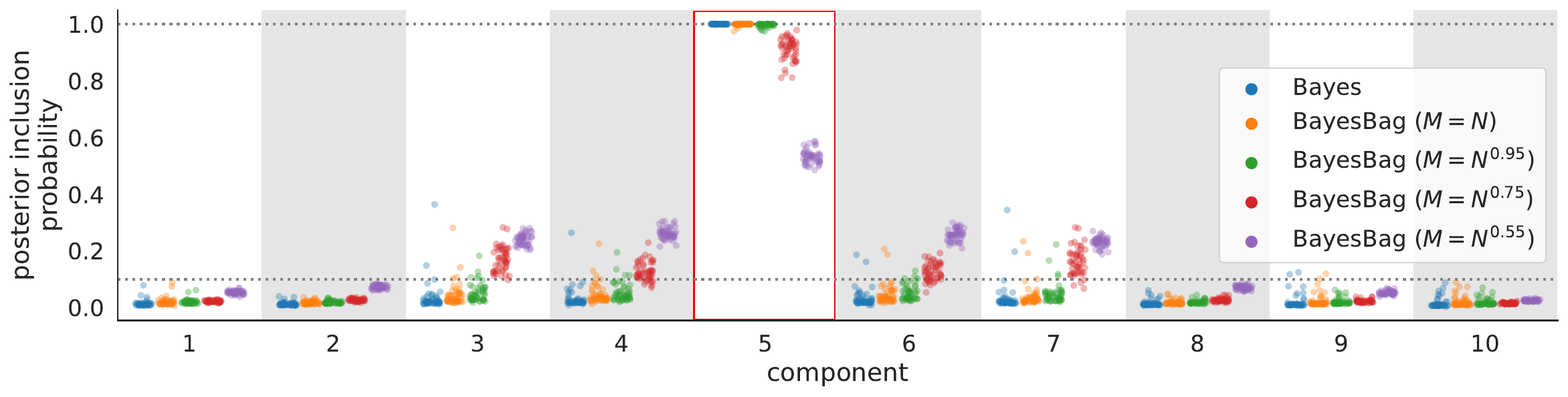}
\caption{\textsf{1-sparse-linear}, $N = 5 \times 10^{3}$}
\end{subfigure} 
\begin{subfigure}[b]{\textwidth}
\centering
\includegraphics[height=1.05in]{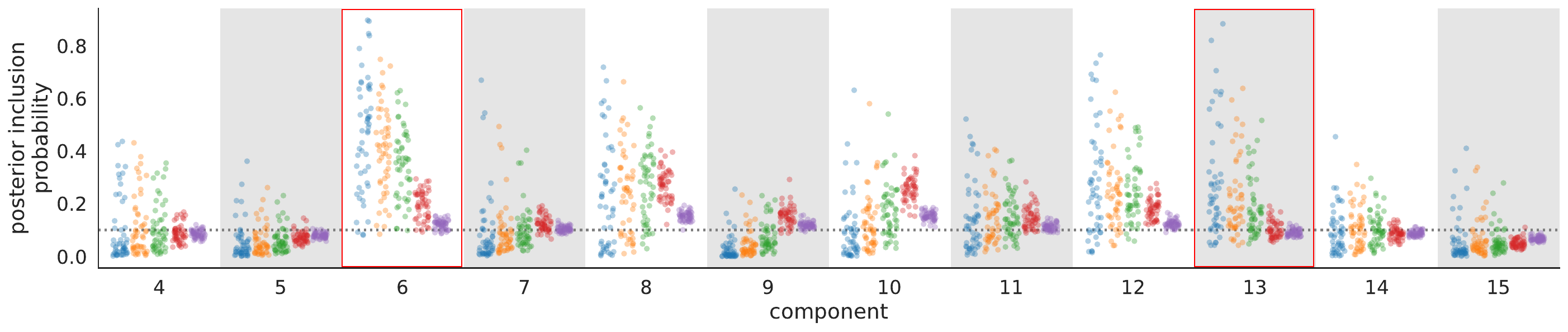}
\caption{\textsf{2-sparse-linear}, $N = 10^{2}$}
\end{subfigure} 
\begin{subfigure}[b]{\textwidth}
\centering
\includegraphics[height=1.05in]{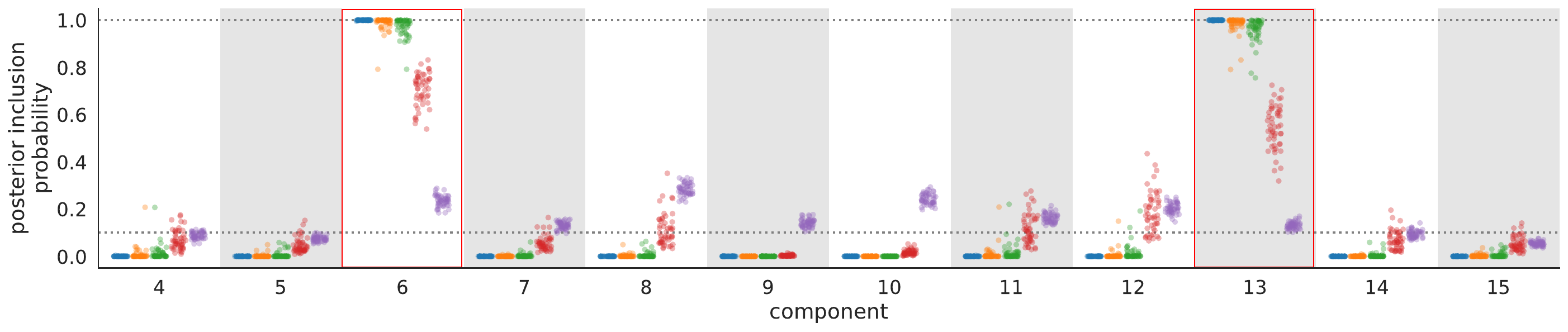}
\caption{\textsf{2-sparse-linear}, $N = 10^{3}$}
\end{subfigure}
\caption{Simulation results for feature selection in linear regression with $k^{\star} = 2$ when the model contains the true distribution.
The BayesBag posterior inclusion probabilities are similar to the Bayes posterior inclusion probabilities, but tend to shrink toward the prior inclusion probability (lower horizontal dotted line).
The data was generated from the assumed model, $Y_{n} = Z_{n}^{\top}\beta_{\dagger} + \eps_{n}$ for $n=1,\ldots,N$, where 
$\eps_{n}\sim\distNorm(0,\sigma^2)$ with $\sigma^2 = 1$,
$Z_{n}\in\reals^D$ is a vector of covariates, and $\beta_{\dagger}\in\reals^D$ is a \textsf{$k$-sparse} vector,
that is, $\beta_{\dagger}$ has $k$ non-zero components. 
The prior on inclusion vectors $\gamma\in\{0,1\}^D$ is proportional to $q_0^{\sum \gamma_d} (1-q_0)^{D - \sum \gamma_d}$, where $q_0 = k/D$,
with the constraint that $\sum_{d=1}^D \gamma_d \leq k^{\star}$.
A conjugate prior is placed on the coefficients and $\sigma^2$ given $\gamma$.
The posterior inclusion probabilities %
were computed by analytically integrating out the parameters and summing over all binary inclusion vectors $\gamma$.
The figure shows results for simulations using 
(a) $D = 10$, $N = 50$, $k=1$, 
(b) $D = 10$, $N = 5{,}000$, $k=1$, 
(c) $D = 20$, $N = 100$, $k=2$, and
(d) $D = 20$, $N = 1{,}000$, $k=2$.
For each of these settings, 50 replicate datasets were generated, and the resulting posterior inclusion probabilities are shown.
Components that were actually nonzero when generating the data are enclosed by red rectangles. 
}
\label{fig:linreg-feature-selection-linear-correlated-pips}
\end{center}
\end{figure}

\begin{figure}[tbp]
\begin{center}
\begin{subfigure}[b]{\textwidth}
\centering
\includegraphics[height=1.25in]{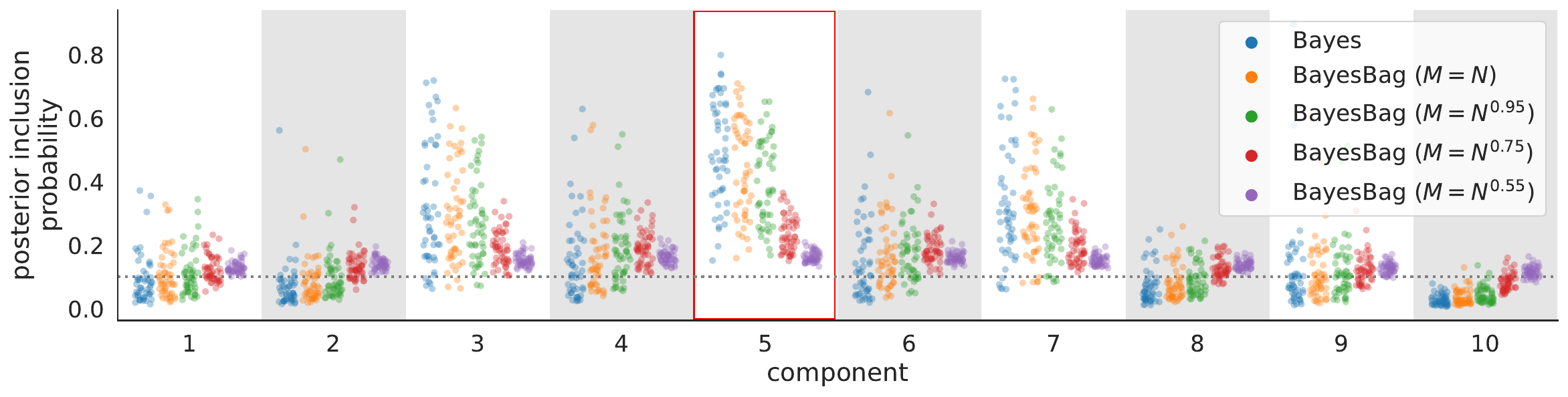}
\caption{$N = 5 \times 10^{1}$}
\end{subfigure}
\begin{subfigure}[b]{\textwidth}
\centering
\includegraphics[height=1.25in]{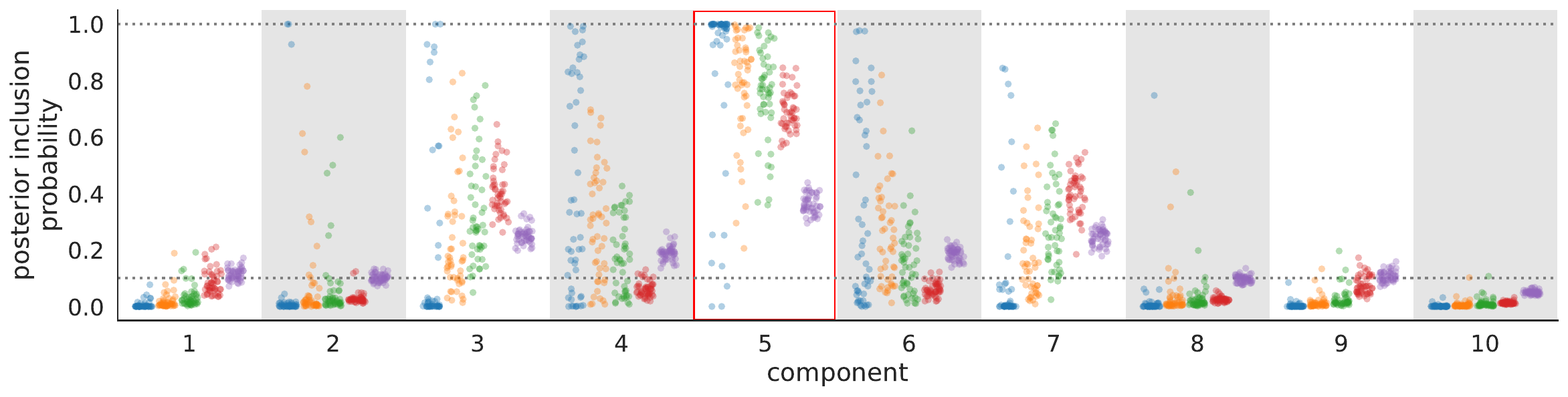}
\caption{$N = 5 \times 10^{2}$}
\end{subfigure} 
\begin{subfigure}[b]{\textwidth}
\centering
\includegraphics[height=1.25in]{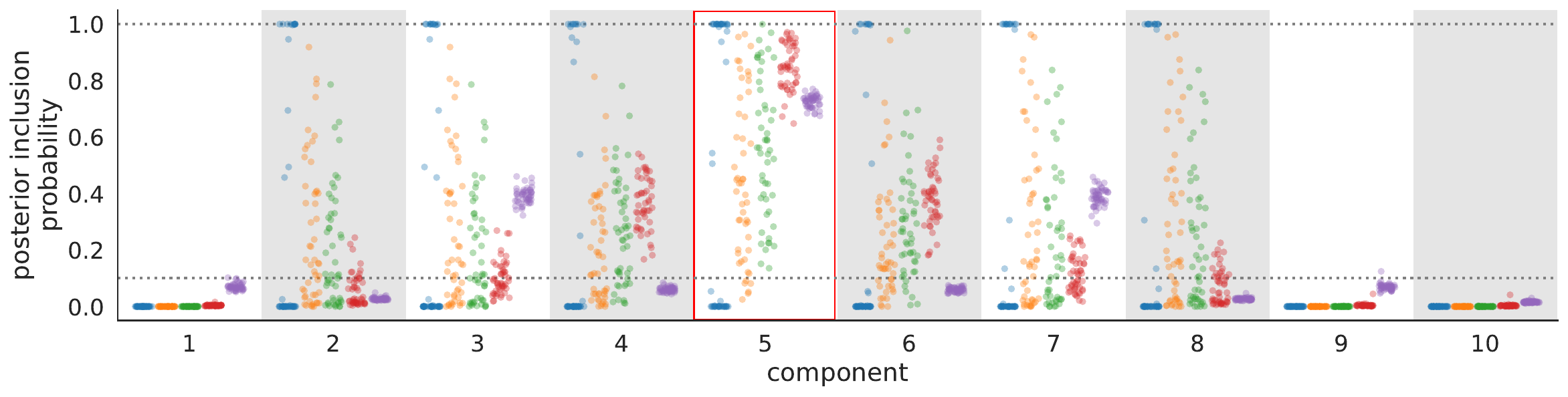}
\caption{$N = 5 \times 10^{3}$}
\end{subfigure}
\begin{subfigure}[b]{\textwidth}
\centering
\includegraphics[height=1.25in]{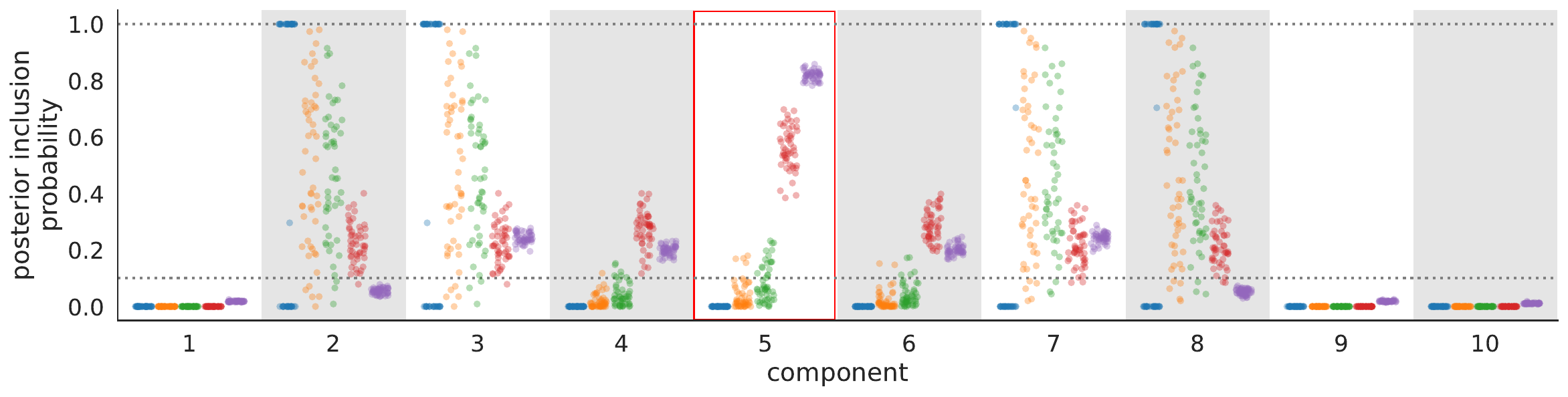}
\caption{$N = 5 \times 10^{4}$}
\end{subfigure}
\caption{Simulation results for feature selection in linear regression when the model is misspecified and $k^{\star} = 2$.
Everything is the same as in \cref{fig:linreg-feature-selection-linear-correlated-pips}(a,b), except that the data was generated using
$Y_{n} = f(Z_{n})^{\top}\beta_{\dagger} + \eps_{n}$, where $f(z) = (z_{1}^{3},\ldots,z_{D}^3)^{\top}$.
Results are shown for (a) $N = 50$, (b) $N = 500$, (c) $N = 5{,}000$, and (d) $N = 50{,}000$.
See the caption of \cref{fig:linreg-feature-selection-linear-correlated-pips} for further explanation.
The Bayes posterior inclusion probabilities show considerable instability both (i) across datasets with $\numobs$ fixed and (ii) as $\numobs$ increases. 
Meanwhile, the BayesBag probabilities are much more stable, particularly for $\bsnumobs = \numobs^{\alpha}$ with $\alpha \le 0.75$. 
The component that was actually nonzero when generating the data is enclosed by a red rectangle; see the text for interpretation.
}
\label{fig:linreg-feature-selection-correlated-pips-1-k-star-2}
\end{center}
\end{figure}

\begin{figure}[tbp]
\begin{center}
\begin{subfigure}[b]{.49\textwidth}
\centering
\hspace{-.1\textwidth}\includegraphics[width=1.1\textwidth]{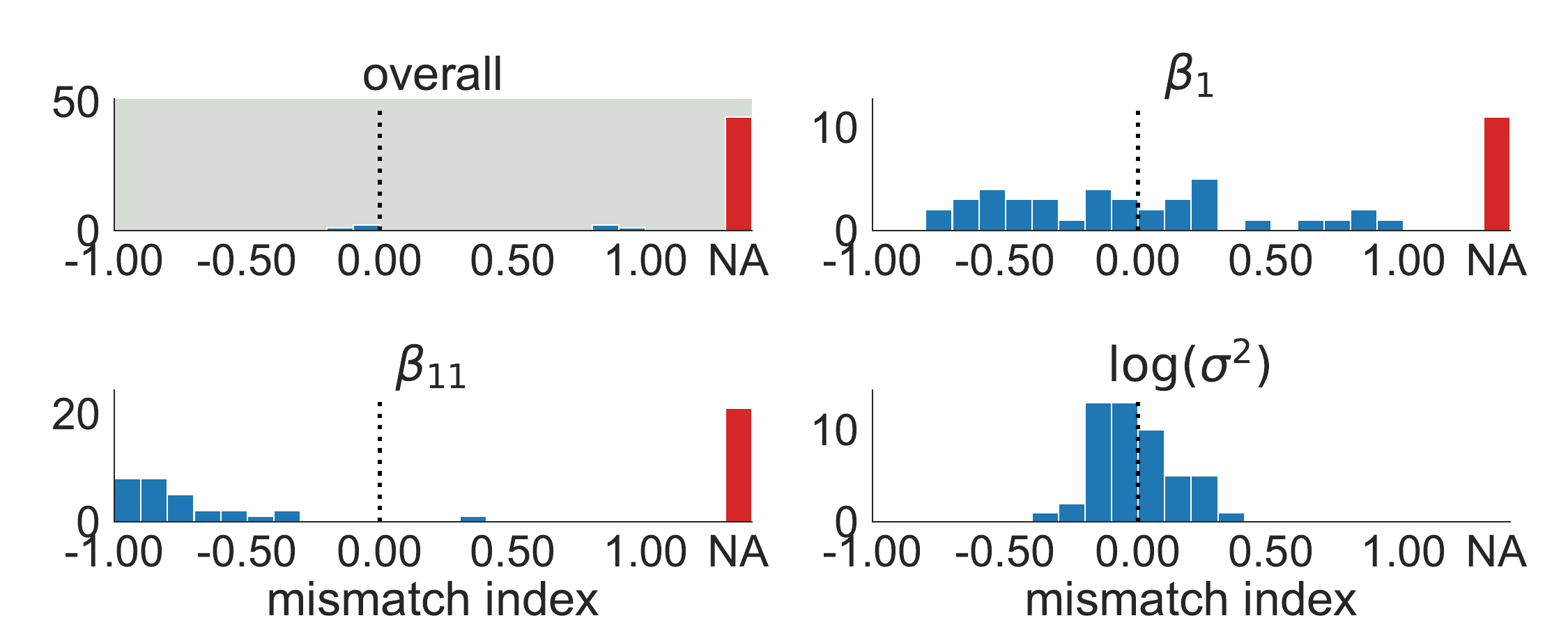}
\caption{\textsf{2-sparse-linear}, $N= 10^{2}$}
\end{subfigure} 
\begin{subfigure}[b]{.5\textwidth}
\centering
\includegraphics[width=1.078\textwidth]{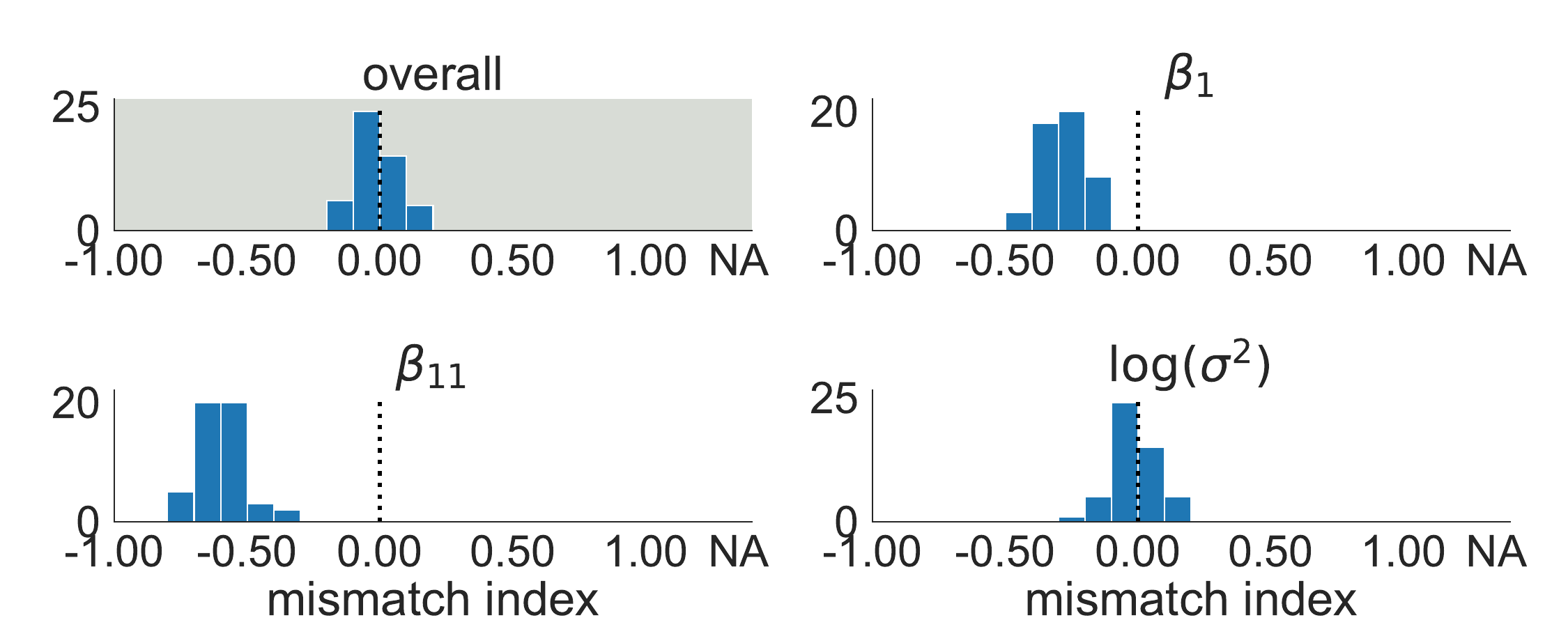}
\caption{\textsf{2-sparse-linear}, $N= 10^{4}$}
\end{subfigure}  \\
\begin{subfigure}[b]{.49\textwidth}
\centering
\hspace{-.1\textwidth}\includegraphics[width=1.1\textwidth]{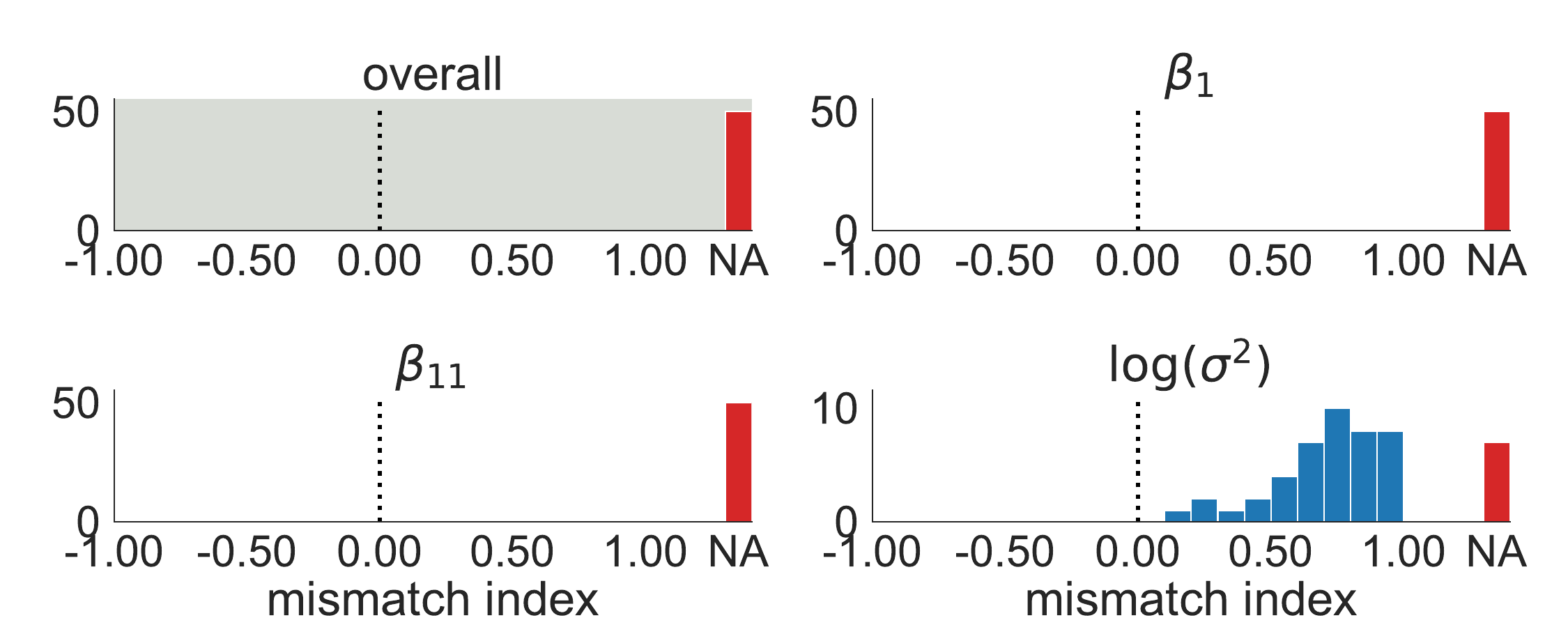}
\caption{\textsf{2-sparse-nonlinear}, $N= 10^{2}$}
\end{subfigure}
\begin{subfigure}[b]{.5\textwidth}
\centering
\includegraphics[width=1.078\textwidth]{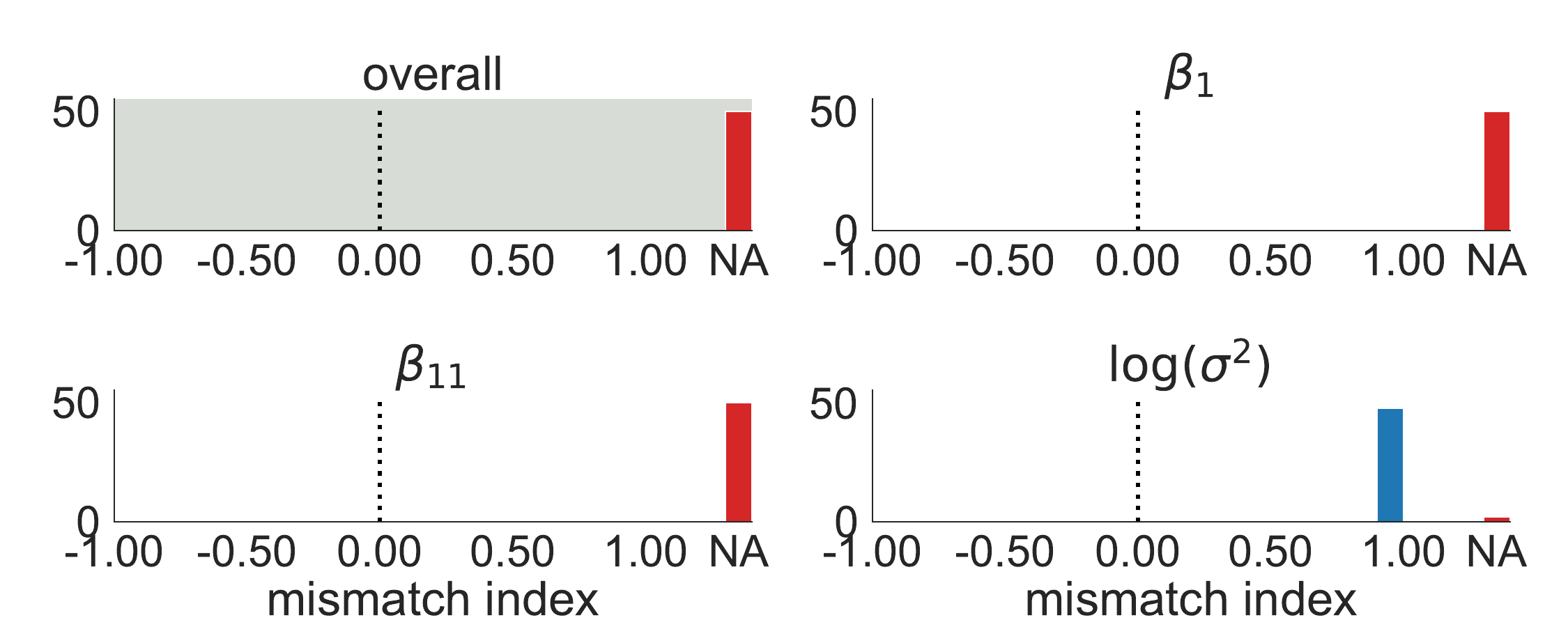}
\caption{\textsf{2-sparse-nonlinear}, $N= 10^{4}$}
\end{subfigure} 
\caption{Model--data mismatch indices $\modelmismatch$ from the simulations on feature selection in linear regression.
The overall $\modelmismatch$ value and the $\modelmismatch$ for selected parameters are shown in the case of \textsf{2-sparse} data ($k = 2$) with $D = 20$ regressors.
The figure shows histograms of $\modelmismatch$ over 50 replicate datasets 
generated using the well-specified linear case of $f(z) = z$ (as in \cref{fig:linreg-feature-selection-linear-correlated-pips}) 
with (a) $N = 100$ and (b) $N = 10{,}000$,
and the misspecified nonlinear case of $f(z) = (z_{1}^{3},\ldots,z_{D}^3)^{\top}$ (as in \cref{fig:linreg-feature-selection-correlated-pips-1-k-star-2})
with (c) $N = 100$ and (d) $N = 10{,}000$.
We only display two components of $\beta$ since the $\modelmismatch$ values follow fairly similar distributions for all components.
The results show that in the well-specified setting, when $N$ is sufficiently large (panel (b)), $\modelmismatch$ tends to be near zero, indicating correct specification as expected.
An exception is that the $\modelmismatch$ value for $\beta_{1 1}$ is closer to $-1$, indicating that the Bayes posterior on $\beta_{1 1}$ may be somewhat underconfident.
When $N$ is small (panel (a)), $\modelmismatch$ is often $\nan$ in these simulations, reflecting the poor identifiability of the coefficients
due to strong correlation in the regressors.
Meanwhile, in the misspecified setting (panels (c) and (d)), $\modelmismatch$ is typically $\nan$ for the coefficients, reflecting that the model is misspecified 
and there may also be identifiability issues.
}
\label{fig:linreg-feature-selection-correlated-model-mismatch-2-sparse}
\end{center}
\end{figure}

\paragraph*{Results} 
We are interested in verifying the theory of \cref{sec:model-selection} in the finite-sample regime,
which suggests that when the model is misspecified, similar models may be assigned wildly varying probabilities under the usual posterior (Bayes),
while the bagged posterior (BayesBag) probabilities will tend to be more balanced.
Figures \ref{fig:linreg-feature-selection-linear-correlated-pips}, \ref{fig:linreg-feature-selection-correlated-pips-1-k-star-2}, and \ref{fig:linreg-feature-selection-correlated-pips-1-k-star-1} to \ref{fig:linreg-feature-selection-weakly-correlated-pips-2}
show the Bayes and BayesBag posterior inclusion probabilities for each component, for all 50 replications. 
First, \cref{fig:linreg-feature-selection-linear-correlated-pips} shows that when the model is correctly specified, the Bayes and BayesBag posteriors with $\alpha \ge 0.95$ behave similarly.
However, BayesBag can be more stable even in this well-specified setting, exhibiting fewer outlier posterior inclusion probabilities. 
As $\alpha$ decreases, BayesBag yields substantially more conservative inferences, in the sense that the posterior inclusion probabilities tend to shrink toward the prior inclusion probability.

In the misspecified setting, the results 
are more interesting and subtle (Figs.~\ref{fig:linreg-feature-selection-correlated-pips-1-k-star-2},  \ref{fig:linreg-feature-selection-correlated-pips-1-k-star-1}-\ref{fig:linreg-feature-selection-weakly-correlated-pips-2}). 
Due to the misspecification and correlated regressors, it no longer holds in general that the components that were actually non-null in the data-generating process 
will be selected (see \cref{app:linreg-feature-selection} and \citealp{Buja:2019:models-1,Buja:2019:models-2}). 
For the \textsf{1-sparse-nonlinear} data, when $k^{\star} = 1$, 
the Bayes and BayesBag posteriors behave quite similarly and concentrate on component 5, which is asymptotically optimal;
see \Cref{fig:linreg-feature-selection-correlated-pips-1-k-star-1}. 
However, when $k^{\star} = 2$, two models are asymptotically optimal and equivalent, 
namely, the models with $\supp(\gamma) = \{2,3\}$ and $\supp(\gamma) = \{7,8\}$, where $\supp(\gamma) \defined \{d : \gamma_d \neq 0\}$.
Meanwhile, $\{4,5\}$ and $\{5,6\}$ are asymptotically equivalent but slightly less-than-optimal. 
As shown in \cref{fig:linreg-feature-selection-correlated-pips-1-k-star-2}, in this case the Bayes posterior is unstable and, for large values of $\numobs$,
concentrates on $\{2,3\}$ or $\{7,8\}$ with equal probability. 
For $\bsnumobs  \in \{\numobs, \numobs^{0.95}\}$, BayesBag places roughly uniformly distributed mass on the same four components for large values of $\numobs$. 
Meanwhile, for $\bsnumobs  \in \{\numobs^{0.75}, \numobs^{0.55}\}$, BayesBag is much more stable and puts more mass on component 4, 5, and 6. 
Thus, we see exactly the behaviors predicted by the asymptotic analyses in \cref{sec:model-selection}.
We defer discussion of the results for \textsf{2-sparse-nonlinear} data (\cref{fig:linreg-feature-selection-correlated-pips-2,fig:linreg-feature-selection-weakly-correlated-pips-2}) 
to \cref{app:additional-figs}.

\Cref{fig:linreg-feature-selection-correlated-model-mismatch-1-sparse,fig:linreg-feature-selection-correlated-model-mismatch-2-sparse} 
show model--data mismatch index values
for the litmus model with $\gamma_{d} = 1$ for all $d = 1,\dots,D$,
on a representative subset of experimental configurations.
For the \textsf{k-sparse-linear} data, the overall mismatch indices were either near zero or were $\nan$, reflecting that the model is correctly specified 
but there are some issues with poor identifiability.
For the \textsf{k-sparse-nonlinear} data, the mismatch indices were nearly all $\nan$, reflecting that the model is misspecified 
and there may also be identifiability issues.

\paragraph{Summary}
Overall, the simulation results are in agreement with our asymptotic theory from \cref{sec:model-selection}: 
the behavior of Bayes can vary dramatically with the 
dataset size and the degree of misspecification, whereas BayesBag is much more stable.
Additionally, the simulations provide insight into the behavior of the bagged posterior when $\bsnumobs$ is sublinear in $\numobs$.
Of particular note is that $\bsnumobs = \numobs^{0.95}$ yields noticeably improved stability with little loss of statistical efficiency.
Meanwhile, for settings with substantial misspecification, taking $\bsnumobs = \numobs^{\alpha}$ with $\alpha \in [0.55, 0.75]$ 
may be preferable -- with the caveat that inferences will tend to be much more conservative. 

\begin{table}[tp]
\caption{Real-world datasets used in experiments. LR = linear regression, PTR = phylogenetic tree reconstruction.
For LR, $\numobs =$ \# samples and $D =$ \# covariates.
For PTR, $\numobs =$ \# features and $D =$ \# species.}
\begin{center}
\begin{tabular}{lccc}
\toprule
Name 				& Model & $\numobs$ 	& $D$ \\
\midrule
California housing		& LR	& {20,650} 	& 8 \\
Boston housing 		& LR 	& 506		& 13 \\
Diabetes 			& LR 	& 442		& 10 \\
Residential building 	& LR 	& 371		& 105 \\
Whale mitochondrial coding DNA & PTR &  10,605 & 14 \\
Whale mitochondrial amino acids & PTR & 3,535 & 14 \\
\bottomrule
\end{tabular}
\end{center}
\label{tbl:datasets}
\end{table}%

\section{Applications} \label{sec:experiments}

\subsection{Feature selection for linear regression}

We compare Bayesian model selection and BayesBag model selection for linear regression on four real-world datasets, summarized in \cref{tbl:datasets}.
Based on our findings in \cref{sec:linreg-model-selection-sim}, for BayesBag we consider $\bsnumobs = \numobs^{\alpha}$ with $\alpha \in \{1.0, 0.95, 0.75\}$. 
We use a prior inclusion probability of $q_{0} = 3/D$ and use $k^{\star} = D$ for the maximum number of nonzero components, except on the residential building dataset, 
where for computational tractability we use $k^{\star} = 3$. 
We set the model hyperparameters to $a_{0} = 2$, $b_{0} = 1$, and $\lambda = 1$.

We expect the parameters to be well-identified for all datasets except the residential building dataset, since
the residential building dataset requires only 58 out of 104 principal components to explain 99\% of the variance, whereas for the other three datasets,
$D$ out of $D$ principal components are needed to explain $99\%$ of the variance. 
The model mismatch indices (for the litmus model with $\gamma_{d} = 1$ for all $d = 1,\dots,D$) 
are in agreement with expectations, since only the residential building dataset has a model mismatch index of $\nan$.
For the other datasets, the mismatch indices are $1.00$ (California housing), $0.62$ (Boston housing), and $0.03$ (Diabetes), 
which suggests that the model is misspecified for the two housing datasets.

\begin{figure}[tbp]
\begin{center}
\centering
\begin{subfigure}[b]{\textwidth}
\centering
\includegraphics[height=1.25in]{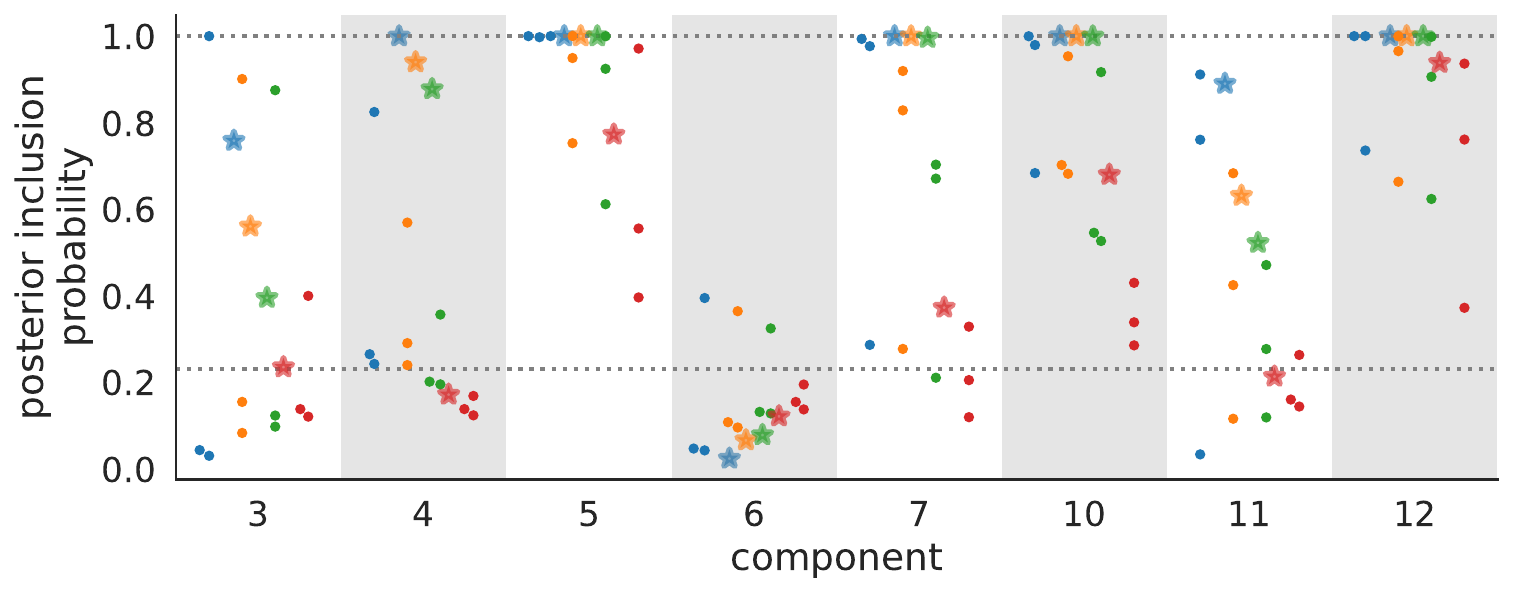}
\caption{Boston housing ($\modelmismatch =  0.62$)}
\end{subfigure}  \\
\begin{subfigure}[b]{.44\textwidth}
\centering
\includegraphics[height=1.25in]{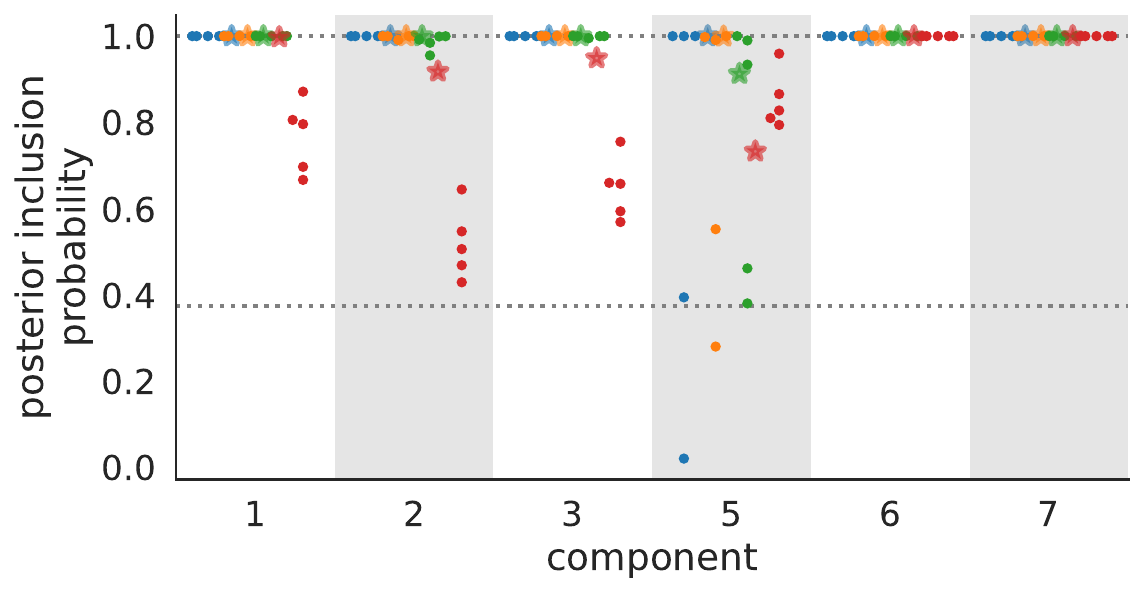}
\caption{California housing ($\modelmismatch = 1.00$)}
\end{subfigure}  
\begin{subfigure}[b]{.44\textwidth}
\centering
\includegraphics[height=1.25in]{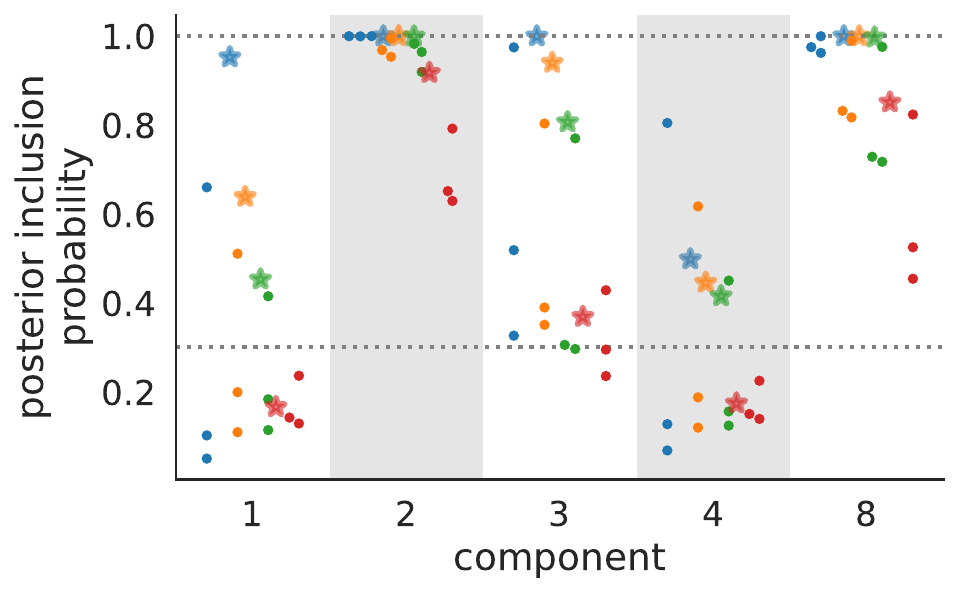}
\caption{Diabetes ($\modelmismatch = 0.03$)}
\end{subfigure} \\
\centering
\begin{subfigure}[b]{\textwidth}
\centering
\includegraphics[height=1.2in]{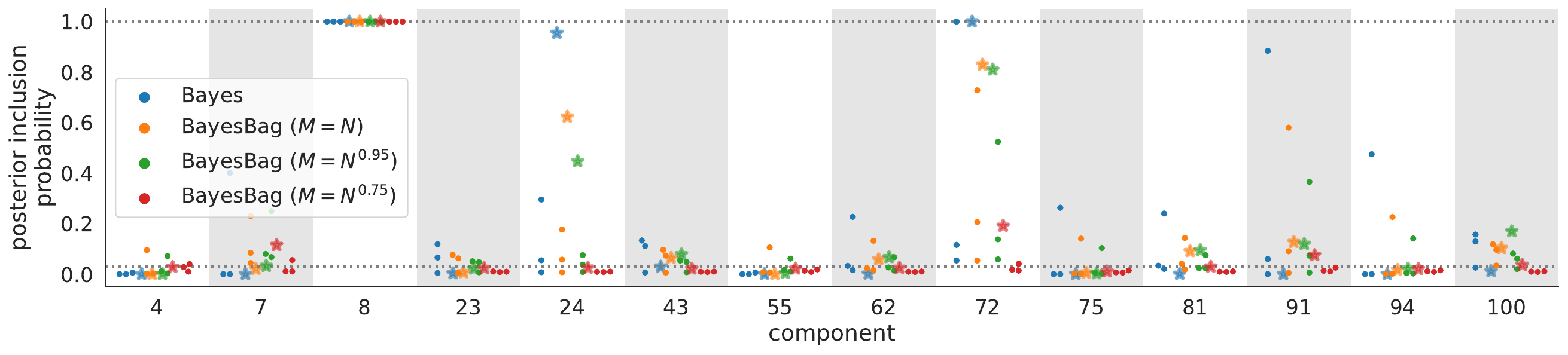}
\caption{Residential building ($\modelmismatch = \nan$)}
\end{subfigure} 
\caption{
Application to feature selection on four real-world datasets; see \cref{tbl:datasets} for dataset details.
The assumed model is the same as in \cref{sec:linreg-model-selection-sim} (see also the caption of \cref{fig:linreg-feature-selection-linear-correlated-pips}),
using a prior inclusion probability of $q_{0} = 3/D$, where $D$ is the number of regressors.
To assess reproducibility, we randomly split each dataset into roughly equally sized parts, and computed the posterior inclusion probabilities for each split separately (indicated by \textbullet) as well as for the full dataset (indicated by $\star$).
As before, the posterior inclusion probabilities are computed by analytically integrating out the parameters and summing out all possible binary inclusion vectors.
For the residential building dataset, we constrain the model to only allow up to three nonzero components, for computational tractability.
For visual readability, we only display the components with posterior inclusion probabilities above three times the prior inclusion probability (lower horizontal dotted line).
The BayesBag posterior inclusion probabilities exhibit greater reproducibility, 
in that (i) the between-split differences tend to be smaller
and (ii) the differences between the split posterior inclusion probabilities and the full data posterior inclusion probabilities 
also tend to be smaller for BayesBag than Bayes.
}
\label{fig:linreg-feature-selection-real}
\end{center}
\end{figure}

\Cref{fig:linreg-feature-selection-real} shows the posterior inclusion probabilities for all four datasets.
To compare the reliability of the methods, we also run each method on subsets of the data obtained by randomly dividing each dataset into roughly equally sized splits (\Cref{fig:linreg-feature-selection-real}).
We use three splits for all datasets except for California housing, for which we use five splits
since $N$ is substantially larger. 
Generally, across splits, BayesBag produced lower-variance, more conservative posterior inclusion probabilities that are more consistent with the 
posterior inclusion probabilities from the full datasets.
BayesBag with $\bsnumobs = \numobs^{0.75}$ is noticeably more conservative than Bayes and BayesBag with  $\bsnumobs \in \{\numobs^{0.95}, \numobs\}$;
for the two datasets with mismatch indices that suggest significant misspecification (California housing and residential building), 
such stability appears particularly desirable. 
These results are in agreement with the simulation results in \cref{sec:linreg-model-selection-sim}.

\begin{figure}[tbp]
\begin{center}
\begin{subfigure}[b]{\textwidth}
\centering
\includegraphics[height=1.25in]{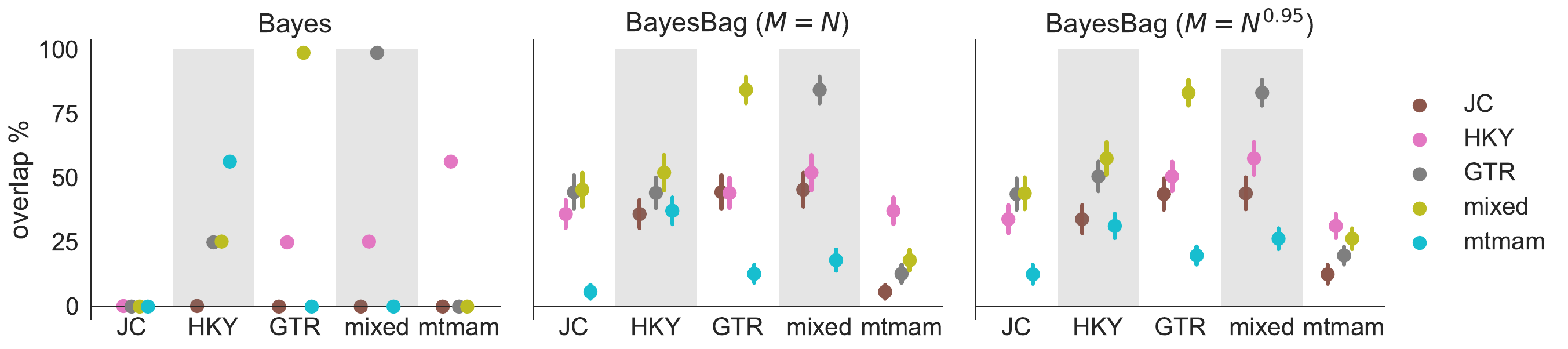}
\caption{Inter-model comparison}
\end{subfigure}
\begin{subfigure}[b]{\textwidth}
\centering
\includegraphics[height=1.25in]{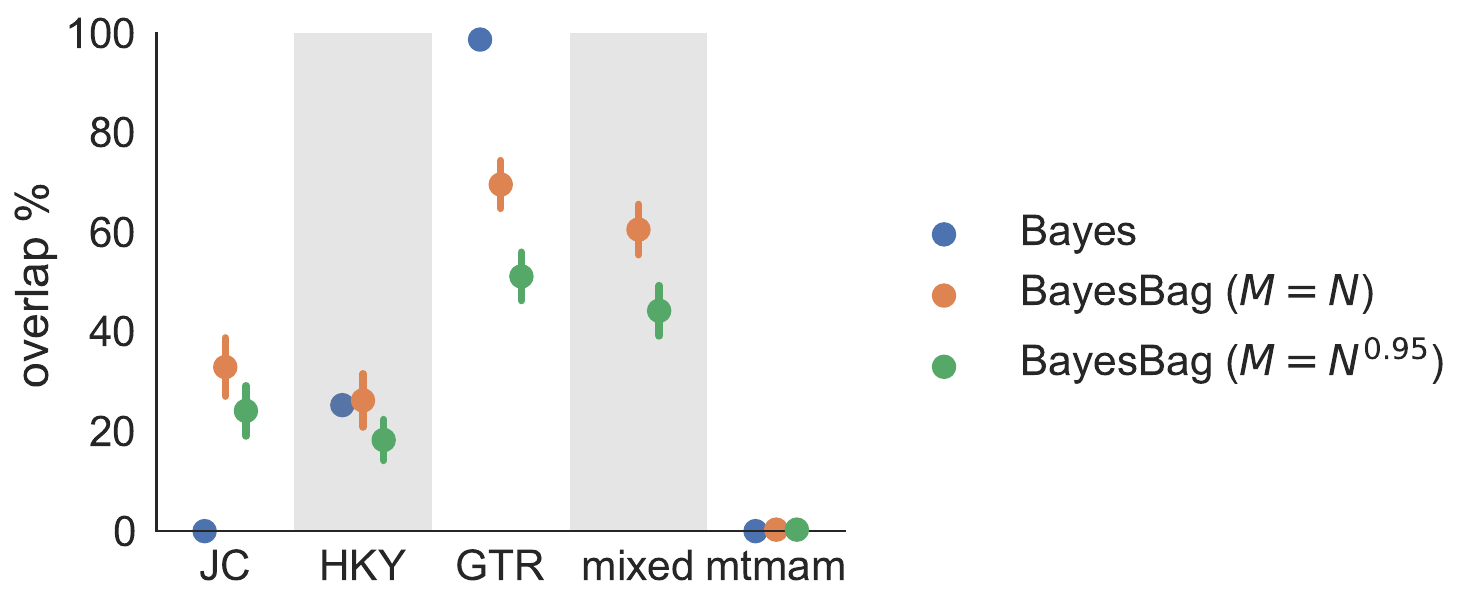}
\caption{Comparison to \textsf{mixed} using usual Bayesian posterior}
\end{subfigure}
\begin{subfigure}[b]{\textwidth}
\centering
\includegraphics[height=1.25in]{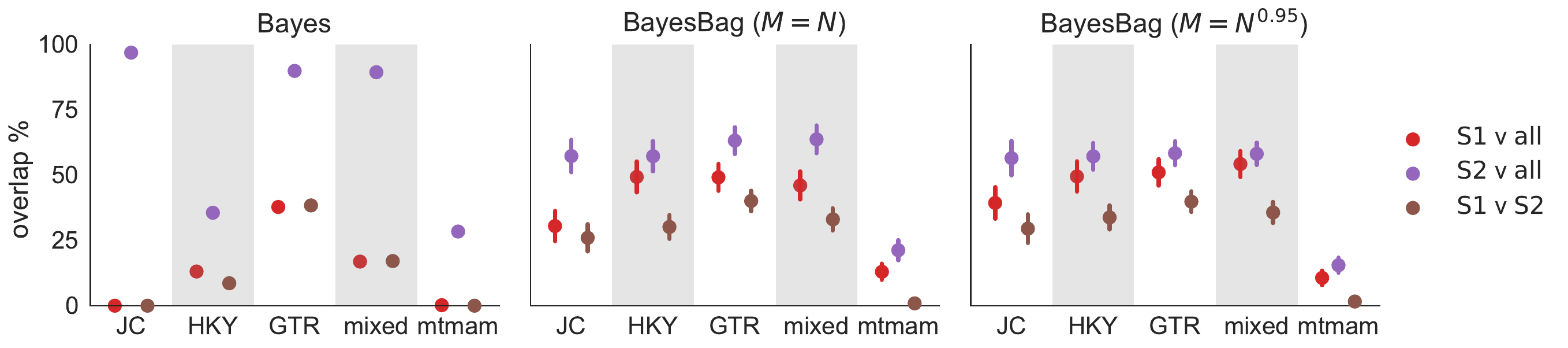}
\caption{Intra-model comparison}
\end{subfigure} 
\caption{
Application to phylogenetic tree inference on a whale genetics dataset.
To assess reproducibility, we computed the posterior under five different evolutionary models 
(\textsf{JC}, \textsf{HKY}, \textsf{GTR}, \textsf{mixed}, \textsf{mtmam}).
We quantify the similarity of posteriors by computing the overlapping probability mass of 99\% highest posterior density regions. 
To quantify uncertainty in the overlap due to Monte Carlo error, 
80\% confidence intervals are shown for the overlaps involving BayesBag.
Panel (a) shows the posterior overlap for each pair of models.
The usual posterior (Bayes) is quite sensitive to the choice of model,
exhibiting $\approx 0\%$ overlap in many cases, for instance, between JC and the other models.
Meanwhile, the bagged posterior (BayesBag) is more robust, exhibiting overlaps in a reasonable range.
Panel (b) shows the overlap between the Bayes posterior for the \textsf{mixed} model, which is the most flexible of the DNA models,
and the Bayes or BayesBag posterior for each other model.
Panel (c) shows the overlap when using the same model on different subsets of the data
--- specifically, splitting the genomic data for each species into two halves (\textsf{S1}, \textsf{S2})
or using \textsf{all} data.
}
\label{fig:whale-comparisons}
\end{center}
\end{figure}

\subsection{Phylogenetic tree reconstruction} \label{sec:phylogenetic-experiments}

Finally, we investigate the use of BayesBag for reconstructing the phylogenetic tree of a collection of species based on their observed characteristics. 
This is an important model selection problem due to the widespread use of phylogeny reconstruction algorithms.
Systematists have exhaustively documented that Bayesian model selection of phylogenetic trees can behave poorly.
In particular, the posterior can provide contradictory results depending on
what characteristics are used (for example, coding DNA or amino acid sequences), what evolutionary model is used, or which outgroups are 
included~\citep{Waddell:2002,Douady:2003,Wilcox:2002,Alfaro:2003,Huelsenbeck:2004,Buckley:2002,Lemmon:2004,Yang:2007}.
We illustrate how BayesBag model selection provides reasonable inferences that are significantly more robust to the choice of data and model. 

\paragraph{Models and data} 
We use the whale dataset from \citet{Yang:2008}, consisting of mitochondrial coding DNA from 13 whale species and the 
hippopotamus (\cref{tbl:datasets}). The hippopotamus is included as an ``outgroup'' species to identify the root of the tree, 
because the assumed evolutionary models are time-reversible and hence the trees are modeled as unrooted.
We consider four DNA models (\textsf{JC}, \textsf{HKY+C+$\Gamma_{5}$}, \textsf{GTR+$\Gamma$+I}, and \textsf{mixed+$\Gamma_{5}$}) and one amino acid
model (\textsf{mtmam+$\Gamma_{5}$});
see \citet{Yang:2008} for more details on these models. 
For brevity, we refer to the models as, respectively, \textsf{JC}, \textsf{HKY}, \textsf{GTR}, \textsf{mixed}, and \textsf{mtmam}.
To approximate the usual posterior (Bayes) and the bagged posterior (BayesBag), we use MrBayes 3.2~\citep{Ronquist:2012} with 2 independent runs, each with 4 coupled chains run 
for 1,000,000 total iterations (discarding the first quarter as burn-in). 
We confirm acceptable mixing using the built-in convergence diagnostics for MrBayes. 
For BayesBag, we take $B=100$ in all experiments and, since the number of models is very large, we only consider $\bsnumobs \in \{\numobs, \numobs^{0.95}\}$.

\paragraph{Evaluation}
Our goal is to investigate whether BayesBag avoids the self-contradictory inferences that Bayes produces.
To this end, we compare the output of different configurations of the data, model, and inference method, as follows.
We compute the set of trees in the 99\% highest posterior density (HPD) regions for each $\langle$data, model, inference method$\rangle$ configuration.
For selected pairs of configurations, 
we then compute the overlap of the two 99\% HPD regions in terms of (a) probability mass and (b) number of trees.
Since the BayesBag posterior is approximated via Monte Carlo as in \cref{eq:bayesbag-approximation}, 
we quantify the uncertainty in each overlap by reporting an 80\% confidence interval for the overlapping mass.
We compute these intervals using standard bootstrap methodology for a Monte Carlo estimate.

\paragraph{Results}
First, we look at the overlap between pairs of models.
As shown in \cref{fig:whale-comparisons}(a) and \cref{tbl:whale-cross-model-comparison}, there is substantially more overlap when using BayesBag. 
The difference is particularly noticeable when comparing \textsf{JC} (the simplest model) or \textsf{mtmam} (the amino acid model) to the other models.
When using Bayes, \textsf{JC} has either 0\% or (in one case) 0.2\% overlap with the other models while \textsf{mtmam} only overlaps with \textsf{HKY}.  
Thus, these pairs of models produce contradictory results when using Bayesian model selection.
On the other hand, when using BayesBag, all pairs of models have nonzero overlap, with typical amounts ranging from 30\% to 50\%. 
Hence, compared to Bayes, BayesBag provides results that are more consistent across models.

However, the good overlap between BayesBag posteriors does not necessarily mean that it is performing well, since it could simply be producing posteriors that are too diffuse, spreading the posterior mass over a very large number of trees.
Notably (as expected), BayesBag with $\bsnumobs = \numobs^{0.95}$ leads to a more diffuse posterior with 5--15 overlapping trees compared to 3--11 trees when $\bsnumobs = \numobs$. 
To further investigate the possibility of the BayesBag posteriors being too diffuse, we consider the overlap of the BayesBag posterior for each model and 
the Bayes posterior for \textsf{mixed}, which is the most complex of the DNA models. 
As shown in \cref{fig:whale-comparisons}(b) and \cref{tbl:whale-cross-model-comparison-to-mixed-G5}, all of the BayesBag posteriors (with the exception of \textsf{mtmam}) put 
substantial posterior probability on the 99\% HPD region of the Bayes \textsf{mixed} posterior.
Moreover, all but BayesBag \textsf{mtmam} has two trees in the overlap, which is the maximum possible since the Bayes \textsf{mixed} 99\% HPD region only
contains two trees. 
Finally, using BayesBag with $\bsnumobs = \numobs^{0.95}$ results in fairly small decreases (relative to BayesBag with $\bsnumobs = \numobs$) 
in the mass on the two trees in the Bayes \textsf{mixed}  99\% HPD region. 

Next, we perform intra-model comparisons by considering three datasets: the complete whale dataset (denoted \textsf{all}) and two additional datasets formed
by splitting the genomic data for each species in half (denoted \textsf{S1} and \textsf{S2}). 
Ideally, for each model, we hope to see substantial overlap when comparing the results across these three datasets (\textsf{all}, \textsf{S1}, and \textsf{S2}). 
However, when using the Bayes posterior, there is little to no overlap in many cases, particularly for the simpler \textsf{JC} model and \textsf{mtmam};
see \cref{fig:whale-comparisons}(c) and \cref{tbl:whale-self-consistency}.
Meanwhile, the BayesBag posteriors typically exhibit overlaps of between 21\% and 56\%, with less (though still nonzero) 
overlap with \textsf{mtmam}. 
These results suggest that BayesBag exhibits superior reproducibility in terms of uncertainty quantification. 

Finally, we compute the mismatch index for each model on the complete whale dataset, obtaining  
0.23 (\textsf{JC}), \nan\ (\textsf{HKY}), 0.47 (\textsf{GTR}), 0.84 (\textsf{mixed}), and 0.34 (\textsf{mtmam}). 
These mismatch indices suggest significant amounts of model misspecification, with the simpler \textsf{JC} model likely underestimating 
the actual degree of misspecification.
Thus, using the BayesBag posterior with $\bsnumobs = \numobs^{0.95}$ appears to be advisable. 

\section{Discussion} \label{sec:discussion}

In this paper, we have developed an approach to overcome the instability of Bayesian model selection when the models are all misspecified. 
This type of misspecification is common in scientific settings where idealized but interpretable models are commonly used (such as in systematics, population and cancer genetics, and economics). 
Our bagged posterior approach is theoretically justified, easy to use, and widely applicable. 
However, we see three potential limitations in practice. 
The first is that bagged posterior model selection tends to be more conservative, with posterior model probabilities farther from the extremes of zero and one.
The recommended workflow discussed in \cref{sec:workflow} is designed to at least partially ameliorate this issue,
however, this conservative behavior may be a necessary price for greater stability and reliability. 
The second limitation is the additional computational cost required for the naive estimation of the bagged model probabilities.
The development of more computationally efficient alternatives is an important direction for future work.
A final limitation is that our asymptotic theory only covers cases where the observations are independent.
Extending the theory to cover important structured models like time-series and spatial models is another valuable direction for future work.

\subsection*{Acknowledgments}

Thanks to Pierre Jacob for bringing P.~B\"uhlmann's BayesBag paper to our attention 
and to Ziheng Yang for sharing the whale dataset and his MrBayes scripts.
Thanks also to Ryan Giordano and Pierre Jacob for helpful feedback on an earlier draft of this paper,
and to Peter Gr\"unwald, Natalia Bochkina, Mathieu Gerber, and Anthony Lee for helpful discussions.
Finally, thanks to the AE and three reviewers for their constructive comments; 
in particular, we thank the third reviewer, who provided numerous insightful comments 
that substantially enhanced the scope and readability of the paper.

\bibliographystyle{imsart-nameyear}

\bibliography{library,../bayesbag}

\newpage

\appendix 

\counterwithin{figure}{section}
\renewcommand{\thefigure}{\Alph{section}.\arabic{figure}}
\counterwithin{table}{section}
\renewcommand{\thetable}{\Alph{section}.\arabic{table}}

\section{Additional figures and tables} \label{app:additional-figs}

\begin{figure}[tbp]
\begin{center}
\begin{subfigure}[b]{\textwidth}
\centering
\includegraphics[height=1.25in]{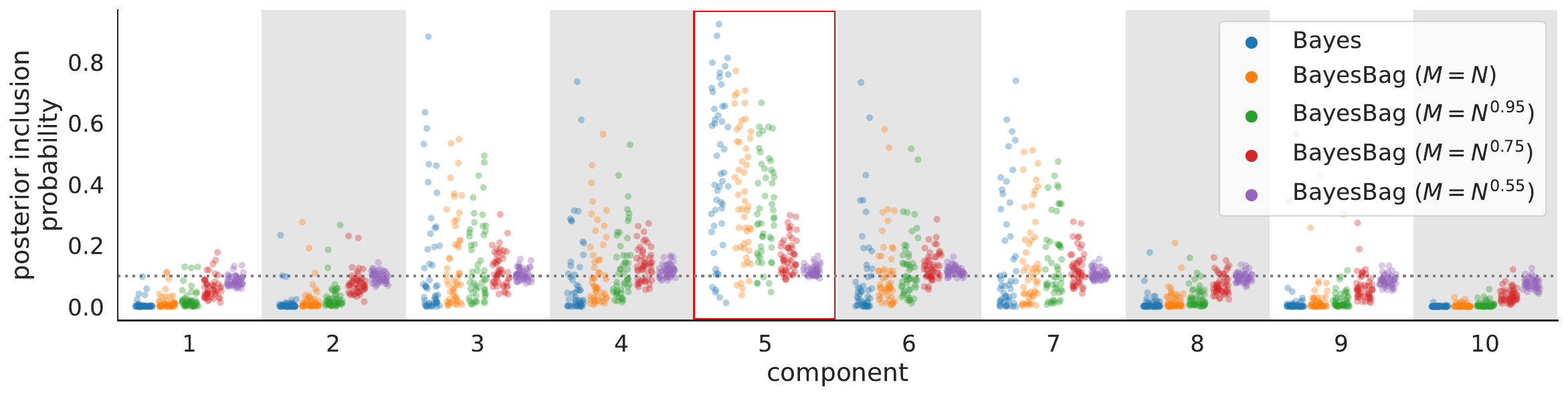}
\caption{$N = 5 \times 10^{1}$}
\end{subfigure}
\begin{subfigure}[b]{\textwidth}
\centering
\includegraphics[height=1.25in]{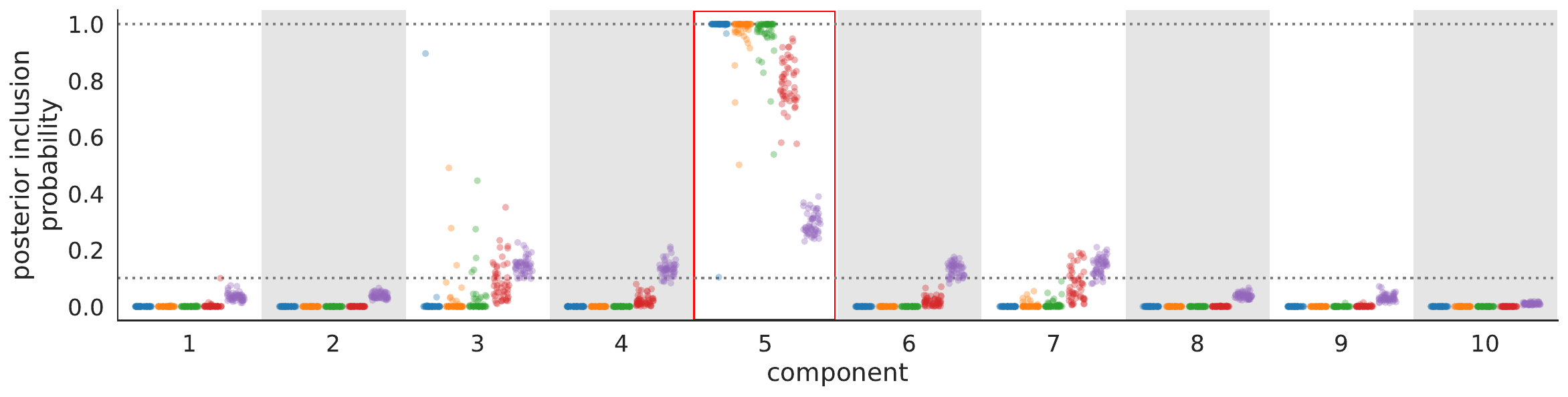}
\caption{$N = 5 \times 10^{2}$}
\end{subfigure} 
\begin{subfigure}[b]{\textwidth}
\centering
\includegraphics[height=1.25in]{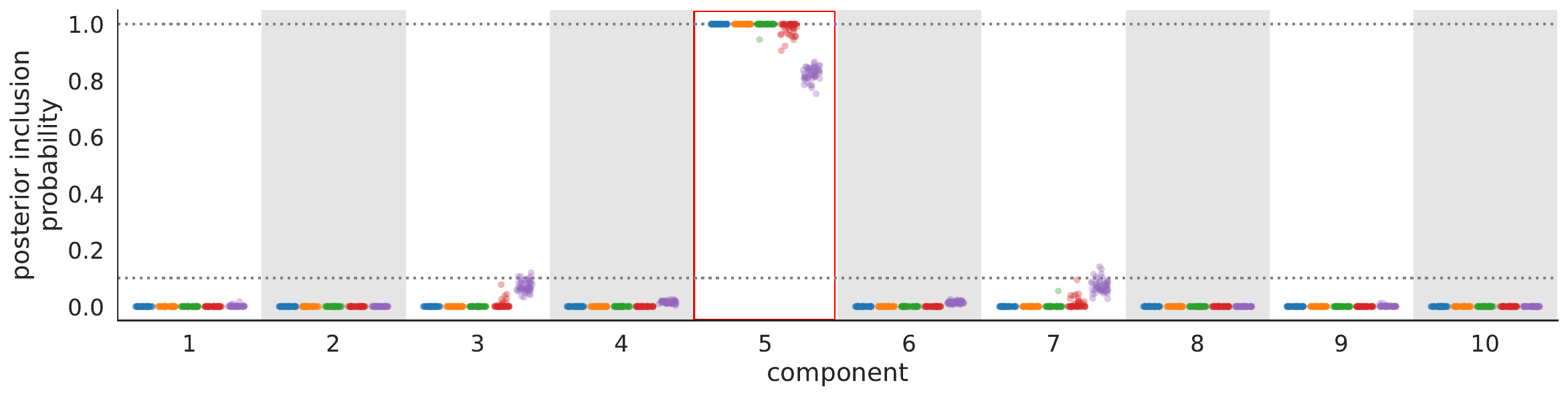}
\caption{$N = 5 \times 10^{3}$}
\end{subfigure}
\begin{subfigure}[b]{\textwidth}
\centering
\includegraphics[height=1.25in]{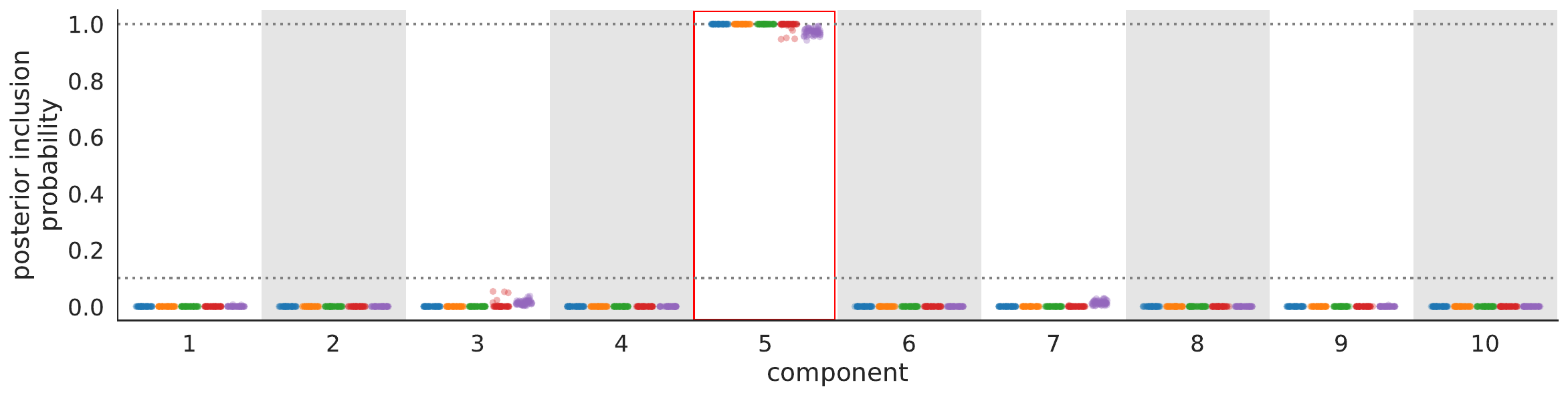}
\caption{$N = 5 \times 10^{4}$}
\end{subfigure}
\caption{Simulation results for feature selection in linear regression when the model is misspecified and $k^{\star} = 1$.
See the caption of \cref{fig:linreg-feature-selection-correlated-pips-1-k-star-2} for further explanation. 
The Bayes posterior inclusion probabilities show some instability both (i) across datasets with $N$ fixed and (ii) as $N$ increases. 
Meanwhile, the BayesBag posterior inclusion probabilities are much more stable.
}
\label{fig:linreg-feature-selection-correlated-pips-1-k-star-1}
\end{center}
\end{figure}

\begin{figure}[tbp]
\begin{center}
\begin{subfigure}[b]{\textwidth}
\centering
\includegraphics[height=1.15in]{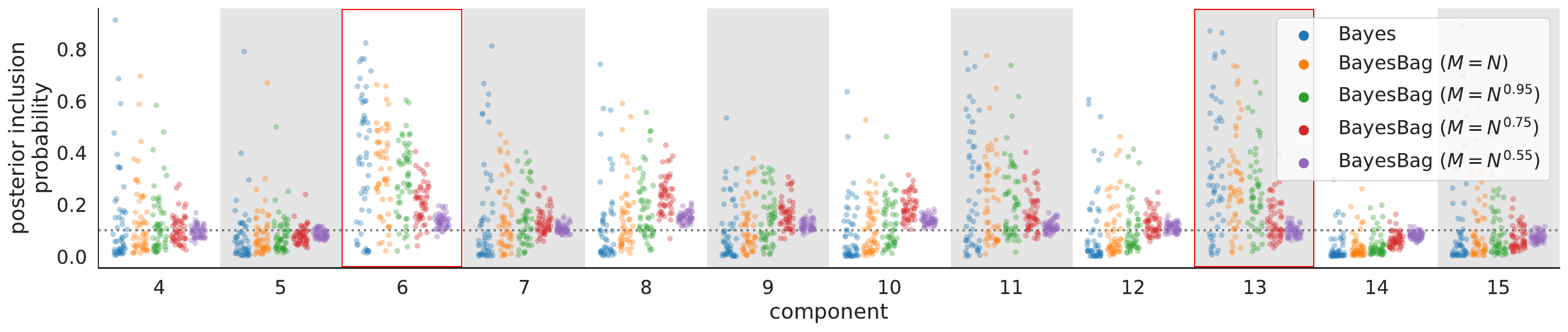}
\caption{$N = 10^{2}$}
\end{subfigure}
\begin{subfigure}[b]{\textwidth}
\centering
\includegraphics[height=1.15in]{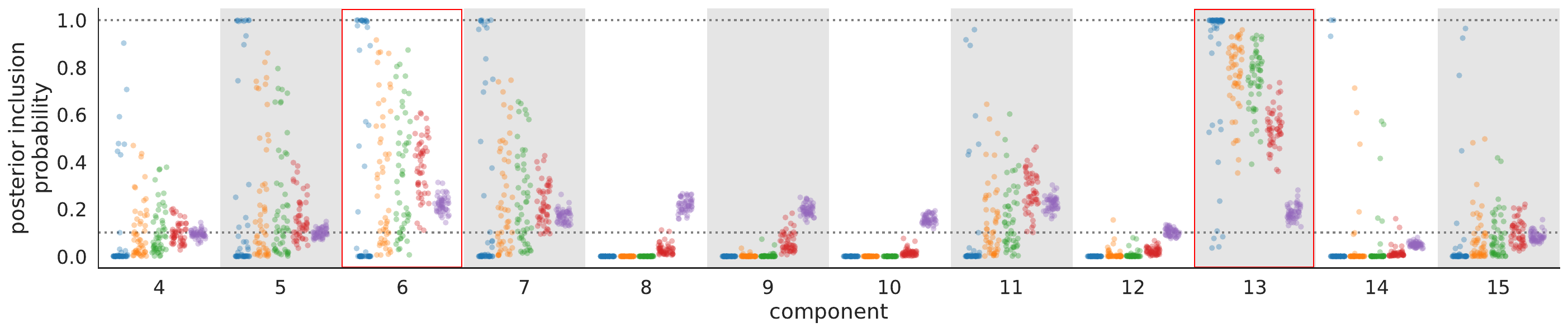}
\caption{$N = 10^{3}$}
\end{subfigure}
\begin{subfigure}[b]{\textwidth}
\centering
\includegraphics[height=1.15in]{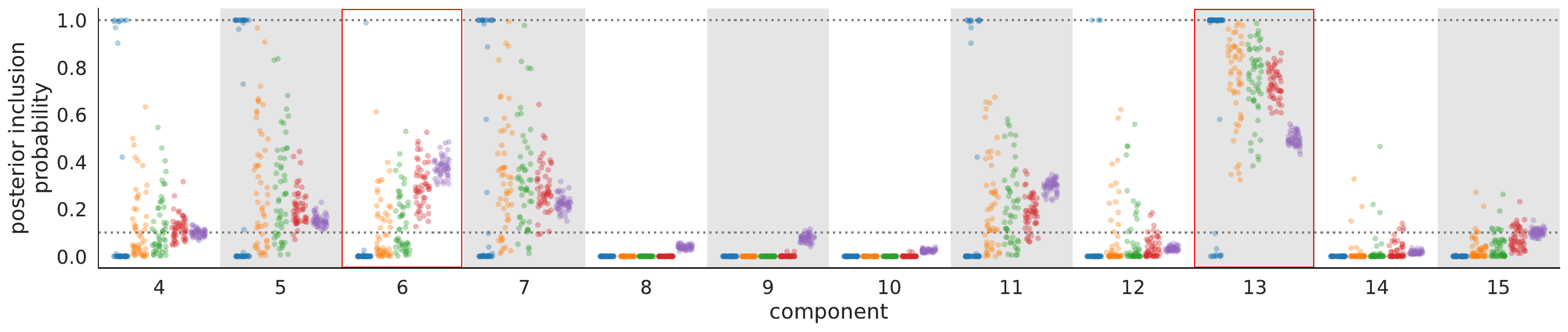}
\caption{$N = 10^{4}$}
\end{subfigure}
\begin{subfigure}[b]{\textwidth}
\centering
\includegraphics[height=1.15in]{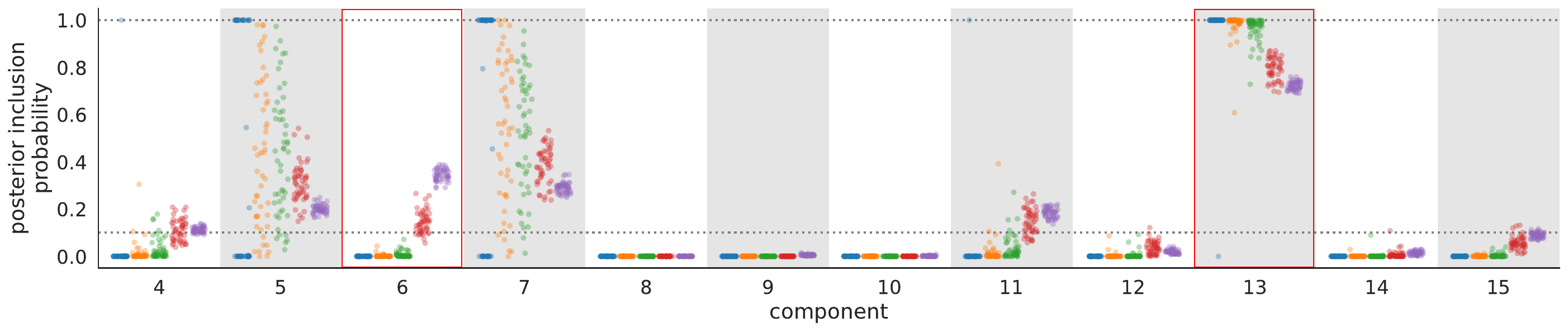}
\caption{$N = 10^{5}$}
\end{subfigure}   
\caption{Posterior inclusion probabilities for misspecified \textsf{2-sparse-nonlinear} data.
See the caption of \cref{fig:linreg-feature-selection-linear-correlated-pips} for further explanation.}
\label{fig:linreg-feature-selection-correlated-pips-2}
\end{center}
\end{figure}

\begin{figure}[tbp]
\begin{center}
\begin{subfigure}[b]{\textwidth}
\centering
\includegraphics[height=1.15in]{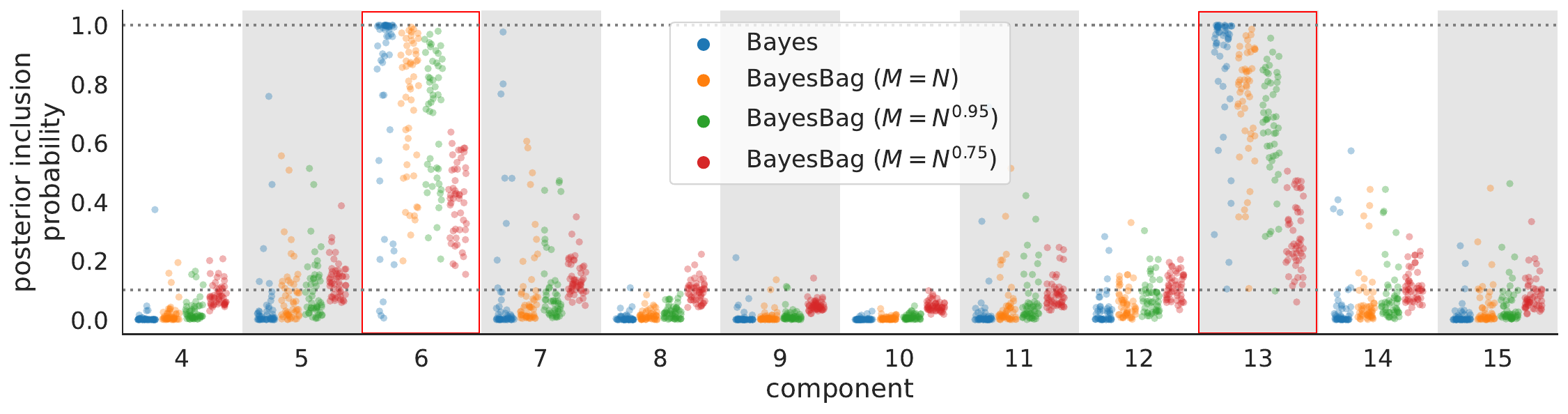}
\caption{$N = 10^{2}$}
\end{subfigure}
\begin{subfigure}[b]{\textwidth}
\centering
\includegraphics[height=1.15in]{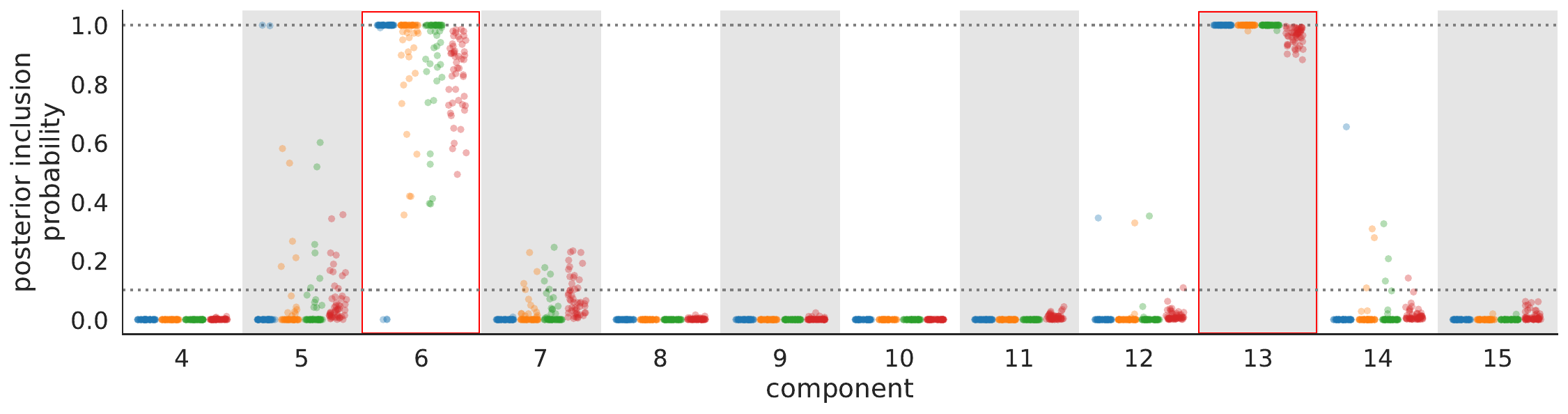}
\caption{$N = 10^{3}$}
\end{subfigure}
\begin{subfigure}[b]{\textwidth}
\centering
\includegraphics[height=1.15in]{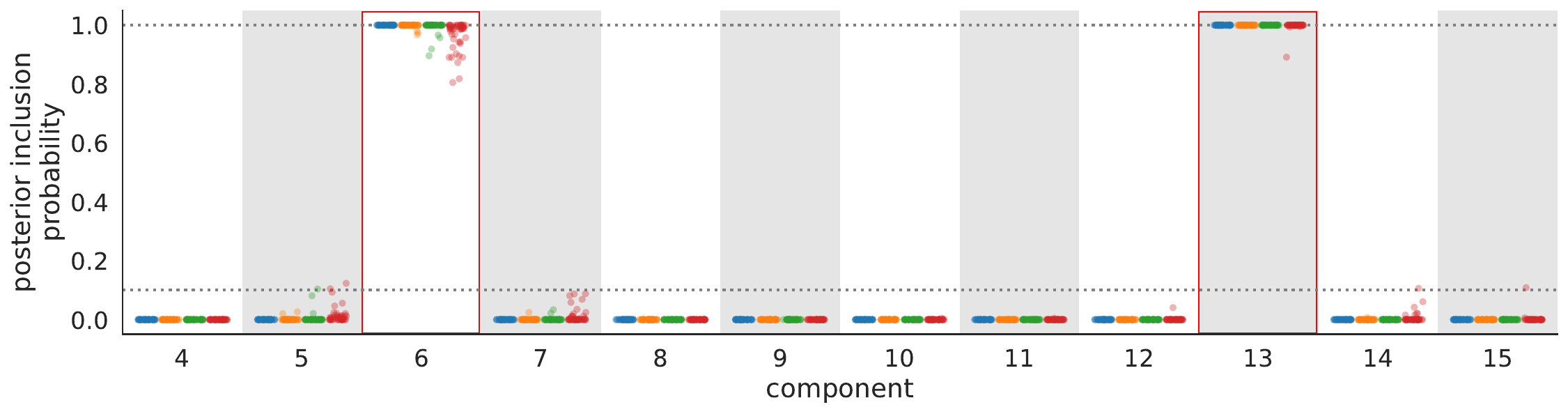}
\caption{$N = 10^{4}$}
\end{subfigure} 
\caption{Posterior inclusion probabilities for misspecified \textsf{2-sparse-nonlinear} data with $\psi = 4$, which results in weaker correlation between the covariates.
Compared to \cref{fig:linreg-feature-selection-correlated-pips-2}, all methods show greater stability and concentrate on the asymptotically optimal components. 
See the caption of \cref{fig:linreg-feature-selection-linear-correlated-pips} for further explanation.}
\label{fig:linreg-feature-selection-weakly-correlated-pips-2}
\end{center}
\end{figure}

\begin{figure}[b]
\begin{center}
\begin{subfigure}[b]{.49\textwidth}
\centering
\hspace{-.1\textwidth}\includegraphics[width=1.1\textwidth]{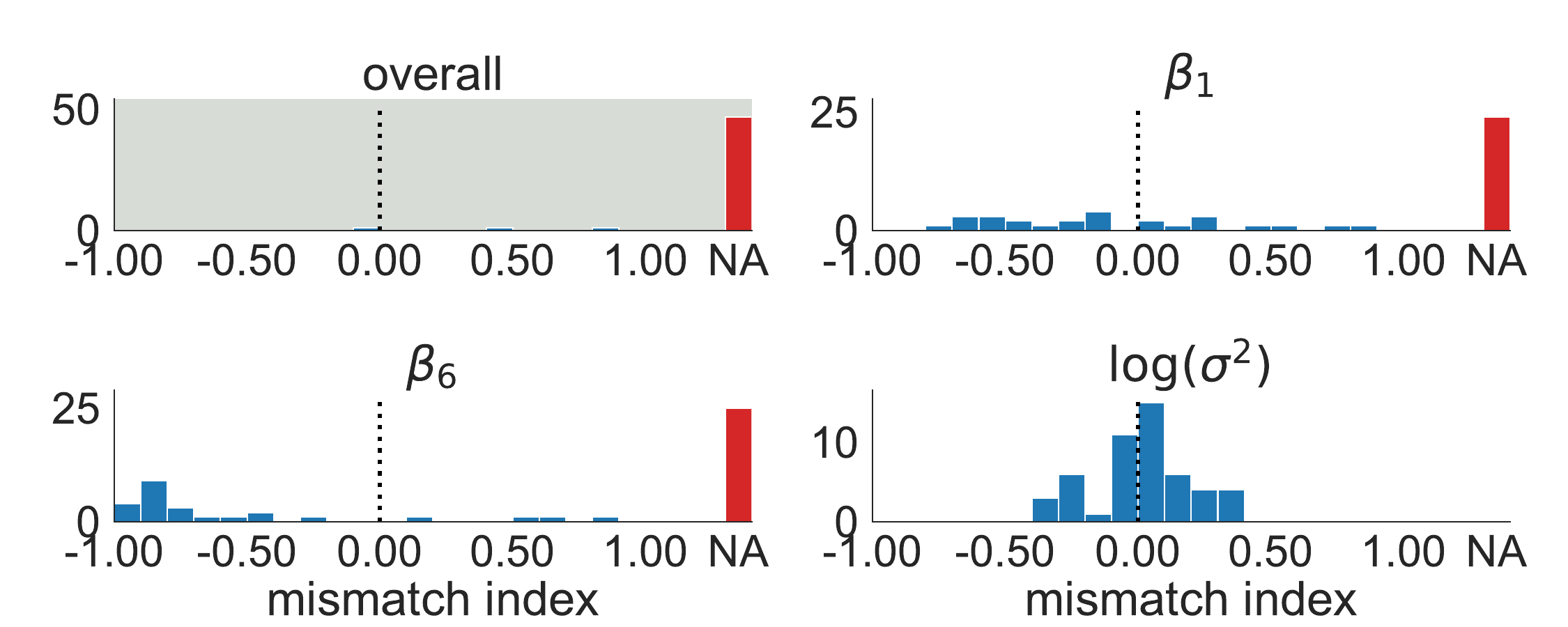}
\caption{\textsf{1-sparse-linear}, $N=5 \times 10^{1}$}
\end{subfigure}
\begin{subfigure}[b]{.5\textwidth}
\centering
\includegraphics[width=1.078\textwidth]{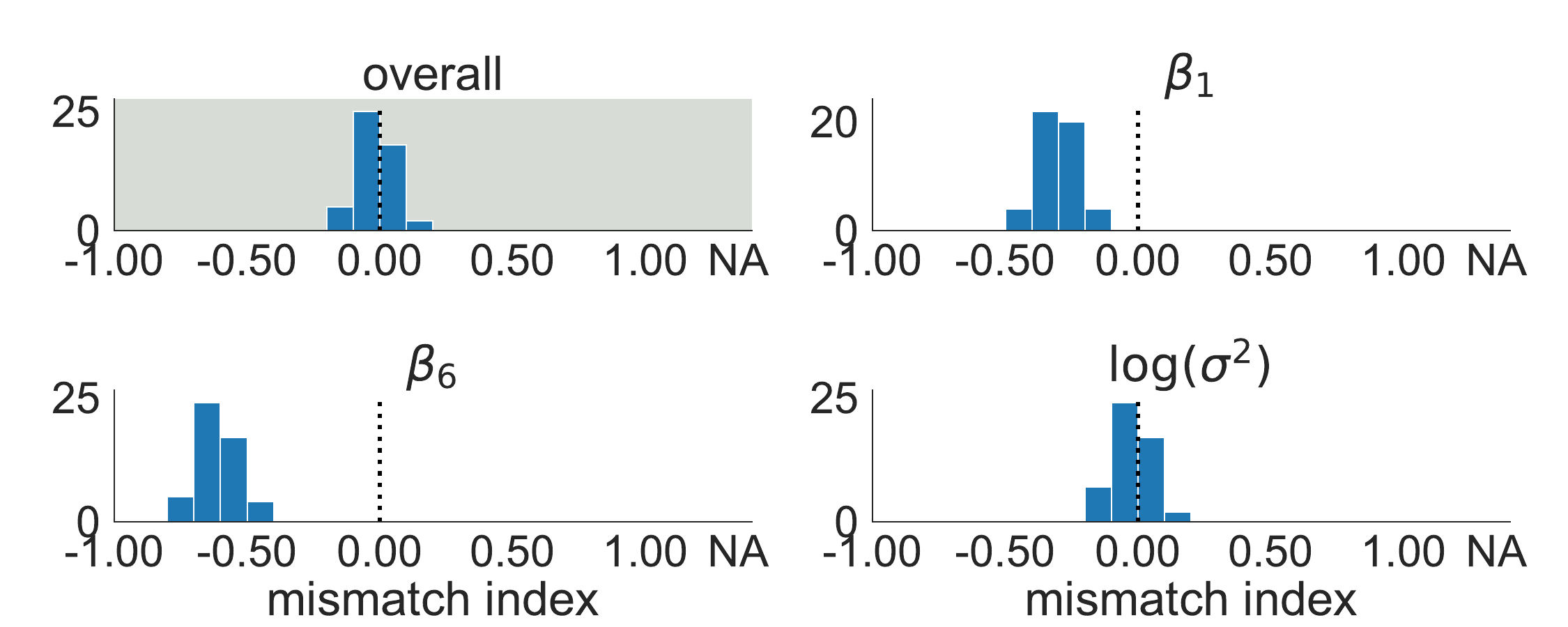}
\caption{\textsf{1-sparse-linear}, $N= 5 \times 10^{3}$}
\end{subfigure} \\
\begin{subfigure}[b]{.49\textwidth}
\centering
\hspace{-.1\textwidth}\includegraphics[width=1.1\textwidth]{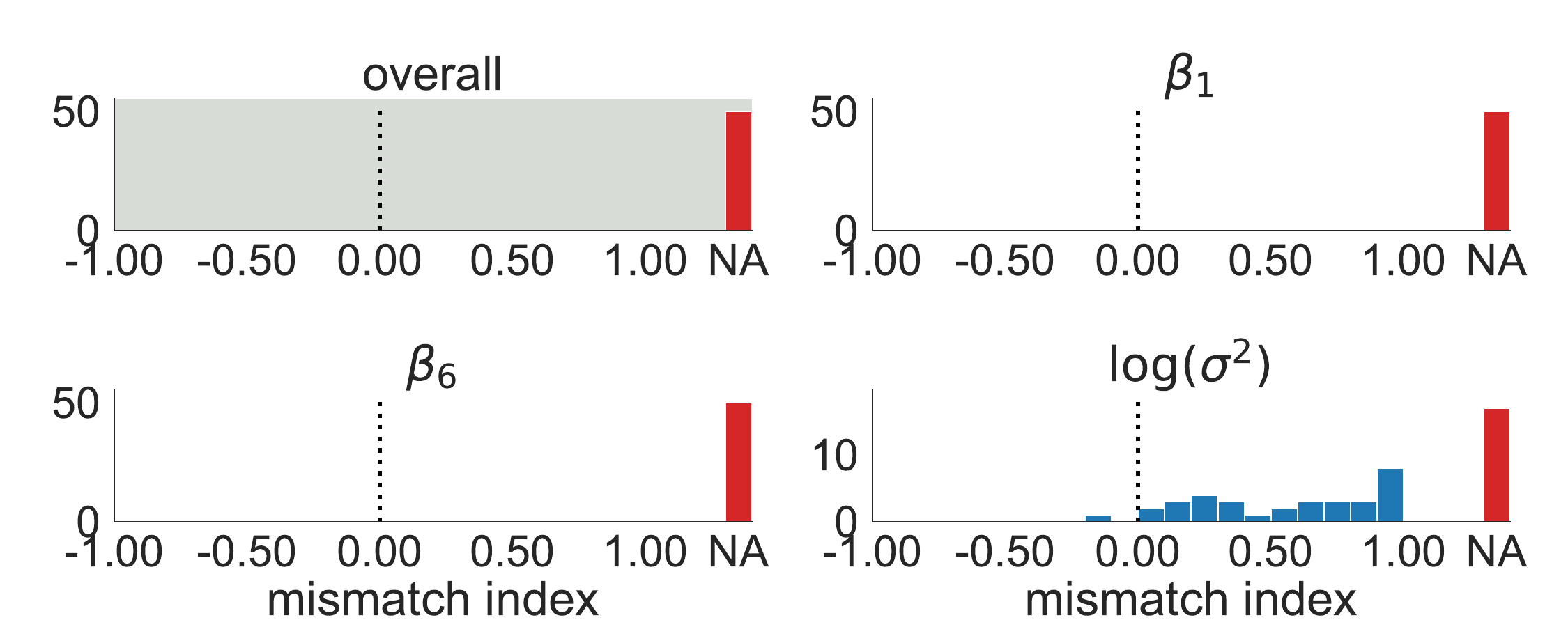}
\caption{\textsf{1-sparse-nonlinear}, $N=5 \times 10^{1}$}
\end{subfigure}
\begin{subfigure}[b]{.5\textwidth}
\centering
\includegraphics[width=1.078\textwidth]{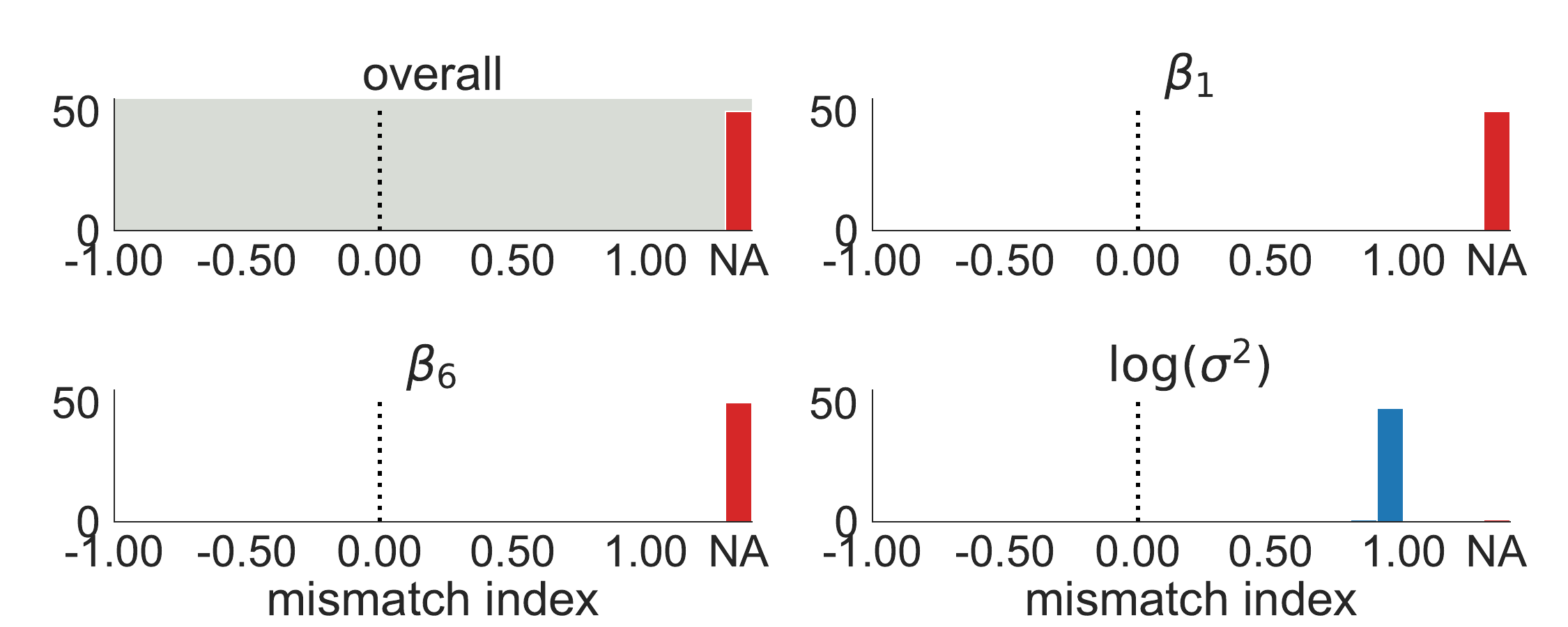}
\caption{\textsf{1-sparse-nonlinear}, $N=5 \times 10^{3}$}
\end{subfigure} 
\caption{Model--data mismatch indices $\modelmismatch$ for selected parameters as well as the overall $\modelmismatch$ value, in the case of \textsf{1-sparse} data with $k^{\star} = 2$.
We only display two components of $\beta$ since the $\modelmismatch$ values follow fairly similar distributions for all components.
}
\label{fig:linreg-feature-selection-correlated-model-mismatch-1-sparse}
\end{center}
\end{figure}

\Cref{fig:linreg-feature-selection-correlated-pips-1-k-star-1} shows results for the \textsf{1-sparse-nonlinear} data
with $k^{\star} = 1$, under the setup of the simulation study in \cref{sec:linreg-model-selection-sim}.
The Bayes and BayesBag posteriors behave quite similarly and concentrate on component 5, which is asymptotically optimal.
\Cref{fig:linreg-feature-selection-correlated-pips-2} shows results for \textsf{2-sparse-nonlinear} data, with
results similar to the \textsf{1-sparse-nonlinear} case with $k^{\star} = 2$ in \cref{fig:linreg-feature-selection-correlated-pips-1-k-star-2}.
In this case components $\{7, 13\}$ are asympotically optimal but there are other models that are nearly optimal, in particular $\{5, 13\}$ and $\{11, 12\}$.
The usual posterior (Bayes) shows large instability when $\numobs$ is small but eventually the posterior inclusion probability of component 13 converges to 1. 
However, even though component 7 is asymptotically optimal, the posterior inclusion probabilities of component 5 and 7 are sometimes 0 and sometimes 1. 
For BayesBag with $\bsnumobs = \numobs^{\alpha}$, the posterior inclusion probabilities become more stable as $\alpha$ decreases, particularly when $\numobs$ is large. 
\Cref{fig:linreg-feature-selection-weakly-correlated-pips-2} shows results with the same settings as \cref{fig:linreg-feature-selection-correlated-pips-2} except
$\psi = 4$, resulting in weaker correlation between the covariates and in model $\{6, 13\}$ being optimal.
In this case all methods are quite stable. 
\cref{fig:linreg-feature-selection-correlated-model-mismatch-1-sparse} 
shows model--data mismatch index values for a representative subset of experimental configurations for \textsf{1-sparse} data with $k^{\star} = 2$. 
For the \textsf{1-sparse-linear} data, the overall mismatch indices were either near zero or, when $\numobs$ is small, were $\nan$, reflecting that the model is correctly specified 
but there are some issues with poor identifiability.
For the \textsf{1-sparse-nonlinear} data, the mismatch indices were nearly all $\nan$, reflecting that the model is misspecified 
and there may also be identifiability issues.

\Cref{tbl:whale-cross-model-comparison,tbl:whale-cross-model-comparison-to-mixed-G5,tbl:whale-self-consistency} provide numerical summaries of the 
results shown in \cref{fig:whale-comparisons}.

\clearpage

\begin{table}[t]
\caption{Overlap between the posteriors for each pair of models, given the whale dataset, when using Bayes or BayesBag.
The ``mass'' column shows the overlap of the two 99\% HPD density regions and ``\# trees'' shows the number of trees in the intersection of the two regions. 
For BayesBag, ``mass'' shows an 80\% confidence interval for the overlap
and ``\# trees'' shows the median number of trees in the intersection.
}
\begin{center}
\begin{tabular}{rclrcrcrc}
\toprule
\multicolumn{3}{c}{Comparison} &  \multicolumn{2}{c}{Bayes} & \multicolumn{2}{c}{BayesBag ($\bsnumobs = \numobs$)} & \multicolumn{2}{c}{BayesBag ($\bsnumobs = \numobs^{0.95}$)} \\
& & & mass & \# trees & mass (80\% CI) & \# trees  & mass (80\% CI) & \# trees  \\
\midrule 
\textsf{JC} 		& vs	& \textsf{HKY} 	& 0.2\%	& 1 	& (31\%, 41\%) 	& 4 	& (29\%, 39\%)	& 7\\
\textsf{JC} 		& vs	& \textsf{GTR}	& 0\%		& 0	& (38\%, 51\%) 	& 3 	& (38\%, 50\%) 	& 7\\
\textsf{JC} 		& vs	& \textsf{mixed} 	& 0\% 	& 0 	& (39\%, 52\%) 	& 4 	& (38\%, 50\%) 	& 7\\
\textsf{JC}		& vs	& \textsf{mtmam}	& 0\%		& 0	& (3.1\%, 8.4\%)	& 3 	& (8.9\%, 16\%) 	& 5\\
\textsf{HKY}	& vs	& \textsf{GTR}	& 29\%	& 1 	& (38\%, 50\%) 	& 10	& (45\%, 56\%)	& 14\\
\textsf{HKY}	& vs	& \textsf{mixed} 	& 25\% 	& 2 	& (45\%, 59\%) 	& 10	& (51\%, 64\%)	& 14 \\
\textsf{HKY}	& vs	& \textsf{mtmam} & 56\% 	& 2	& (32\%, 42\%) 	& 8 	& (27\%, 36\%)	& 13\\
\textsf{GTR}	& vs	& \textsf{mixed} 	& 99\% 	& 1 	& (79\%, 89\%) 	& 11	& (78\%, 88\%) 	& 15 \\
\textsf{GTR}	& vs	& \textsf{mtmam} & 0\%	& 0	& (9.3\%, 16\%)	& 8 	& (16\%, 23\%) 	& 13\\
\textsf{mixed}	& vs	& \textsf{mtmam} & 0\%	& 0	& (14\%, 22\%) 	& 9  	& (22\%, 30\%) 	& 13\\
\bottomrule
\end{tabular}
\end{center}
\label{tbl:whale-cross-model-comparison}
\end{table}%

\begin{table}[t]
\caption{Overlap between the Bayes posterior for the \textsf{mixed} model and the Bayes or BayesBag posterior for each model, given the whale dataset.
The form of each data entry is the same as \cref{tbl:whale-cross-model-comparison}.}
\begin{center}
\begin{tabular}{crccrccrc}
\toprule
Model &  \multicolumn{2}{c}{Bayes} & & \multicolumn{2}{c}{BayesBag ($\bsnumobs = \numobs$)} & & \multicolumn{2}{c}{BayesBag ($\bsnumobs = \numobs^{0.95}$)} \\
 & mass & \# trees && mass (80\% CI) & \# trees && mass (80\% CI) & \# trees  \\
\midrule 
\textsf{JC} 		& 0\%		& 0 	&& (27\%, 39\%) 	& 2 	&& (19\%, 29\%)	& 2\\
\textsf{HKY}	& 25\%	& 2 	&& (21\%, 32\%) 	& 2 	&& (14\%, 22\%) 	& 2\\
\textsf{GTR}	& 99\% 	& 1 	&& (65\%, 74\%)	& 2 	&& (46\%, 56\%) 	& 2 \\
\textsf{mixed}	& 99\%	& 2	&& (56\%, 66\%) 	& 2  	&& (39\%, 49\%) 	& 2\\
\textsf{mtmam}	& 0\%		& 0	&& (0.3\%, 0.3\%)	& 1	&& (0.3\%, 0.3\%)	& 1  \\
\bottomrule
\end{tabular}
\end{center}
\label{tbl:whale-cross-model-comparison-to-mixed-G5}
\end{table}%

\begin{table}[t]
\caption{Overlap between the posteriors for each pair of whale data subsets (\textsf{all}, \textsf{S1}, and \textsf{S2}), when using Bayes or BayesBag.
The form of each data entry is the same as \cref{tbl:whale-cross-model-comparison}.}
\begin{center}
\begin{tabular}{ccrcrcrc}
\toprule
Model & Comparison &  \multicolumn{2}{c}{Bayes} & \multicolumn{2}{c}{BayesBag ($\bsnumobs = \numobs$)} & \multicolumn{2}{c}{BayesBag ($\bsnumobs = \numobs^{0.95}$)} \\
& & mass & \# trees & mass (80\% CI) & \# trees & mass (80\% CI) & \# trees  \\
\midrule 
\textsf{JC} 		& \textsf{S1} vs \textsf{S2} 	& 0\%		& 0 	& (21\%, 32\%) 	& 4  & (24\%, 35\%) & 5 \\
			& \textsf{all} vs \textsf{S1} 	& 0\%		& 0 	& (25\%, 36\%) 	& 5  & (33\%, 45\%) & 9 \\
			& \textsf{all} vs \textsf{S2} 	& 97\%	& 1 	& (51\%, 65\%) 	& 4  & (50\%, 63\%) & 7 \\
\textsf{HKY}	& \textsf{S1} vs \textsf{S2} 	& 9\%		& 3 	& (26\%, 35\%) 	& 9  & (29\%, 38\%) & 22 \\
			& \textsf{all} vs \textsf{S1} 	& 13\%	& 4 	& (43\%, 55\%) 	& 9  & (44\%, 55\%) & 16 \\
			& \textsf{all} vs \textsf{S2} 	& 36\%	& 4 	& (52\%, 63\%) 	& 11 & (52\%, 62\%) & 17 \\
\textsf{GTR}	& \textsf{S1} vs \textsf{S2} 	& 38\% 	& 2 	& (36\%, 44\%) 	& 14 & (36\%, 44\%) & 37 \\
			& \textsf{all} vs \textsf{S1} 	& 38\%	& 1 	& (44\%, 54\%) 	& 12 & (46\%, 56\%) & 16 \\
			& \textsf{all} vs \textsf{S2} 	& 90\%	& 1 	& (58\%, 68\%) 	& 11 & (54\%, 63\%) & 15 \\
\textsf{mtmam}	& \textsf{S1} vs \textsf{S2} 	& 0\%		& 0	& (0.3\%, 1.5\%)	& 9   & (0.9\%, 2.3\%) & 22 \\
			& \textsf{all} vs \textsf{S1} 	& 0.2\%	& 1 	& (10\%, 16\%) 	& 24 & (7.9\%, 13\%) & 59\\
			& \textsf{all} vs \textsf{S2} 	& 28\%	& 4 	& (17\%, 25\%) 	& 24 & (13\%, 18\%) & 53\\
\bottomrule
\end{tabular}
\end{center}
\label{tbl:whale-self-consistency}
\end{table}%
\clearpage

\section{Feature selection in linear regression} \label{app:linreg-feature-selection}

We derive the KL-optimal linear regression parameters for data generated as in our simulation studies (\cref{sec:linreg-model-selection-sim}).
Assuming a linear regression model and assuming the data follow \cref{eq:simulated-linreg-data}, we have
\[
\lefteqn{\EE\{\log p(Y_{n} \given \gamma, Z_{n}, \beta, \sigma^{2})\}} \\
&= -\frac{1}{2\sigma^{2}}\EE\{(Y_{n} - Z_{\gamma,n}^{\top}\beta)^{2}\} - \frac{1}{2}\log(2 \pi \sigma^{2})\\ 
&= -\frac{1}{2\sigma^{2}}\EE\{(f(Z_{n})^{\top}\beta_{\dagger} +\eps_{n} - Z_{\gamma,n}^{\top}\beta)^{2}\}  - \frac{1}{2}\log(2 \pi \sigma^{2}) \\
\begin{split}
&= -\frac{1}{2\sigma^{2}}\EE\{\beta_{\dagger}^{\top}f(Z_{n})f(Z_{n})^{\top}\beta_{\dagger} + \beta^{\top}Z_{\gamma,n}Z_{\gamma,n}^{\top}\beta - 2\beta_{\dagger}^{\top}f(Z_{n})Z_{\gamma,n}^{\top}\beta\} -\frac{1}{2\sigma^{2}} - \frac{1}{2}\log(2 \pi \sigma^{2}).
\end{split}
\]
Thus, 
\[
\sigma^{2}\grad_{\beta}\EE\{\log p(Y_{n} \given \gamma, Z_{n}, \beta, \sigma^{2})\}
&= -\EE(Z_{n}Z_{n}^{\top})\beta + \EE\{Z_{n}f(Z_{n})^{\top}\}\beta_{\dagger},
\]
so the optimal coefficient vector is 
\[\label{eq:optimal-coefficient-vector}
\beta_{\optsym} = \EE(Z_{\gamma, n}Z_{\gamma, n}^{\top})^{-1}\EE\{Z_{\gamma, n}f(Z_{n})^{\top}\}\beta_{\dagger}.
\]
Thus, when $f$ is not the identity and the regressors are not independent, in general $\beta_{\optsym}$ will be dense even if $\beta_{\dagger}$ is sparse. 

Let $\Sigma_{ZZ} \defined  \EE(Z_{\gamma, n}Z_{\gamma, n}^{\top})$, $\Sigma_{Zf} \defined \EE\{Z_{\gamma, n}f(Z_{n})^{\top}\}$, and $\Sigma_{ff} \defined \EE\{f(Z_{n})f(Z_{n})^{\top}\}$.
For the optimal coefficient vector, we have
\[
\lefteqn{\EE\{\log p(Y_{n} \given Z_{n}, \beta_{\optsym}, \sigma^{2})\}} \\
\begin{split}
&= -\frac{1}{2\sigma^{2}}\left[\beta_{\dagger}^{\top}\Sigma_{ff}\beta_{\dagger} + \beta_{\dagger}^{\top}\Sigma_{Zf}^{\top}\Sigma_{ZZ}^{-1}\Sigma_{Zf}\beta_{\dagger} - 2\beta_{\dagger}^{\top}\Sigma_{Zf}^{\top}\Sigma_{ZZ}^{-1}\Sigma_{Zf}\beta_{\dagger}\right] -\frac{1}{2\sigma^{2}} - \frac{1}{2}\log(2 \pi \sigma^{2})
\end{split} \\
&= -\frac{1}{2\sigma^{2}}\beta_{\dagger}^{\top}\Sigma_{ff}\beta_{\dagger} +  \frac{1}{2\sigma^{2}}\beta_{\dagger}^{\top}\Sigma_{Zf}^{\top}\Sigma_{ZZ}^{-1}\Sigma_{Zf}\beta_{\dagger}  - \frac{1}{2\sigma^{2}} - \frac{1}{2}\log(2 \pi \sigma^{2}).
\]
Thus, the optimal variance is 
\[
\sigma^{2}_{\optsym} = \left(1 + \beta_{\dagger}^{\top}\Sigma_{ff}\beta_{\dagger} - \beta_{\dagger}^{\top}\Sigma_{Zf}^{\top}\Sigma_{ZZ}^{-1}\Sigma_{Zf}\beta_{\dagger}\right)_{+}.
\]
Now plugging in the optimal variance, we have
\[
\lefteqn{\EE\{\log p(Y_{n} \given \gamma, Z_{n}, \beta_{\optsym}, \sigma^{2}_{\optsym})\}} \\
&= \begin{cases} 
-\frac{1}{2}\log(2e\pi) - \frac{1}{2}\log \left(1 + \beta_{\dagger}^{\top}\Sigma_{ff}\beta_{\dagger} - \beta_{\dagger}^{\top}\Sigma_{Zf}^{\top}\Sigma_{ZZ}^{-1}\Sigma_{Zf}\beta_{\dagger}\right) & \sigma^{2}_{\optsym} > 0 \\
\infty & \sigma^{2}_{\optsym} = 0.
\end{cases}
\]

We now derive the formulas for the expected log-likelihood in the setting of our simulation studies so we can compute the asymptotically optimal model(s). 
We have 
\[
\Sigma_{ff,dd'} 
&= \EE(Z_{d}^{3}Z_{d'}^{3}) 
= \EE\{\EE(Z_{d}^{3}Z_{d'}^{3} \mid \xi) \}
= 3\,\EE(3\Sigma_{dd'} + 2\Sigma_{dd'}^{3})
\]
and
\[
\Sigma_{Zf,dd'}
&= \EE(Z_{d}Z_{d'}^{3}) 
= \EE\{\EE(Z_{d}Z_{d'}^{3} \mid \xi) \}
= 3\,\EE(\Sigma_{dd'}\Sigma_{d'd'}).
\]
Letting $\mcK_{dd'} = \exp(-(d-d')^{2}/64)$, we have
\[
\Sigma_{ZZ,dd'} = \EE(\Sigma_{dd'}) 
&= \mcK_{dd'} \times 
	\begin{cases}
		1 & d ~(\textrm{mod}~2) = d' ~(\textrm{mod}~2) \\
		35\sqrt{\pi}/64 & \text{otherwise}
	\end{cases}, \\
\Sigma_{Zf,dd'} = 3\EE(\Sigma_{dd'}\Sigma_{d'd'}) 
&= 3\mcK_{dd'} \times 
	\begin{cases}
		1 & d ~(\textrm{mod}~2) = d' ~(\textrm{mod}~2) = 0 \\
		4/3 & d ~(\textrm{mod}~2) = d' ~(\textrm{mod}~2) = 1 \\
		5\sqrt{\pi}/8 & d ~(\textrm{mod}~2) = 0, d' ~(\textrm{mod}~2) = 1 \\
		35\sqrt{\pi}/64 & d ~(\textrm{mod}~2) = 1, d' ~(\textrm{mod}~2) = 0 
	\end{cases},
\]
and 
\[
\EE[\Sigma_{dd'}^{3}] = \mcK_{dd'}^{3} \times 
	\begin{cases}
		1 & d ~(\textrm{mod}~2) = d' ~(\textrm{mod}~2) = 0 \\
		8/3 & d ~(\textrm{mod}~2) = d' ~(\textrm{mod}~2) = 1 \\
		5\sqrt{\pi}/8 & \text{otherwise}. 
	\end{cases}
\]
Using these formulas, we compute $\EE\{\log p(Y_{n} \given \gamma, Z_{n}, \beta_{\optsym}, \sigma^{2}_{\optsym})\}$  numerically. 
A Python notebook to reproduce these results is included in the Supplementary Materials. 

\section{Proofs}

We use $\convP$ to denote convergence in probability and $\convPouter$ for convergence in outer probability.

\subsection{Proof of \cref{prop:min-B}}

First observe that since for $\numobs > 1$, 
\[
(\numobs - 1/2)\log(1 - 1/\numobs)
&= (\numobs - 1/2) \sum_{i=1}^{\infty} \frac{-1}{i\numobs^{i}}
= -1 - \sum_{i=1}^{\infty} \left(\frac{1}{i+1} - \frac{1}{2i}\right) \frac{1}{\numobs^{i}}
\le -1,
\]
we have the bound  
\[
1 - 1/\numobs \le \exp\{-1/(\numobs - 1/2)\}.\label{eq:prob-bound}
\]
For each $n$, the probability that datapoint $n$ is not included in any of $B$ bootstrap datasets of size $M$ is $(1 - 1/\numobs)^{\bsnumobs B}$. 
Therefore, by a union bound and \cref{eq:prob-bound}, 
\[
\Pr\left(\bigcup_{n=1}^{\numobs}  \bigcap_{b=1}^B \{\text{$n$ not in dataset $b$}\}\right) \le \numobs (1 - 1/\numobs)^{\bsnumobs B} \le \numobs \exp\{-\bsnumobs B / (\numobs - 1/2)\} \le \delta,
\]
where the last inequality follows from the assumption that $B \ge (\numobs - 1/2)\log(\numobs/\delta)/\bsnumobs$.
Therefore, $\Pr(\bigcap_{n=1}^{\numobs}  \bigcup_{b=1}^B \{\text{$n$ in dataset $b$}\}) \ge 1-\delta$, as claimed.

\subsection{Proof of \cref{thm:many-model-selection-asymptotic-posteriors}} \label{sec:proof-of-many-model-selection-asymptotic-posteriors}

We first state some preliminary definitions and results needed for our proof of \cref{thm:many-model-selection-asymptotic-posteriors}.
For a random variable $\xi$, let $\mcL(\xi)$ denote its law (that is, its distribution). 
For vector-valued random variables $\xi, \xi' \in \reals^{L}$, let $\dsupconvex(\mcL(\xi), \mcL(\xi')) \defined \sup_{C\in\mathcal{C}}|\Pr(\xi \in C) - \Pr(\xi' \in C)|$, where $\mathcal{C}$ is the set of measurable convex subsets of $\reals^{L}$. 
\begin{nlem} \label{lem:dK-convergence-in-dist}
If $\dsupconvex(\mcL(\xi_{n}), \mcL(\xi)) \to 0$ then $\xi_{n} \convD \xi$. 
\end{nlem}
\begin{proof}
Let $\eps(n) \defined \dsupconvex(\mcL(\xi_{n}), \mcL(\xi))$. 
For $O$ open, let $\mcO_{n}$ denote a set of $\lfloor 1/\eps(n)^{1/2}\rfloor$ disjoint convex subsets of $O$
such that $\bigcup_n (\bigcup_{C \in \mcO_{n}} C) = O$.
Then
\[
\liminf_{n} \Pr\left(\xi_{n} \in O\right)
&\ge \liminf_{n} \Pr(\xi_{n} \in {\textstyle\bigcup_{C \in \mcO_{n}}} C ) \\
&=  \liminf_{n}\sum_{C \in \mcO_{n}} \Pr(\xi_{n} \in C) \\
&\ge   \liminf_{n}\sum_{C \in \mcO_{n}}\left\{ \Pr(\xi \in C) - |\Pr(\xi \in C) - \Pr(\xi_{n} \in C)|\right\} \\
&\ge   \liminf_{n} \Pr(\xi \in {\textstyle\bigcup_{C \in \mcO_{n}}} C ) - \eps(n)^{1/2} \\
&=  \Pr\left(\xi \in O\right),
\]
so the result follows from the Portmanteau theorem \citep[Theorem 4.25]{Kallenberg:2002}.
\end{proof}
Note that $\dsupconvex(\cdot,\cdot)$ satisfies the properties of a distance metric and is invariant under invertible affine transformations --
that is, $\dsupconvex(\mcL(\xi), \mcL(\xi')) = \dsupconvex(\mcL(A\xi + b), \mcL(A\xi' + b))$ if $A^{-1}$ exists. 
For $t \in \reals^{L}$, let $\ind(t > 0) \defined \ind(t_{1} > 0, \dots, t_{L} > 0)$ and 
$\logistic_{\numobs}(t) \defined \{1 + \sum_{\ell=1}^{L}\exp(-\numobs^{1/2}t_{\ell})\}^{-1}$.
\bnlem \label{lem:indicator-approx-error}
For $t \in \reals^{L}$, if $|t_\ell|>\eps$ for all $\ell  \in \theset{1,\dots,L}$, then $|\phi_{\numobs}(t) - \ind(t > 0)| \le L\exp(- \numobs^{1/2}\eps)$. 
\enlem
\bprf
If $t_{\ell} > 0$ for all $\ell \in \theset{1,\dots,L}$, then $t_{\ell} > \eps$ for all $\ell$, and so 
\[
1 - \phi_{\numobs}(t)
&\le 1 - \frac{1}{1 + L\exp(-\numobs^{1/2}\eps)}
\le L\exp(-\numobs^{1/2}\eps).
\]
Otherwise, $t_{\ell} \le 0$ for some $\ell$, in which case $t_{\ell} < -\eps$ and so
\[
\phi_{\numobs}(t)
&\le \frac{1}{1+ \exp(\numobs^{1/2}\eps)}
\le \exp(-\numobs^{1/2}\eps).
\]
\eprf

Finally, we have the following uniform central limit theorem for (bootstrapped) triangular arrays, 
the proof of which is deferred to \cref{sec:proof-of-uniform-multivariate-CLT}.
\bnprop \label{prop:uniform-multivariate-CLT}
For each $\numobs = 1,2,\ldots$, let $\xi_{\numobs 1},\ldots,\xi_{\numobs \bar\numobs} \dist P_{\numobs}$ independently,
where $P_{\numobs}$ is a distribution on $\reals^{L}$ and $\bar\numobs = \bar\numobs(\numobs) \to \infty$ as $\numobs\to\infty$.
Suppose that as $\numobs\to\infty$,
\begin{enumerate}[label=(\roman*)]
\item $\bar\numobs^{1/2}\EE(\xi_{\numobs 1}) \to \mu \in \reals^{L}$ and
\item $\cov(\xi_{\numobs 1}) \to \Sigma$ positive definite.
\end{enumerate}
\textup{(A)}
If $\bar\numobs^{-1/2} \EE(\|\xi_{\numobs 1}\|_{2}^{3}) \to 0$
then $W_{\numobs} \defined \bar\numobs^{-1/2}\sum_{n=1}^{\bar\numobs} \xi_{\numobs n}$ satisfies
\[
\lim_{\numobs \to \infty} \dsupconvex(\mcL(W_{\numobs}), \distNorm(\mu, \Sigma)) = 0.
\]
\textup{(B)} 
For each $\numobs = 1,2,\ldots$, let $\xi_{\numobs 1}^{\bbsym},\ldots,\xi_{\numobs \bsnumobs}^{\bbsym}|\xi \dist \hP_{\numobs}$ independently,
given the array $\xi = (\xi_{\numobs n} : \numobs=1,2,\ldots;\, n=1,\ldots,\bar\numobs)$, where 
$\hP_{\numobs} \defined \bar{\numobs}^{-1} \sum_{n=1}^{\bar{\numobs}} \delta_{\xi_{\numobs n}}$,
$\bsnumobs = \bsnumobs(\numobs) \to \infty$, and $\lim_{\numobs \to \infty} \bsnumobs/\bar\numobs \in [0,\infty)$.
If $\limsup_{\numobs} \EE(\|\xi_{\numobs 1}\|_{2}^{6}) < \infty$
then $W_{\numobs}^{\bbsym} \defined \bsnumobs^{-1/2} \sum_{m=1}^{\bsnumobs} \xi_{\numobs m}^{\bbsym}$ satisfies
\[
\dsupconvex(\mcL(W^{\bbsym}_{\numobs} - (\bsnumobs/\bar{\numobs})^{1/2}W_{\numobs} \mid \xi),\, \distNorm(0, \Sigma)) \convP 0.
\]
\enprop

\bprf[Proof of \cref{thm:many-model-selection-asymptotic-posteriors} Part (1)]
For notational convenience, let $Y_{\numobs 0, k} \defined \log\modelpriordist(\model_k)$ denote the log prior probability.
For $n = 0,1,\ldots,\numobs$, define the log-likelihood ratios $Z_{\numobs n,k} \defined Y_{\numobs n,1} - Y_{\numobs n,k+1}$, and let $Z_{\numobs n} = (Z_{\numobs n,1},\ldots,Z_{\numobs n,K-1})^\top \in\reals^{K-1}$.
Notice that for the matrix $A \in \reals^{K-1 \times K}$ with entries $A_{i,j} = \ind(j = 1) - \ind(j=i+1)$, we can write $Z_{\numobs n} = AY_{\numobs n}$.
Therefore, 
\[
\lim_{\numobs \to \infty} \numobs^{1/2}\EE(Z_{\numobs 1}) = A \lim_{\numobs \to \infty} \numobs^{1/2}\EE(Y_{\numobs 1})  = A \mu_{\infty}' = \mu_{\infty}
\]
and 
\[
\lim_{\numobs \to \infty} \cov(Z_{\numobs 1}) = A \left\{\lim_{\numobs \to \infty} \cov(Y_{\numobs 1})\right\}A^{\top}  = A \Sigma_{\infty}' A^{\top} = \Sigma_{\infty}.
\]
In addition, since $A$ has full rank, $\cov(Z_{\numobs 1}) = A\Sigma_{\numobs}'A^{\top}$ is positive definite if $\Sigma_{\numobs}'$ is, 
and $\Sigma_{\infty} = A \Sigma_{\infty}' A^{\top}$ is positive definite as well.
Define $W_{\numobs} \defined \numobs^{-1/2}\sum_{n=0}^{\numobs}Z_{\numobs n}$,
and let $W_{\infty} \dist \distNorm(\mu_{\infty}, \Sigma_{\infty})$. 

It follows from \cref{prop:uniform-multivariate-CLT}(A) with $\xi_{\numobs n} = Z_{\numobs n} + Z_{\numobs 0}/\numobs$ and $\bar\numobs = \numobs$
that
\[
\lim_{\numobs \to \infty}\dsupconvex(\mathcal{L}(W_{\numobs}), \mathcal{L}(W_{\infty})) = 0. \label{eq:W-UMCLT}
\]
In particular, by \cref{lem:dK-convergence-in-dist}, \cref{eq:W-UMCLT} implies that $W_{\numobs} \convD W_{\infty}$.
We can write the posterior probability of model $1$ as $\modelpostdistfull{\model_{1}}{\datarvarg{\numobs}} = \logistic_{\numobs}(W_{\numobs})$.
By \cref{lem:indicator-approx-error}, $\logistic_{\numobs}$ converges pointwise to $\ind(t > 0)$ on the set $\{t\in\reals^{K-1} \mid t_{k}\neq 0 \text{ for all } k\}$, 
so it follows from the continuous mapping theorem \citep[Theorem 4.27]{Kallenberg:2002} that 
\[
\modelpostdistfull{\model_{1}}{\datarvarg{\numobs}} = \logistic_{\numobs}(W_{\numobs}) \convD \ind(W_{\infty} > 0) = \ind(-W_{\infty} < 0) \dist \distBern(\Phi_{-\mu_{\infty}, \Sigma_{\infty}}(0))
\]
since $-W_{\infty} \dist \distNorm(-\mu_{\infty}, \Sigma_{\infty})$. 
\eprf

\bprf[Proof of \cref{thm:many-model-selection-asymptotic-posteriors} Part (2)]
Let $(\bscount{\numobs,1},\ldots,\bscount{\numobs,\numobs}) \dist \distMulti(\bsnumobs, 1/\numobs)$ independently of $(\obsrv{1},\obsrv{2},\ldots)$, and define
\[
W^{\bbsym}_{\numobs} \defined \bsnumobs^{-1/2}\left(Z_{\numobs 0} + \sum_{n=1}^{\numobs}\bscount{\numobs n}Z_{\numobs n}\right).
\]
Further, let $\Delta^{\bbsym}_{\numobs} \defined W^{\bbsym}_{\numobs} - (\bsnumobs/\numobs)^{1/2}W_{\numobs}$ and,
independently of $(\obsrv{1},\obsrv{2},\ldots)$, let $\Delta^{\bbsym}_{\infty} \dist \distNorm(0, \Sigma_{\infty})$.
It follows from \cref{prop:uniform-multivariate-CLT}(B) with
$\xi_{\numobs n} = Z_{\numobs n} + Z_{\numobs 0}/\bsnumobs$ and $\bar\numobs = \numobs$ that 
\[
\kappa_{\numobs}^{\bbsym} \defined \dsupconvex(\mathcal{L}(\Delta^{\bbsym}_{\numobs}\given\datarvarg{\numobs}), \mathcal{L}(\Delta^{\bbsym}_{\infty})) \convP 0.  \label{eq:Delta-UMCLT}
\]
We can write the bagged posterior probability of model $1$ as 
\[ \label{eq:bagged-posterior-prob-1}
\bbmodelpostdistfull{\model_{1}}{\datarvarg{\numobs}} 
= \EE\{\logistic_{\bsnumobs}(W^{\bbsym}_{\numobs}) \given \datarvarg{\numobs}\}
= \EE\{\logistic_{\bsnumobs}(\Delta^{\bbsym}_{\numobs} + (\bsnumobs/\numobs)^{1/2}W_{\numobs}) \given \datarvarg{\numobs}\}.
\]
Let $\mcI_{\numobs} = \bigcup_{k=1}^{K-1} \mcI_{\numobs,k}$, where 
$\mcI_{\numobs,k} \defined \reals^{k-1} \times [-\eps_{\numobs}, \eps_{\numobs}] \times \reals^{K-k-1}$ is a convex set
with $\eps_{\numobs} = \bsnumobs^{-1/4}$.
Since the marginal densities of the components $\Delta^{\bbsym}_{\infty,1}, \dots, \Delta^{\bbsym}_{\infty,K-1}$ of $ \Delta^{\bbsym}_{\infty}$ are
bounded by a constant $b$, it follows that for any $\alpha \in \reals^{K-1}$, 
\[
\Pr(\Delta^{\bbsym}_{\numobs} + \alpha \in \mcI_{\numobs} \mid \datarvarg{\numobs}) 
\le \sum_{k=1}^{K-1} \big(\kappa_{\numobs}^{\bbsym} + \Pr(\Delta^{\bbsym}_{\infty} + \alpha \in \mcI_{\numobs,k})\big)
\le (K-1) (\kappa_{\numobs}^{\bbsym} + 2 b \eps_{\numobs}).
\]
By \cref{lem:indicator-approx-error}, $|\phi_{\bsnumobs}(t) - \ind(t > 0)| \le (K-1)\exp(- \bsnumobs^{1/4})$ for all $t \notin\mcI_{\numobs}$.
Thus,
\[ 
&\Big\vert\EE\{\logistic_{\bsnumobs}(\Delta^{\bbsym}_{\numobs} + (\bsnumobs/\numobs)^{1/2}W_{\numobs}) \given \datarvarg{\numobs}\}
- \EE\{\ind(\Delta^{\bbsym}_{\numobs} +  (\bsnumobs/\numobs)^{1/2}W_{\numobs} > 0) \given \datarvarg{\numobs}\}\Big\vert \notag\\
&\le (K-1)\exp(- \bsnumobs^{1/4}) +  (K-1) (\kappa_{\numobs}^{\bbsym} + 2 b \eps_{\numobs}) = \littleoP(1). \label{eq:E-difference-1}
\]
Moreover,
\[ 
&\Big\vert\EE\{\ind(\Delta^{\bbsym}_{\numobs} +  (\bsnumobs/\numobs)^{1/2}W_{\numobs} > 0) \given \datarvarg{\numobs}\}
- \EE\{\ind(\Delta^{\bbsym}_{\infty} +  (\bsnumobs/\numobs)^{1/2}W_{\numobs} > 0) \given \datarvarg{\numobs}\}\Big\vert  \notag\\
&\le \kappa_{\numobs}^{\bbsym} = \littleoP(1). \label{eq:E-difference-2}
\]
Combining \cref{eq:bagged-posterior-prob-1,eq:E-difference-1,eq:E-difference-2}, we have
\[
\bbmodelpostdistfull{\model_{1}}{\datarvarg{\numobs}} 
&= \EE\{\logistic_{\bsnumobs}(\Delta^{\bbsym}_{\numobs} + (\bsnumobs/\numobs)^{1/2}W_{\numobs}) \given \datarvarg{\numobs}\} \\
&= \EE\{\ind(\Delta^{\bbsym}_{\infty} +  (\bsnumobs/\numobs)^{1/2}W_{\numobs} > 0) \given \datarvarg{\numobs}\} + \littleoP(1)  \\
&= \EE\{\ind(-\Delta^{\bbsym}_{\infty} \le  (\bsnumobs/\numobs)^{1/2}W_{\numobs}) \given \datarvarg{\numobs}\} + \littleoP(1)  \\
&= \Phi_{0,\Sigma_{\infty}}((\bsnumobs/\numobs)^{1/2}W_{\numobs}) + \littleoP(1)  \\
&\convD \Phi_{0,\Sigma_{\infty}}(\bsscale^{1/2}W_{\infty}),
\]
where the last equality follows from the definition of $\Delta^{\bbsym}_{\infty}$,
and convergence in distribution follows from the assumption that $\bsnumobs/\numobs \to \bsscale$, \cref{eq:W-UMCLT}, and Slutsky's theorem.
\eprf

\subsection{Proof of \cref{cor:exchangeable-model-selection-asymptotic-posteriors}} \label{sec:proof-of-exchangeable-model-selection-asymptotic-posteriors}

Note that $\modelpostdistfull{\model_{1}}{\datarvarg{\numobs}} = \phi(\lmldiff{\datarvarg{\numobs}})$ and $\bbmodelpostdistfull{\model_{1}}{\datarvarg{\numobs}} = \EE\{\phi(\lmldiff{\bsdatarvarg{\bsnumobs}})\given \datarvarg{\numobs}\}$
where $\phi(t) = \{1 + \exp(-t)\}^{-1}$.
Under assumptions (i)-(iv), we have the asymptotic expansion~\citep{Clarke:1990,Dawid:2011}
\[
\lmldiff{\datarvarg{\numobs}} = \frac{1}{2}(D_{2} - D_{1}) \log \numobs + \sum_{n=1}^{\numobs} \log \frac{\lik{\obsrv{n} \given 1}{\optmodelparam{1}}}{\lik{\obsrv{n} \given 2}{\optmodelparam{2}}} + \bigoP(1).
\]
Letting $Z_{n} \defined \log\lik{\obsrv{n} \given 1}{\optmodelparam{1}} - \log\lik{\obsrv{n} \given 2}{\optmodelparam{2}}
= \loglik{\obsrv{n}}{1,\optmodelparam{1}} - \loglik{\obsrv{n}}{2,\optmodelparam{2}}$,
the conclusions follow as in the proof of \cref{thm:model-selection-asymptotic-posteriors}, although the argument is somewhat simplified by the fact that $\obsrv{1},\obsrv{2},\ldots$ are i.i.d., 
so we do not need to reason about triangular arrays and can use standard multivariate Berry--Esseen bounds \citep{Gotze:1991}.

\subsection{Proof of \cref{prop:uniform-multivariate-CLT}} \label{sec:proof-of-uniform-multivariate-CLT}

\bprf[Proof of Part \textup{(A)}]
Define $\mu_{\numobs} \defined \bar\numobs^{1/2}\EE(\xi_{\numobs 1})$ and $\Sigma_{\numobs} \defined  \cov(\xi_{\numobs 1})$.
By assumption, for all $\numobs$ sufficiently large, $\mu_{\numobs}\in\reals^{L}$ and $\Sigma_{\numobs}$ is positive definite;
for such $\numobs$, let $\txi_{\numobs n} \defined \Sigma_{\numobs}^{-1/2}\{\xi_{\numobs n} - \EE(\xi_{\numobs n})\}$,
otherwise, let $\txi_{\numobs n} \sim \distNorm(0, I)$.
Then $\EE(\txi_{\numobs n}) = 0$ and $\cov(\txi_{\numobs n}) = I$, the identity matrix. 
Letting $\tilde{W}_{\numobs} \defined  \bar\numobs^{-1/2}\sum_{n=1}^{\bar\numobs} \txi_{\numobs n}$,
\citet[][Theorem 1.1]{Raic:2019} shows that
\[ \label{eq:raic-bound}
\dsupconvex(\mcL(\tilde{W}_{\numobs}), \distNorm(0, I)) 
&\le (42 L^{1/4} + 16)\bar\numobs^{-1/2} \EE(\| \txi_{\numobs 1} \|_{2}^{3}).
\]
By Cauchy--Schwarz and the bound $\|x - m\|^3 \le (2 \max\{\|x\|,\|m\|\})^3 \le 8(\|x\|^3+\|m\|^3)$,
\[ \label{eq:third-moment-bound}
\EE(\| \txi_{\numobs 1} \|_{2}^{3}) \le \|\Sigma_{\numobs}^{-1/2}\|_{2}^{3}\, \EE(\|\xi_{\numobs 1} - \EE(\xi_{\numobs 1})\|_{2}^{3})
\le 16\|\Sigma_{\numobs}^{-1/2}\|_{2}^{3}\, \EE(\|\xi_{\numobs 1}\|_{2}^{3})
\]
when $\mu_{\numobs}\in\reals^{L}$ and $\Sigma_{\numobs}$ is positive definite.
Combining \cref{eq:raic-bound,eq:third-moment-bound} with (ii) and the assumption that $\bar\numobs^{-1/2} \EE(\|\xi_{\numobs 1}\|_{2}^{3}) \to 0$, we conclude that 
\[ \label{eq:intermediate-uniform-multivariate-CLT}
\lim_{\numobs \to \infty} \dsupconvex(\mcL(\tilde{W}_{\numobs}), \distNorm(0, I))  = 0. 
\]
Now, by the triangle inequality,
\[ \label{eq:triangle-ineq-multivariate-CLT}
\dsupconvex(\mcL(W_N),\distNorm(\mu,\Sigma)) \leq \dsupconvex(\mcL(W_N),\distNorm(\mu_N,\Sigma_N)) + \dsupconvex(\distNorm(\mu_N,\Sigma_N),\distNorm(\mu,\Sigma)).
\]
Note that $\tilde{W}_{\numobs} = \Sigma_{\numobs}^{-1/2} (W_{\numobs} - \mu_{\numobs})$ when $\mu_{\numobs}\in\reals^{L}$ and $\Sigma_{\numobs}$ is positive definite.
Thus, for the first term in \cref{eq:triangle-ineq-multivariate-CLT}, we have
\[ \label{eq:affine-trans-limit-multivariate-CLT}
\lim_{\numobs \to \infty} \dsupconvex(\mcL(W_N),\distNorm(\mu_N,\Sigma_N)) = \lim_{\numobs \to \infty} \dsupconvex(\mcL(\tilde{W}_N), \distNorm(0,I)) = 0
\]
by \cref{eq:intermediate-uniform-multivariate-CLT}
and the fact that $\dsupconvex$ is invariant under invertible affine transformations.
For the second term in \cref{eq:triangle-ineq-multivariate-CLT}, we have
\[ \label{eq:KL-bound-multivariate-CLT}
\dsupconvex(\distNorm(\mu_N,\Sigma_N),\distNorm(\mu,\Sigma)) 
&\le d_{\textsc{TV}}(\distNorm(\mu_N,\Sigma_N),\distNorm(\mu,\Sigma)) \\
&\le 2^{-1/2} \textsc{kl}(\distNorm(\mu_N,\Sigma_N),\distNorm(\mu,\Sigma))^{1/2}
\xrightarrow[\numobs \to\infty]{} 0,
\]
where $d_{\textsc{TV}}$ denotes total variation distance and $\textsc{kl}$ denotes Kullback--Leibler divergence.
The second step in \cref{eq:KL-bound-multivariate-CLT} is by Pinsker's inequality,
and the limit in \cref{eq:KL-bound-multivariate-CLT} is zero by the formula for the KL divergence between multivariate Gaussians,
along with assumptions (i) and (ii).
\eprf

\bprf[Proof of Part \textup{(B)}]
First, it is straightforward to verify that part (A) holds ``in probability'' in the following sense: If each $P_\numobs$ is a random distribution
and the assumed limits hold in probability, then $\dsupconvex(\mcL(W_{\numobs}), \distNorm(\mu, \Sigma)) \convP 0$.
Note that when verifying this extension of the proof of (A), one must handle the event $E_{\numobs} = \{\mu_{\numobs} \in \reals^{L},\, \Sigma_{\numobs} \text{ is positive definite}\}$;
this can be done by using $\Pr(E_{\numobs}^c) \to 0$ and in place of \cref{eq:affine-trans-limit-multivariate-CLT},
use the fact that for all $\varepsilon > 0$,
\[
\Pr(\dsupconvex(\mcL(W_N),\distNorm(\mu_N,\Sigma_N)) > \varepsilon) \le \Pr(\dsupconvex(\mcL(\tilde{W}_N), \distNorm(0,I)) > \varepsilon) + \Pr(E_{\numobs}^c)
\xrightarrow[\numobs \to\infty]{} 0.
\]

Thus, to prove part (B), we can apply part (A) 
with $\txi^{\bbsym}_{\numobs m} \defined \xi^{\bbsym}_{\numobs m} - \bar{\numobs}^{-1}\sum_{n=1}^{\bar{\numobs}} \xi_{\numobs n}$
in place of $\xi_{\numobs n}$ and $\bsnumobs$ in place of $\bar\numobs$,
since $\bsnumobs^{-1/2} \sum_{m=1}^{\bsnumobs} \txi_{\numobs m}^{\bbsym} = W^{\bbsym}_{\numobs} - (\bsnumobs/\bar{\numobs})^{1/2}W_{\numobs}$. 
To apply part (A), we need to establish that the assumed limits hold in probability.
Specifically, we show that
\begin{enumerate}
\setlength{\itemindent}{1em}
\item[(B.i)] $\bsnumobs^{1/2}\EE(\txi^{\bbsym}_{\numobs 1}\mid \xi) \convP 0$,
\item[(B.ii)] $\cov(\txi^{\bbsym}_{\numobs 1}\mid \xi) \convP \Sigma$, and 
\item[(B.iii)] $\bsnumobs^{-1/2} \EE(\|\txi^{\bbsym}_{\numobs 1}\|_{2}^{3}\mid \xi) \convP 0$.
\end{enumerate}

First, (B.i) holds trivially since $\EE(\txi^{\bbsym}_{\numobs m}\mid \xi) = 0$.
For (B.ii), note that $\cov(\txi^{\bbsym}_{\numobs m}\mid \xi) = \bar\numobs^{-1}\sum_{n=1}^{\bar\numobs} \xi_{\numobs n}\xi_{\numobs n}^\top - \bar\xi_{\numobs} \bar\xi_{\numobs}^\top$,
where $\bar\xi_{\numobs} \defined \bar{\numobs}^{-1}\sum_{n=1}^{\bar{\numobs}} \xi_{\numobs n}$.
Further, the assumption that $\limsup_{\numobs} \EE(\|\xi_{\numobs 1}\|_{2}^{6}) < \infty$ 
implies $\limsup_{\numobs} \EE(\|\xi_{\numobs 1}\|_{2}^{k}) < \infty$ for $1\le k \le 6$ as well.
Thus, $\bar\xi_{\numobs} \convP 0$ by the weak law of large numbers \citep[Theorem 2.24]{Durrett:2011}
applied to $\txi_{\numobs n} \defined \xi_{\numobs n} - \EE(\xi_{\numobs 1})$,
since $\limsup_{\numobs} \EE(\|\txi_{\numobs 1}\|_{2}^{2}) < \infty$ and assumption (i) implies $\EE(\xi_{\numobs 1}) \to 0$.
Similarly, $\bar\numobs^{-1}\sum_{n=1}^{\bar\numobs} \xi_{\numobs n}\xi_{\numobs n}^\top \convP \Sigma$ 
by the weak law of large numbers applied to the entries of $\xi_{\numobs n}\xi_{\numobs n}^\top - \EE(\xi_{\numobs n}\xi_{\numobs n}^\top)$,
using (i), (ii), and $\limsup_{\numobs} \EE(\|\xi_{\numobs 1}\|_{2}^{4}) < \infty$.
Therefore, we have $\cov(\txi^{\bbsym}_{\numobs m}\mid \xi) \convP \Sigma$.

Finally, to see (B.iii), observe that
\[
\bsnumobs^{-1/2} \EE(\|\txi^{\bbsym}_{\numobs 1}\|_{2}^{3}\mid \xi) 
\le 16 \bsnumobs^{-1/2} \EE(\|\xi^{\bbsym}_{\numobs 1}\|_{2}^{3}\mid \xi)
= 16 \bsnumobs^{-1/2} \bar\numobs^{-1} \sum_{n=1}^{\bar\numobs} \|\xi_{\numobs n}\|_{2}^{3}
\convP 0
\]
by the weak law of large numbers, since $\limsup_{\numobs} \EE(\|\xi_{\numobs 1}\|_{2}^{6}) < \infty$ and 
$\bsnumobs = \bsnumobs(\numobs) \to \infty$.
\eprf

\end{document}